\documentclass{ieeeaccess}
\usepackage{cite}
\usepackage{amsmath,amssymb,amsfonts}
\usepackage[english]{babel}
\usepackage[utf8]{inputenc}
\usepackage{algorithm}
\usepackage{arevmath}     
\usepackage[noend]{algpseudocode}
\ifCLASSINFOpdf
  \usepackage[pdftex]{graphicx}
  \graphicspath{{../figs/}}
  \DeclareGraphicsExtensions{.pdf,.jpeg,.png,.eps,.jpg}
\else
\fi
\usepackage{textcomp}
\usepackage{array}
\usepackage{longtable}
\usepackage{multirow}
\usepackage{supertabular}
\usepackage{flushend}
\usepackage{tabu}
\usepackage{lmodern}
\usepackage[utf8]{inputenc}
\usepackage[T1]{fontenc}
\usepackage{makecell}
\usepackage{lmodern}
\usepackage{adjustbox}
\usepackage{textcomp}
\usepackage{array}
\usepackage{tabularx,tabu}
\usepackage{multirow}
\usepackage{subcaption}
\usepackage{stfloats}
\usepackage{threeparttable}
\usepackage{comment}
\usepackage{array}
\usepackage{booktabs}
\usepackage{lscape}
\newcommand{\hak}[1]{\textcolor{black}{#1}}

\usepackage{subcaption}  
\usepackage{caption,setspace}
\def\BibTeX{{\rm B\kern-.05em{\sc i\kern-.025em b}\kern-.08em
    T\kern-.1667em\lower.7ex\hbox{E}\kern-.125emX}}
    
\begin{document}
\history{Date of publication xxxx 00, 0000, date of current version xxxx 00, 0000.}
\doi{10.1109/ACCESS.2021.3093442}

 \title{A Systematic Review of Bio-Cyber Interface Technologies and Security Issues for Internet of Bio-Nano Things}
\author{
    \uppercase{
                Sidra Zafar\authorrefmark{1}, 
                Mohsin Nazir\authorrefmark{1}, 
                Taimur Bakhshi\authorrefmark{2},
                Hasan Ali Khattak\authorrefmark{3},}
                    \IEEEmembership{Senior Member, IEEE},
        \uppercase{
                Sarmadullah Khan\authorrefmark{4},
                Muhammad Bilal\authorrefmark{5,*},}
                    \IEEEmembership{Senior Member, IEEE},
        \uppercase{
                Kim-Kwang Raymond Choo\authorrefmark{6}, }       \IEEEmembership{Senior Member, IEEE}
        \uppercase{ 
                Kyung-Sup Kwak\authorrefmark{7,*}} \IEEEmembership{Life Senior Member, IEEE} and
        \uppercase{
                Aneeqa Sabah\authorrefmark{8}
            }
    }
    
\address[1]{Department of Computer Science, Lahore College for Women University, Lahore 54000, Pakistan (email: sidra.zafar@lcwu.edu.pk, mohsin.nazir@lcwu.edu.pk)}

\address[2]{
Center for Information Management \& Cyber Security, National University of Computer \& Emerging Sciences, Lahore, Pakistan. (email: taimur.bakhshi@nu.edu.pk)}

\address[3]{School of Electrical Engineering and Computer Science, National University of Sciences and Technology (NUST), Islamabad, Pakistan (email: hasan.alikhattak@seecs.edu.pk)}

\address[4]{School of Computer Science and Informatics, De Montfort University Leicaster, LE1 9BH, UK.; (email: sarmadullah.khan@dmu.ac.uk)}

\address[5]{Deptartment of Computer Engineering, Hankuk University of Foreign Studies, Yongin-si, Gyeonggi-do, 17035, Korea.; (email: mbilal@hufs.ac.kr)}

\address[6] {Department of Information Systems and Cyber Security, University of Texas at San Antonio TX 78249, USA. (email: raymond.choo@fulbrightmail.org)}

\address[7]{Department of Information and Communication Engineering, Inha University, Incheon, 22212, Korea (email: kskwak@inha.ac.kr)}

\address[8]{Department of Physics, Lahore College for Women University,Lahore,54000, Pakistan (email:aneeqas29@gmail.com)}

\markboth
{S. Zafar, M. Nazir, T. Bakshi, HA. Khattak \headeretal:}
{Security in the Internet of Bio-Nano Things: Challenges and Research Opportunities}

\tfootnote{This work was supported by National Research Foundation of Korea-Grant funded by the Korean Government (Ministry of Science and ICT)-NRF-2020R1A2B5B02002478).}

\corresp{Corresponding authors:  Muhammad Bilal and Kyung-Sup Kwak}

\begin{abstract}
Advances in synthetic biology and nanotechnology have contributed to the design of tools that can be used to control, reuse, modify, and re-engineer cells’ structure, as well as enabling engineers to effectively use biological cells as programmable substrates to realize Bio-NanoThings (biological embedded computing devices). Bio-NanoThings are generally tiny, non-intrusive, and concealable devices that can be used for \textit{in-vivo} applications such as intra-body sensing and actuation networks, where the use of artificial devices can be detrimental. Such (nano-scale) devices can be used in various healthcare settings such as continuous health monitoring, targeted drug delivery, and nano-surgeries. These services can also be grouped to form a collaborative network (i.e., nanonetwork), whose performance can potentially be improved when connected to higher bandwidth external networks such as the Internet, say via 5G. However, to realize the IoBNT paradigm, it is also important to seamlessly connect the biological environment with the technological landscape by having a dynamic interface design to convert biochemical signals from the human body into an equivalent electromagnetic signal (and vice versa). This, unfortunately, risks the exposure of internal biological mechanisms to cyber-based sensing and medical actuation, with potential security and privacy implications. This paper comprehensively reviews bio-cyber interface for IoBNT architecture, focusing on bio-cyber interfacing options for IoBNT like biologically inspired bio-electronic devices, RFID enabled implantable chips, and electronic tattoos. This study also identifies known and potential security and privacy vulnerabilities and mitigation strategies for consideration in future IoBNT designs and implementations.
\end{abstract}


\titlepgskip=-15pt

\maketitle

\section{Introduction}
With recent pandemics and the associated lockdown regime, there has been renewed, if not accelerated, interest, in exploring electronic and remote delivery of healthcare services. One of the recent trends is in the Internet of Bio-Nano Things (IoBNT) systems, which comprise biological nanonetworks that sense biological and chemical changes in the environment (i.e., human body) and send the collected data over the Internet to data centers for further processing. These nanonetworks can also perform medical actuation from the commands sent remotely by the relevant healthcare providers. Typically, biological nanonetworks consist of nanosized computing devices (also referred to as 'Bio-Nano Things'), which work collaboratively to achieve sensing and actuation tasks within the deployed environment\cite{steager2013automated,kim2009microengineered,ergeneman2008magnetically,nakamura2008capsule,dubach2007fluorescent,chen2005development,carlsen2014bio,aylott2003optical}. 

Supported tasks include targeted drug delivery (TDD), continuous health monitoring, tissue engineering, and tumor detection
\cite{timko2010remotely,pan2011swallowable,fernandez2007magnetic,liao2010indications,li2003cholesterol,kawahara2013chip,fluckiger2007ultrasound,freitas2006pharmacytes}. Conventional wireless communication technologies are generally not suitable to support communications within the nanonetworks due to limitations in transceiver size, computation capability, and propagation channel of the bio-nano things. This necessitates the design of novel communication technologies for these nanonetworks \cite{akyildiz2008nanonetworks, marzo2019nanonetworks}, and examples include molecular communication (MC) \cite{felicetti2014tcp,nakano2012molecular}, nano electromagnetic (EM) communication at Tera Hz band, acoustic and nanomechanical communication \cite{akyildiz2010electromagnetic,akyildiz2010internet}.

To ensure the smooth operation of IoBNT devices and systems, seamless connections of intra body nanonetworks and external networks (such as the Internet) are essential in supporting more complex, real-world healthcare applications. External access to intrabody nanonetworks requires a hybrid interface device that understands both paradigms' communication protocols, specifically biochemical signals generated by intrabody nanonetworks and electromagnetic signals received from the Internet. Bio-cyber interfacing can be summarily defined as a set of operations performed in sequence to convert biochemical signals received from intra-body nanonetworks into electrical signals for the cyber domain of Internet and \textit{vice versa} \cite{velloso2011web}. In other words, the design and modeling of the bio-cyber interface are crucial yet challenging in IoBNT implementation.

Interdisciplinary research efforts devoted to the development of these bio-electronic devices have resulted in frameworks such as second skin \cite{michel2017optogenerapy}, bio-cyber interface \cite{akyildiz2015internet}, and bio driver \cite{kirichek2016live}. Bio-electronic interfaces may be composed of materials characterized by electromagnetic properties, which can alter
their formation in the presence of biological molecular complexes and accordingly modulate current in the
electrical circuit. Popular examples of bio-electronic interfacing include wireless chemical sensors, RFID devices, and electronic tattoos.

\begin{table*}[!bp]
    \centering
    \caption{The comparison of existing literature and surveys on IoBNT}
    \label{tab:tabsurvey}
    \setlength{\extrarowheight}{10pt}

    \setlength{\tabcolsep}{3pt}

    \begin{tabu} to \linewidth {X[0.4]X[2.3]X[1.8]X[1.5]X[1]X[4]}
    \hline
\textbf{Ref} &
    \textbf{Networking Domain} &
    \textbf{Main Focus} &
    \textbf{Bio Cyber Interfacing} &
    \textbf{Security} &
    \textbf{Description} \\ \hline
    \cite{kulakowski2020nano} &
  Nano Communication Networks, Body Area Networks &
  Interfacing options for nano-micro communication &
  \checkmark &
  X &
  Identifies potential materials and molecules that can be used for interfacing between Nano and micro communication technologies. \\
\cite{usman2018security} &
  Wireless Body Area Networks &
  Security &
  X &
  \checkmark &
  Description of Four-tiered architecture of WBAN, presentation of security issues in each layer. \\
\cite{loscri2014security} &
  Molecular Communication(MC) &
  Security in MC &
  X &
  \checkmark &
  Comprehensive discussion on security and privacy issues of Molecular Communication and their potential countermeasures. \\
\cite{akyildiz2010internet} &
  IoNT &
  Roadmap paper &
  X &
  X &
  Early investigation of IoNT to explore enabling technologies, architecture, challanges and opportunities in IoNT \\
\cite{yang2019comprehensive} &
  IoNT and body-centric networking &
  Survey paper &
  \checkmark &
  * &
  Overview of IoNT technology, and introduction of hybrid communication scheme for nano micro interfacing. \\
\cite{al2020intelligence} &
  5G oriented IoNT &
  Artificial Intelligence and security in IoNT &
  X &
  \checkmark &
  This article provides a critical overview of the IoNT considering the main application areas, architecture, limitations, and design factors. Comprehensive overview of employing Artificial Intelligence for IoNT. Moreover, a brief discussion of attack types and attackers on IoNT. \\
\cite{akyildiz20206g} &
  6G and next generation wireless networks &
  Roadmap paper &
  * &
  X &
  Usecases,enabling technologies and key drivers for 6g and beyound technologies i.e., IoNT, IoBNT and quantum computing \\
\cite{silva2020domiciliary} &
  IoMT &
  Review paper &
  * &
  X &
  The technological challenges \& solutions for wearable bio-electronics for patient monitoring and domiciliary hospitalization \\
\cite{williams2020electronic} &
  Electronic Tattoos &
  Review paper &
  * &
  X &
  The materials and engineering requirements, fabrication developments, and sensing and therapeutic advances in electronic tattoos \\
Our study &
  IoBNT &
  Review paper &
  \checkmark &
  \checkmark &
  Interdisciplinary research on bio cyber interfacing options for IoBNT. -Security for IoBNT\\ \hline
  \multicolumn{5}{p{275pt}}{-Legends \checkmark $=$ Completely addressed X $=$ Not addressed * $=$ Partially addressed  }\\

    \end{tabu}
\end{table*}

Wireless chemical sensors sense the changes in levels of chemical substrates of the object and send them wirelessly over the Internet to processing applications \cite{oh2019second,kassal2013wireless,kassal2018wireless}. Similarly, inkjet printable electronics which use nanomaterials based ink, have been utilized to facilitate fabrication of RFID enabled implantable devices that sense bodily parameters and send them to some nearby devices (e.g., mobile devices) via radio-frequency identification (RFID) \cite{singh2017inkjet}. Moreover, electronic tattoo based stick-able sensors can be skin-mounted to collect epidermal and sweat gland information, which can be used to facilitate continuous patient monitoring \cite{bandodkar2015tattoo}.

One of the many challenges in designing bio-cyber interfaces is how to accurately model molecular information collected by intra-body sensors into an equivalent electromagnetic signal. Several factors are limiting the wider adoption of these devices, for example, data security and user safety. For example, the capability to access our human body through IoBNT applications can be exploited to carry out "bio-cyber terrorism" \cite{akyildiz2015internet}, where attackers can identify and exploit vulnerabilities in these IoBNT applications and/or their underpinning infrastructure to directly manipulate human body functionality, (covertly) exfiltrate user information, or cause the human body to malfunction. The feasibility of conventional, as well as novel security measures, needs to be established for these (resource-constrained) devices, and lack of trust in security and privacy will limit the adoption of such systems. This necessitates the design of security measures to minimize the impact of a cyber-attack and unlawful profiling and surveillance of individuals. This work is a step towards providing a roadmap showing what has been done in IoBNT security and what needs to be done.

\subsection{EXISTING LITERATURE REVIEWS AND SURVEYS} \label{subsection:EXISTING LITERATURE REVIEWS AND SURVEYS}
The idea of integrating biological cells into the communication engineering perspective was first proposed by Akyldiz et al. \cite{akyildiz2015internet}. The authors described the architecture of IoBNT and its potential applications, such as collecting vital physiological parameters from the human body and transmitting them to the remote healthcare provider. They also described the enabling technologies to realize the IoBNT paradigm and focused on molecular communication as the most promising intrabody communication technology. Furthermore, the challenges in developing safe and efficient techniques for the exchange of information and the development interface between the biological and cyber world (i.e., bio-cyber interface). Challenges and opportunities were presented.

{ Another article \cite{kulakowski2020nano} overviews Nano communications, BAN (Body Area Network) communications, and the potential mechanisms that can be used for the interfacing between nano-micro-macro communications. This survey article identifies specific materials and molecules such as Light stimulated channelrhodopsins, BRET(Bioluminescence Resonance Energy Transfer), ATP(Adenosine Triphosphate), Photodetectors, and SPR (Surface Plasmon Resonance), which can be used as part of bio cyber interface to achieve the overall interfacing between nano and micro networking domains. Unlike our work, this survey article does not provide any insights into the bio cyber interface's complete design and prospective components. Moreover, this article does not discuss the security perspective of the communication between nano-micro and macro devices.

 A roadmap paper has been presented in \cite{ akyildiz20206g}, which surveys the key enabling technologies of next generation wireless communications. This paper provides a detailed insight into the vital technologies that will play an integral role in 6G networks and beyond, provides use cases of applications that will be enabled by 6G and the open research challenges in this domain. Moreover, this paper also review the early stage technologies such as Internet of Nano Things (IoNT), Internet of Bio Nano Things (IoBNT), and quantum communication, which are expected to benefit from 6G and beyond communications. This paper presents an overall picture of the next generation wireless technologies, discussing every enabling technology in general, whereas our research specifically focuses on a detailed and comprehensive overview of one of the 6G enabling technologies i.e.,  IoBNT.}
 
Other surveys from related fields include those focusing on Internet of Nano-Things (IoNT) \cite{akyildiz2010internet}. The authors introduced the IoNT architecture, and explained the individual components and functionality from an information-theoretic perspective. Also, it presents research challenges in terms of channel modeling, information encoding and protocols for nanonetworks and IoNT were presented. 

Another survey on IoNT in the context of body centric communication was presented in \cite{yang2019comprehensive}. Specifically, the authors described the IoNT architecture and proposed a novel hybrid communication medium for body centric networks. The hybrid communication scheme proposed using both MC and EM communication in parallel, for nano micro interfacing, to achieve efficient results. The authors also briefly discussed existing security approaches for IoNT.

Qadri et al \cite{qadri2020future} discussed the possible emerging technologies for healthcare IoT (H-IoT), the options for enabling H-IoT (e.g., IoNT was identified as one of the enablers for healthcare applications). Another review on healthcare applications of IoNT was presented by Pramanik et al. \cite{pramanik2020advancing}, who presented a taxonomy of nanotechonolgy, nanoparticles and nanozymes, types and fabrication options for biosensors, and bionanosensors for healthcare applications of IoNT.

{ In the field of electronic tattoos, a review paper \cite{williams2020electronic} presents the materials and engineering requirements, fabrication developments, and sensing and therapeutic advancements of electronic tattoos. According to the authors, there are three components of a theragnostic-based electronic tattoo i.e., Sensing components for diagnostics, supporting electronics for data transmission, and drug delivery component.} 

{Another survey paper \cite{bernal2019security} that presents state-of-the-art in security and safety issues of the Brain Computer Interaction. This work thoroughly identifies some novel cyberattacks, thier impact on security and safety and their countermeasures. Some of the attacks like misleading stimuli are strictly related to BCI as it indicates alteration of the nuerosignals generated by the patient. Other attacks and counter measures are generic for the fields of IoT, IMD, and even IoBNT. This paper is provides exceptional insights into the security aspect. The main difference in our study and this survey paper is that the purpose of our study is to cover some very important architectural and design aspects of IoBNT, specifically bio cyber interfaces, along with security issues of IoBNT.} 
 
To the best of the authors' knowledge, there is no comprehensive review of IoBNT security solutions in the literature; nevertheless, the following research from similar fields such as IoMT and BCI (Brain Computer Interaction) provide some preliminary insights into the security challenges. 
{Keeping in view the recent pandemic and COVID-19 situation a review paper\cite{silva2020domiciliary} has been published recently, which discusses the technological challenges and solutions for wearable bio-electronics for patient monitoring and domiciliary hospitalization. In the end, they have presented a case study application of the Internet of Medical Things(IoMT) for domiciliary hospitalization of COVID 19 patients.}

However, there have been no unified works that have presented a unified view of the protocol stack models \cite{qadri2020future}.
A survey on WBAN (Wireless Body Area Networks) was presented in \cite{usman2018security}, which focuses only on the security aspects. The authors proposed a four-tier architecture for WBAN and identified communication technologies for each tier. The communication technologies and devices of WBAN are similar to that of IoBNT. 
 
\begin{figure*}[!bp]
\includegraphics[width=0.8\linewidth]{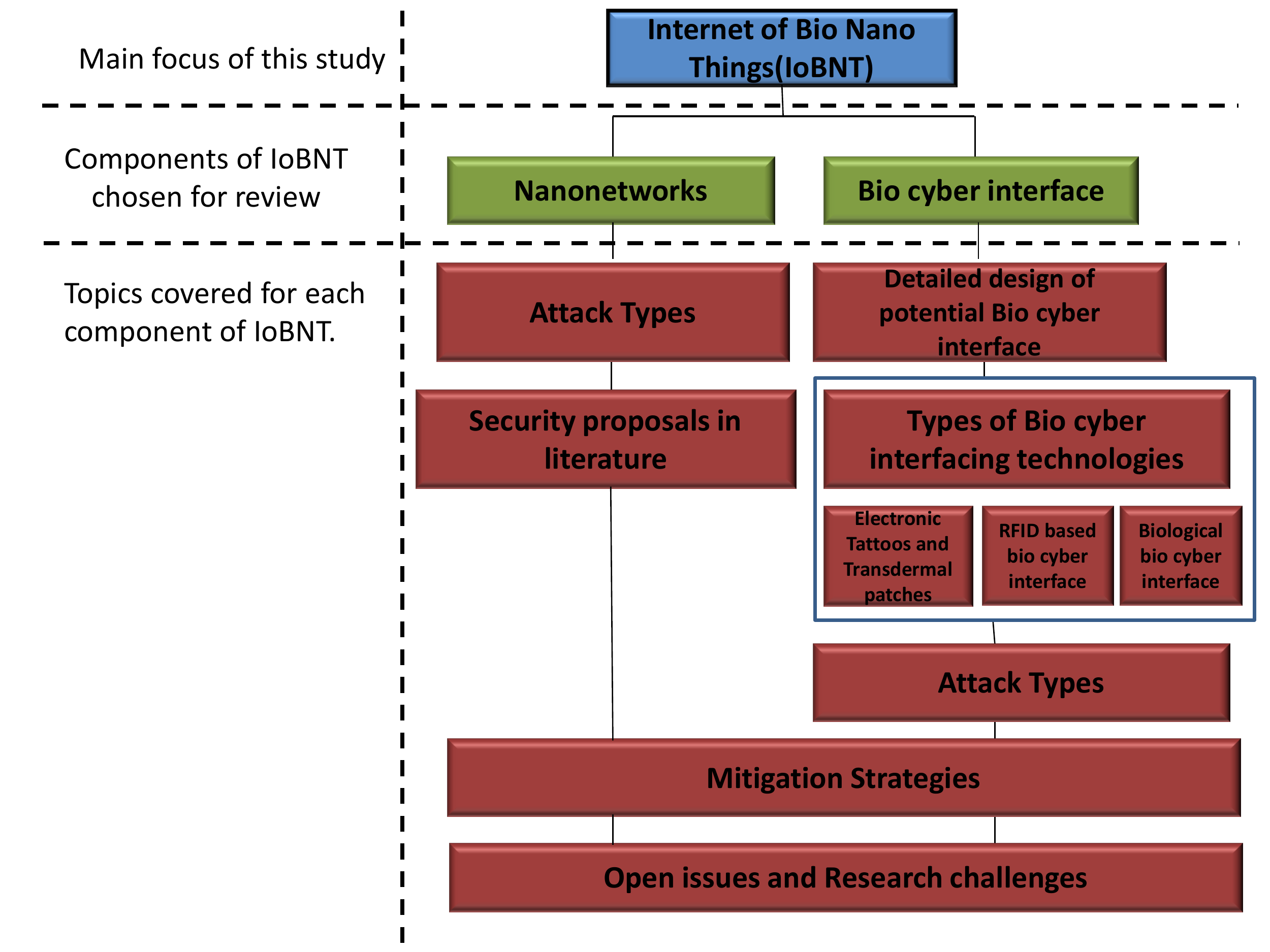}
\centering
\caption{\hak{Overview of topics covered in this article.}}
\label{fig:overview}
\end{figure*} 

Molecular communication is the key and most promising enabling technology for the healthcare applications of IoBNT \cite{dissanayak2021exact}. A comprehensive review paper is presented by authors in \cite{loscri2014security}. This paper presents a layered architecture of Molecular Communication and identifies the possible attack types and possible counter measures in each layer.

In \cite{al2020intelligence}, the authors studied the security of big 5G-oriented IoNT and the potential utility of machine learning to deal with the (big) data generated by the large number of IoNT devices. The authors outlined the potential security attackers and attack types. The comparison of the existing surveys and this study is presented in Table \ref{tab:tabsurvey}.

\subsection{PAPER SCOPE AND CONTRIBUTIONS}
{This paper provides a comprehensive review of published research on IoBNT and the security and privacy considerations that it raises. To do so, relevant keywords were used to search published papers in databases such as IEEExplore, Science Direct, ACM Digital Library, and Google Scholar. The main keywords or combination of keywords used were "Internet of Bio Nano Things", "Bio cyber Interface", "Electronic Tattoos", "Wearable Internet of Things (IoT)", "Radio Frequency Identification (RFID)", "IoBNT Security, and privacy", "Cyberattacks in IoT", and "Nano networks security", published within 2010-2020 timeline. This resulted in around 80 papers after filtering by categories, topic relevance, time of publication, and contributions.

The distribution of papers across journals shows that the papers were mainly published in journals that cover interdisciplinary topics such as electrochemistry, biotechnology, wireless communication, theragnostics, and optogenetics.}
While there have been a number of survey and review articles on closely related fields (i.e., IoNT,IoMT and WBAN) (see Section \ref{subsection:EXISTING LITERATURE REVIEWS AND SURVEYS}), as well as those on the underlying communication engineering principles (e.g., channel modeling, modulation techniques, enabling and communication technologies, and
other networking protocols which are applicable to IoBNT)  \cite{yang2019comprehensive, al2020intelligence, pramanik2020advancing}, there has been no survey or review article that focuses on IoBNT challenges and research opportunities.

This paper identifies and highlights the key design issues in the IoBNT implementations (e.g., interface design options between the biological and cyber world) and the related security challenges.  A lot of prominent researches in the field of IoBNT and next generation technologies have emphasized on interdisciplinary research efforts to make these novel systems holistic and practical \cite{akyildiz2015internet,akyildiz20206g,kulakowski2020nano}. Therefore, this article's main contributions are to bring together insights from disciplines like physics, biology, optogenetics, and electrochemitry to explore more potential bio interfacing mechanisms that are being used in these fields and their applicability to IoBNT domain.

\hak{
 In summary this work aims to make following contributions:}
\begin{itemize}
\color{black}
\item To provide a detailed design of the potential bio cyber interfacing device for IoBNT applications.
\item To survey three different bio cyber interfacing technologies namely, biological bio cyber interface, electronic tattoos and RFID based bio cyber interface.
\item To discuss various security issues and concerns that are specific to each bio cyber interfacing technology.
\item To provide detailed discussion on security of IoBNT and individuating attacks and threats in IoBNT components i.e., nanonetworks and bio cyber interface.
\item To provide novel and potential mitigation strategies for the security of IoBNT.
\item To present open issues, challenges, and future research directions involving IoBNT and its security.
\end{itemize}

A detailed elaboration of the contributions and covered topics in this article is depicted in Figure \ref{fig:overview}.

\begin{figure*}[!bp]
\includegraphics[width=\linewidth]{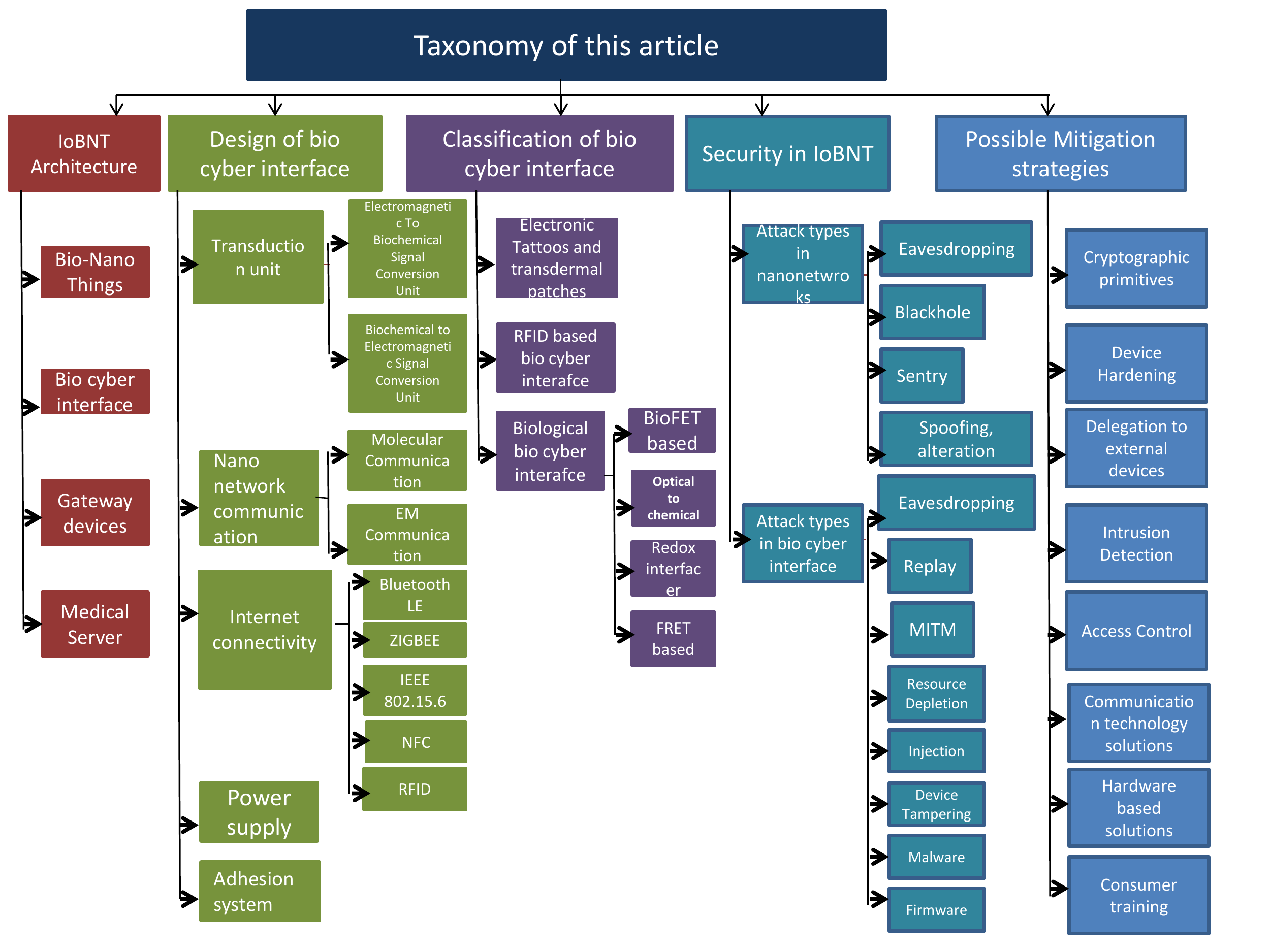}
\centering
\caption{\hak{Taxonomy of Systematic Review of Bio-Cyber Interface Technologies and Security Issues in the Internet of Bio-Nano Things.}}
\label{fig:taxonomy}
\end{figure*}

\subsection{ARTICLE STRUCTURE}
The rest of the paper is organized as follows. The architecture of IoBNT is described in the next section. The bio-cyber interface and the taxonomy of bio-cyber interface devices are introduced in Sections  \ref{section:BIO CYBER INTERFACE ARCHITECTURE} and \ref{section:CLASSIFICATION OF BIO CYBER INTERFACE}, respectively. The potential security threats and mitigation techniques connected with IoBNT are discussed in Sections \ref{section:SECURITY IN IoBNT} and \ref{section:POTENTIAL MITIGATION STRATEGIES}, respectively. The last two sections address the problems and future research prospects, as well as the paper conclusion. Figure \ref{fig:taxonomy} depicts the overall taxonomy of this paper.

\section{INTERNET of BIO NANO THINGS ARCHITECTURE}


\begin{figure*}[!ht]
\centering
\includegraphics[width=0.80\textwidth]{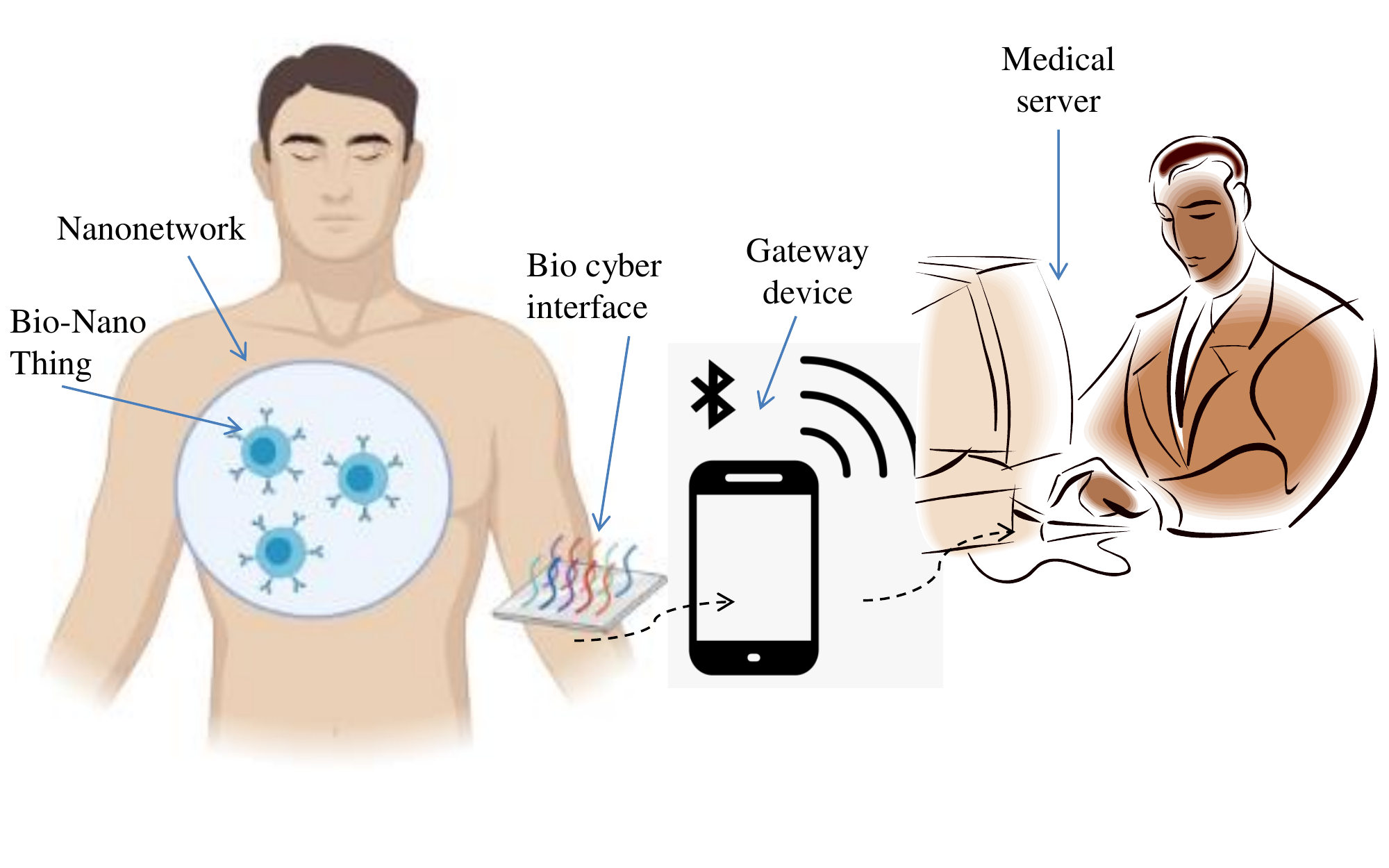}
\caption{\hak{A Typical Internet of Bio-Nano Things Architecture, where a bio-chemical signal from inside the human body is converted into electromagnetic signal \textit{via} bio cyber interface, and transmitted through Bluetooth or equivalent technology towards medical server for further analysis and processing}} \label{fig:IoBNT Architecture}
\end{figure*}

As previously discussed, the IoBNT network helps to sense biological and chemical changes around the environment and send the aggregated data to the data center for further processing \cite{al2017cognitive}. To realize IoBNT, heterogeneous devices need to collaboratively work together at nano- to macro-scale, via some interfaces between the electrical domain of the Internet and the biochemical domain of the IoBNT networks \cite{akyildiz2015internet}. For example, as shown in Fig. \ref{fig:IoBNT Architecture}, the major components in a IoBNT communication network include bio-nano things, nanonetwork, bio-cyber interface, gateway devices, and some medical server(s). The components of IoBNT architecture are elaborated below:

\subsection{BIO-NANO THINGS}
Bio-nano things are nanodevices that are not only a computing machine that can be reduced to a few nanometers in size, but also a device that uses the unique properties of nanocells and nanoparticles to detect and measure new types of phenomena at nanoscale. For example, nanodevices can detect chemical compounds in a fraction of a billion.
 \cite{roman2004single,schedin2007detection}, or the presence of different infectious agents such as virus or harmful bacteria \cite{tallury2010nanobioimaging,yeh2009real}.
 
Nanodevice is essentially comprise of a number of hardware constituents and all the software and programming of the nanodevice is included in the information processing unit.
There are two types of Nanodevices, electronic nanodevice, and biological nanodevice. Electronic nanodevices use novel nanotechnology materials like Carbon Nano Tubes (CNT) \cite{buther2017formal,pereira2011electronic} and Graphene Nano Ribbons \cite{akyildiz2008nanonetworks} for device construction. Biological nanodevices are built using the tools from nanotechnology and synthetic biology.  

Biological nanodevices can be fabricated by reprogramming biological materials \cite{yoo2011bio,nishimura2014genetic} like cells \cite{tan2015cell}, viruses \cite{lockney2011viruses,esfandiari2016new}, bacteria \cite{steidler2004live}, bacteriophage \cite{yacoby2007targeted},
erythrocytes (i.e., a red blood cell), leukocytes (a type of
white blood cell) and stem cells \cite{batrakova2011cell,su2015design} or by artificially synthesizing biomolecules like liposome, nanosphere, nanocapsule, micelle, dendrimer, fullerene and deoxyribonucleic (DNA) capsule. Moreover, hybrid nanodevices can be fabricated by applying both the above-mentioned approaches.  The size of nano devices may range from macro molecule to typical biological cell.

\begin{figure*}[!htbp]
\centering
\includegraphics[width=\textwidth]{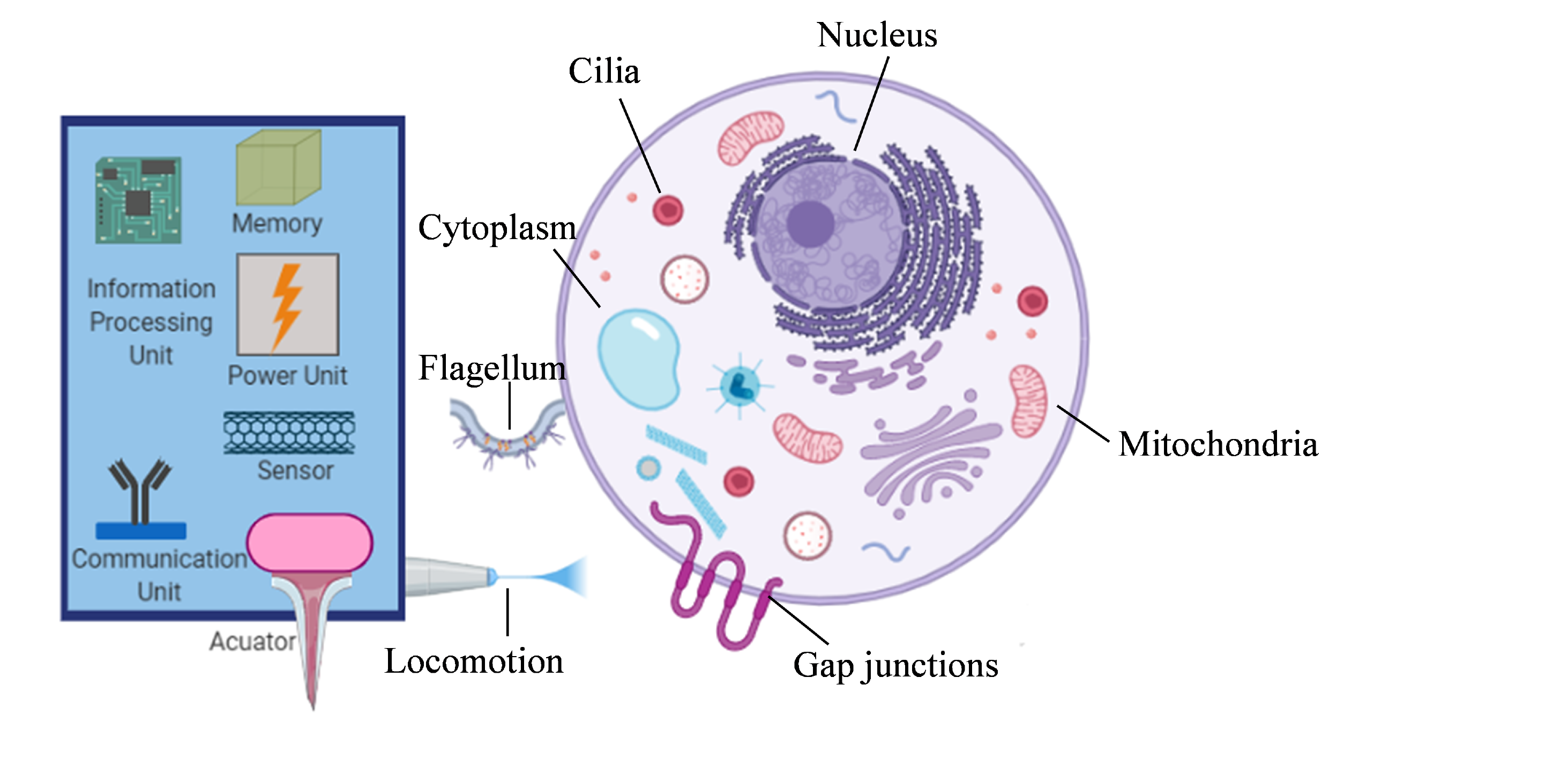}
\caption{Bio-nano thing : A mapping between the components of a typical IoT embedded computing device, and the elements of a biological cell.}
\label{fig:cell}
\end{figure*}

The material used to compose nano devices can be only biological \cite{strobel2021richard} (proteins, DNA sequences, lipids, biological cell) or they can be synthesized with non-biological materials such as magnetic particles and gold nanorods. There are a number of naturally occurring nano devices e-g protein motors that bind certain types of molecule on cargo transport them through filaments and unbind them at destination, liposomes capable of storing and releasing certain types of molecules, biological cells coated with non-biological material for non-cell native functions(e-g absorbing mercury). An envsioned bio-nano thing is presented in Figure \ref{fig:cell}.

\subsection{NANO NETWORK}
Nano network is comprised of several nano-scale devices such as nano transmitter, nano receiver, nano router, and other specialized nanodevices to perform exclusive tasks like sensing, actuation, monitoring, and control. Intra-body nanonetwork is generally deployed in the environment (e.g.,  human body orally or through injection) to realize \textit{invivo} biomedical applications. The nanonetwork consists of devices (e.g., nanomachine, nano router, and nano micro interface) at the nanoscale (1-100 nm), which work collaboratively to achieve sensing and actuation tasks \cite{akyildiz2008nanonetworks}. Nanomachines can be interconnected to execute collaborative tasks in a distributed manner. The size of nanodevices makes them feasible for \textit{in vivo} applications, where these non-invasive machines can easily be placed in hard-to-access areas (e.g., deep inside tissue) and perform therapeutic operations. 

\subsection{BIO CYBER INTERFACE}

Nano micro interface is mediating device between nano and micro scale communication; it is referred to as bio-cyber interface and bio electronics device throughout this article to keep the generality. Basically, this device converts electrical signals received as commands from the medical server of healthcare providers into biochemical signals understandable by intra-body nanonetworks and \textit{vice versa}\cite{koucheryavy2021review}. A detailed description of bio cyber interface design is presented in Section \ref{section:BIO CYBER INTERFACE ARCHITECTURE}
    
\subsection{GATEWAY DEVICES}
Gateway devices are now an integral part of most IoT applications, where they function as a relay device between sensors and the Internet. Gateway devices are computing devices such as smartphones, tablets, laptops, or mini-computers. A gateway device supports the mobility of patients and ensures efficient signal reception of low transmission range technologies (NFC, RFID, Bluetooth LE, etc). The wearable bio-electronic devices are usually designed to be compatible with the operating system requirements of smartphones \cite{oh2019second,kassal2017smart}. For RFID enabled applications,smartphones integrated with the RFID readout unit, which eliminates the need for external RFID tag readers. Similarly, for bio-electronic devices enabled with Bluetooth LE as the communication modality, smartphones are the ideal choice.
 Access point can be a cellular base station or a WiFi access  point, which helps to route the body sensor‘s traffic to the medical server.
 
 \subsection{MEDICAL SERVER}
  Medical server acts as a repository in which all the sensory data collected and sent from the patient‘s body is stored, analyzed and processed. This server may act as a terminal for real-time and continuous health monitoring, where emergency situations can be mitigated by sending alert messages. Only authorized entities can access this server to send commands and receive collected data, due to the critical nature of biomedical applications.

\section{A DETAILED DESIGN OF POTENTIAL BIO CYBER INTERFACE DEVICE} \label{section:BIO CYBER INTERFACE ARCHITECTURE}
Bio-cyber interface is the hybrid and most sophisticated device, that is capable of communicating between nanoscale and microscale devices \cite{akyildiz2010internet,akyildiz2015internet,al2017cognitive}. This device receives aggregated data from nanonetwork, processes nanoscale data in its transduction unit to convert it into a format suitable for the conventional network (e-g Internet) and sends it microscale devices. The components of envisioned bio-cyber interface are described in detail
below:

\subsection{TRANSDUCTION UNIT}
The transduction unit in the bio-cyber interface performs the operation of converting electrical signals from external devices into biochemical signals readable by Bio-Nano Things \cite{chude2016biologically}. Transduction properties in bio-cyber interface can be achieved by engineering devices using biomaterials, artificially synthesized biomaterials, non-bio materials or by combining all the approaches to create hybrid design \cite{nakano2014externally}. The transduction unit is further divided into two constituent parts, Electro-bio transduction unit, and Bioelectro transduction unit.

\subsubsection{ELECTROMAGNETIC TO BIOCHEMICAL SIGNAL CONVERSION UNIT}
Electro – Bio Transduction Unit converts the electrical signals, received from external devices into biochemical signals, and transmits them to intra-body  nanonetwork for further processing. This unit consists of a decoder, a drug storage unit, an external physical effect source, and an injection chamber. The decoder receives the signal transmitted from an external device and derives logic gates. The logic gates are binary commands that produce some physical effect in the environment like thermal, optical or magnetic radio-frequency (RF) signals. The physical effect source (thermal, optical or magnetic field) placed around the drug storage unit, stimulates the nanomachines to release their content in response to external changes in the environment. The injection chamber injects the released molecules into the blood vessel network. 

\begin{figure}[!htpb]
\includegraphics[width=\linewidth]{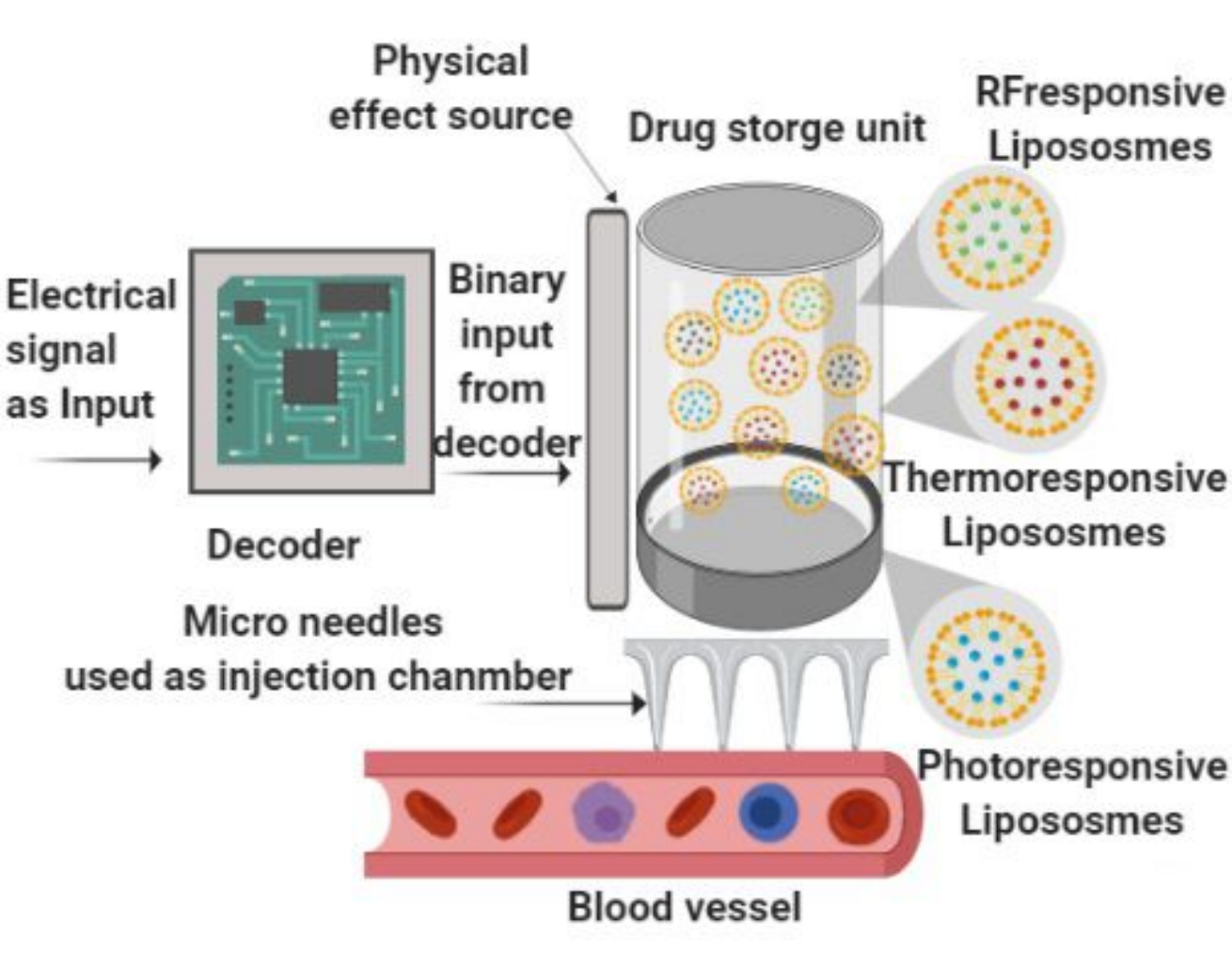}
\centering
\caption{Schematic illustration of Electromagnetic to Biochemical signal conversation unit. The binary code is received from medical server, the decoder converts the binary signal and pass it to physical effect source which in turn activates the drug storage unit and release the drug.  }
\label{fig:electrobio}
\end{figure}

The drug storage compartment of the bio-cyber interface contains nanomachines that are fabricated with materials sensitive to external physical effects like changes in light, temperature, pH or enzymes\cite{hu2014enzyme}. The injected nanomachines traverse the blood vessel network and are anchored at the targeted site due to high affinity. The nanomachines are equipped with ligands (i.e., signaling molecules in the MC channel) that bind to reciprocal receptors (i.e., receiving molecule) and are only expressed at the targeted site \cite{chude2016biologically}. The process of electromagnetic to bio chemical signal conversion is depicted in Fig \ref{fig:electrobio} \cite{chude2016biologically}. Bio and non-bio materials can be used to engineer nanomachine, like liposomes that are drug nanocarriersfabricated with a coating that is sensitive to external environmental factors \cite{nakano2014externally}.

Photosensitive materials release encapsulated molecules when stimulated by light at certain wavelength emitted from an external laser source \cite{wu2009genetically}. The process of photoisomerization destabilizes the bilayer membrane of liposome upon light illumination and allows the release of photoresponsive molecules. For example, caged compounds release molecules by bond breaking \cite{ellis2007caged} and gold nanorods generate heat as a response to their conformational change \cite{dykman2012gold,chude2016biologically}.
    
Temperature-sensitive materials release their contents upon a nonlinear sharp change in temperature of the environment such as temperature-sensitive liposomes \cite{needham2000new,dicheva2013cationic} and dendrimers \cite{kono2007temperature}. Such a sharp change in temperature triggers the thermo-responsive liposomes to release encapsulated molecules in the environment.  Thermoresponsive liposomes should ideally maintain their load at body temperature (~37°C) and deliver the encapsulated drug only upon an increase intra-body temperature \cite{chude2016biologically}.

Magnetic particles release their contents in response to magnetic radio-frequency signals generated by an external source. For example, gold nanocrystals attached to DNA molecules induce the hybridization of the DNA molecules \cite{hamad2002remote}, generating double-stranded or signal stranded DNA molecules. An architecture for bio-cyber interface has been proposed in \cite{chude2016biologically} which uses two kinds of liposomes that react to variations in light and temperature. According to the proposal when the decoded binary command from the external device is 011, thermal responsive liposomes release their contents. When the decoded command is 111 photosensitive liposomes release their contents. Another proposal\cite{nakano2014externally} has performed a wet laboratory experiment to engineer artificially synthesized materials (ART) using polystyrene bead, to operate as in-messaging nanomachine to forward messages to intra-bodynanonetwork The functionality in the ART based nano device is not yet implemented and is reserved for future research work.

An information-theoretic model for glucose monitoring as an application of IoBNT has been proposed in \cite{abbasi2017information}. The model utilizes pancreatic beta-cell to transmit
insulin molecules into the intra-body network via blood vessel channel according to commands received from the healthcare provider. The feedback of glucose level is transmitted by muscle cells towards the beta cell to be transmitted towards the healthcare provider. The model is currently designed for intra-body nanonetworks in this proposal and can be made online for continuous health monitoring by using cyber-interfaced beta cells.
\subsubsection{BIOCHEMICAL TO ELECTROMAGNETIC SIGNAL CONVERSION UNIT}
The bio electro transduction unit detects the concentration of received molecules and converts the molecular signals into electrical form. According to \cite{chude2016biologically}, this unit employs a cell structure to be used as a bioluminescence bioreporter. The bioreporter works on the principle bioluminescence reaction that produces a reporter protein upon excitation from an analyte.

The analyte here can be transmitted information molecules, moving towards the receiving nanomachines and the reporter protein is Luciferase (LU). Bioluminescence is the production and emission of light by living organisms as a result of a chemical reaction during which chemical energy is converted into light energy. When the two chemical enzymes Luciferin (L) and Luciferase (LU), catalyze oxygen in the presence of ATP, a chemical reaction releases energy in the form of light. Fluorescence molecules such as rhodamine derivatives are another form of luminescent materials that can be used to produce light energy \cite{nakano2014externally}. 

These materials emit fluorescence upon excitation from
microbiological conditions, which can be detected by external devices with the help of fluorescence microscopy. The emitted light energy is detected by a nanoscale light-sensitive sensor, which produces an equivalent electric signal in response. The electrical signal drives the transmitter to send the derived information through the wireless channel. Research contributions towards the realization of bio-cyber interface have resulted in some excellent theoretical frameworks for the Bio electrotransduction unit. The transduction unit for bio-cyber interface can be designed using natural/synthesized biological materials or novel nanotechnology materials like carbon and graphene.

\subsection{COMMUNICATION UNIT}
The connection of the human body with the Internet is the basic notion behind the IoBNT paradigm. bio-cyber interface can be used as a module for connecting the human body to the Internet. bio-cyber interface must possess the capability to wirelessly transmit the aggregated intra-body data to the healthcare provider. Bio cyber interface is responsible for communication with the intra body nano networks as well as with the internet. Therefore, it must contain two communication units i.e.,one for intra body communication and one for external communication with the internet.

\subsubsection{NANO NETWORK COMMUNICATION}
As discussed earlier the intrabody nano network communication cannot be realized through the traditional wireless communication technologies due to the minute size of nano transceivers inside the human body. Therefore, novel enabling technologies have been investigated that are feasible for communication inside the human body and provides an interface to access the human body. These technologies include molecular communication\cite{dissanayak2021exact} and nano-electromagnetic communication at the THz band. Both of these technologies are discussed in detail below:

\paragraph{MOLECULAR COMMUNICATION (MC)}
Molecular communication performs the exchange of messages between bio-nano machines through information encoded molecules \cite{nakano2012molecular,nakano2014molecular}. To generalize MC into a communication engineering perspective, researchers have defined an architecture of MC that consists of key communication concepts and processes \cite{nakano2014molecular}. The basic components of MC architecture include the sender bio nanomachine, receiver bio nanomachine, and the propagation medium. The sender bio nanomachine encodes the information molecules, usually in the form of molecular concentration (i.e., the number of information molecules per unit volume of solvent molecules) modulated over time or some other form according to application. The encoded information is then released into the environment through unbinding from the sender bio nanomachine. 

The propagation medium for \textit{in-vivo} applications is an aqueous medium that allows the flow of molecules towards the destination. The process of molecule propagation can be active or passive. In active or walkway based propagation, molecules are encapsulated in vesicles and are transported
through molecular motors towards the destination. Whereas in passive or diffusion-based propagation\cite{rudsari2021tdma}, information molecules diffuse away into the environment and are transported towards the destination through guide molecules. The receiver bio nanomachine captures the information molecules form the propagation medium and
decodes them into some chemical reaction. Chemical reactions may include the production of some signal for other molecules, performing some simple task or producing other molecules.

In \cite{nakano2012molecular} state-of-the-art in MC have been presented, discussing architecture, theoretical and physical modeling and challenges and opportunities in the development of MC based systems. Some researchers have also related MC with traditional networking protocols such as \cite{felicetti2014tcp} proposed a
TCP like molecular communication which is a connectionoriented protocol. Moreover, another proposal \cite{nakano2014molecular} presents
an OSI like layered architecture for MC. The functionality of each layer and relevant research challenges and opportunities according to each have been discussed in detail. 

In \cite{farsad2016comprehensive}, a comprehensive survey on recent advancements on MC according to communication, perspective has been presented, which provides a detail discussion on the transmitter, receiver and propagation medium of MC systems.
Some other works on the development of MC systems
include the transmitter and receiver design \cite{kuscu2019transmitter,kuscu2016modeling,kuscu2016physical} modulation techniques \cite{kim2013novel,kuran2010energy,garralda2011diffusion,kuran2011modulation}, channel modeling and noise analysis \cite{farsad2014channel,llatser2011exploring,chou2013extended, noel2014improving,pierobon2010physical}. The most prominent application
area of MC is healthcare and nanomedicine. In this direction, a comprehensive survey paper\cite{felicetti2016applications}on the medical applications of MC has been proposed that discusses a taxonomy of potential medical applications of MC, challenges, and opportunities in realizing them and future directions. Among the medical applications of MC, Targeted Drug Delivery (TDD) has got the most attention from the research community. A very comprehensive
survey paper on the MC based TDD that discusses system models and design requirements/ challenges for MC based TDD systems have been conducted by authors in \cite{chude2017molecular}.

Other work on MC based TDD can be found in \cite{chude2016molecular, chude2019nanosystems,chahibi2014antibody,chahibi2014molecular,chahibi2015molecular,chahibi2015pharmacokinetic}.
MC is the most promising IoBNT technology due to its bio-inspired nature \cite{akyildiz2015internet} and suitability for bio medical applications. Therefore most of the work in nanocommunication and IoBNT domain is dominanted by MC.

\paragraph{NANO ELECTROMAGNETIC (EM) COMMUNICATION}
Nano EM communication at THz band is another communication option for intr-abody communication, whose feasibility was first discussed in \cite{akyildiz2008nanonetworks}. In this technology, EM waves are used as an information carrier between source and destination. The advancement in novel nanomaterials such as Carbon nanotubes (CNT) and graphene derivatives such as graphene nanoribbons (GNR) has opened up doors for EM communication at the nanoscale \cite{akyildiz2010electromagnetic}. The THz spectrum is less vulnerable to propagation effects such as scattering, which makes it ideal
for intrabody communication due to its safety for biological tissue i.e., non ionization \cite{piro2015terahertz}. 

The state-of-the-art in EM nano communication was proposed by Akyildiz et al \cite{akyildiz2010electromagnetic}, which discusses the components, architecture and
manufacturing possibilities of nanosensors for nano EM communication. Moreover, this paper also discusses the potential application areas of nano EM networks and challenges in realizing them. A new channel modeling scheme for nano EM networks based on radiative transfer theory is proposed in \cite{jornet2011channel}. This paper also investigated the channel capacity of THz band based EM communication using different power allocation schemes. A state-of-the-art review paper that discusses possible biomedical
applications of THz EM communication, current models and possible antenna designs, is proposed by Abbassi et al \cite{abbasi2016nano}.

\subsubsection{COMMUNICATION WITH THE INTERNET}
For better signal reception and to increase the mobility of patients with bio-cyber implants, smartphone devices are used as gateway devices to the internet \cite{chude2016biologically}. Smartphone devices have now become an integral part of many IoT applications due to its ubiquity, advanced computational capabilities and opensource \cite{zhang2016biosensors}. Wireless Body Area Networks (WBANs), Wireless Chemical Sensors (WCS) and other body area sensors are now being developed to be compatible with the operating system of the smartphone. 

This section includes wireless technologies that are feasible for the connection between bio-cyber interface and smartphone device. Wireless Body Area Networks (WBANs) is a similar field of IoBNT which utilize implanted sensor devices to wirelessly transmit, measured human body parameters to healthcare applications via the Internet \cite{movassaghi2012wireless}. Research in WBANs is relatively mature as compared to IoBNT, therefore, a literature survey from WBANs has been included in this section. Moreover, researchers in the field of chemical engineering have been working on wireless chemical sensors (WCS) for more than a decade now \cite{kirichek2016live}.

WCSs are wearable patches that are capable of transmitting sensed bodily chemical values to the connected smartphone for analysis and processing by healthcare applications. The wireless communication technologies that are suitable with bio-cyber interface design are identified through a literature review of relevant domains such as WBANs and WCSs. This work also go through some of the important factors to consider when designing a bio-cyber interface, such as size, power supply requirements, data rate, real-time vs. on-demand data transfer, transmission range, and warning capability. Selection of suitable wireless technology depends upon the bio-cyber interface application system requirements and the underlying mode of operation of the wireless technology.

There are a number of active and passive wireless technologies available today to establish human-Internet connectivity. Active wireless technologies contain transponders able to transmit and receive radio frequency waves at high data rates and long-distance. Active wireless technologies require a continuous battery supply for transmission and to power the circuitry. Examples of active wireless technologies include Zigbee and Bluetooth, and IEEE 802.15.6. While passive wireless technologies include RFID and Near Field Communication (NFC). These technologies are briefly explained below and are illustrated in Table \ref{tab:communicationMechanims}.

\paragraph{BLUETOOTH LOW ENERGY (LE) COMMUNICATION}
Bluetooth (LE) Low Energy technology was introduced to wirelessly connect small low power devices to mobile terminals. This technology is ideal to be integrated with implants for healthcare applications as it supports ultra-low power consumption \cite{strey2013bluetooth}. It contains tiny Bluetooth radio to actively send and receive messages from nearby smartphone devices. The transmission range of Bluetooth LE is up to 10 m, the data rate is 1 Mbps and the frequency band is 2.4 GHz ISM \cite{georgakakis2010analysis}. Pairing time with other devices is in milliseconds, which is appropriate for alarm and emergency conditions in healthcare applications \cite{cao2009enabling}.

\paragraph{ZIGBEE}
ZigBee has an active transmitter able to communicate with mobile devices over a distance of 10-100 m. Zigbee is considered the most cost-effective technology due to its low power and low data rates. ZigBee can operate on three ISM bands with data rates from 20 Kbps to 250 Kbps \cite{cao2009enabling}.
Although ZigBee provides a large transmission range, it is not a good candidate for continuous health monitoring applications due to its low data rates \cite{movassaghi2012wireless}.

\paragraph{IEEE 802.15.6 STANDARD}
The rapid popularity of Wireless Body Area Networks (WBANs) and Consumer Electronics called for a standard
communication medium to address the special needs of the WBANs such as low power, low cost, low complexity, high throughput and short-range wireless communication in and around the human body. IEEE standards association established the IEEE 802.15 Task Group 6 for the standardization of WBAN \cite{kurunathan2015study, ieee2010ieee}. It quotes: ‘‘The IEEE 802.15 Task Group 6 (BAN) is developing a communication standard optimized for low power devices and operation on, in or around the human body (but not limited to humans) to serve a variety of applications including medical, consumer electronics personal entertainment and other’’[IEEE 802.15 WPAN Task Group 6 Body Area Networks. 2011]. IEEE.802.15.6 has different frequency bands in different countries ranging from 16- 27 MHz \cite{lewis2010ieee}. 

Medical Implant Communications Service (MICS) band is a licensed band used for implant communication and has the same frequency range (402-405 MHz) in most of the countries \cite{kurunathan2015study}. MICS band is suitable to be integrated for bio-cyber interface as it defines
protocols that are compatible with system requirements of bio-cyber interface. IEEE.802.15.6 can operate within a transmission range of 3 m with data rates up to 10 Mbps \cite{mile2018hybrid}.

IEEE.802.15.6 has also been recognized to be used as a standard for Human Body Communication (HBC). HBC uses the human body as signal propagation medium, therefore certain concerns must be taken into account such as increased mobility of the patient, lower power
consumption, small battery size yet life to span around a time several months, and management to aggregate burst
data in presence of continuous triggers from physiological data. Considering these HBC concerns \cite{nie2015statistical} proposed a medium access method for statistical frame-based time division multiple access (S-TDMA) protocol that
demonstrates lower data latency, lower power consumption, and higher transmission efficiency.

\paragraph{NEAR FIELD TECHNOLOGY (NFC)}
NFC is a passive communication designed to operate in the ultra-low transmission range of up to 20 cm. NFC technology contains NFC tag similar to the RFID tag, which is powered by a readout device. Communication in NFC requires devices to touch each other or be in close vicinity to each other. Smartphone transmits power wirelessly to resonant circuits through inductive coupling
which turns “ON” the NFC tag. Once the transmission session is complete, the tag is marked as “OFF” and
becomes unreadable \cite{jena2019wireless}.

This technology is ideal for implantable devices with a low power source as the communication is powered by the
reader device (in our case smartphone). Ultra-Low transmission range in NFC is not a drawback, rather it is a
strength as it provides rapid connection \cite{movassaghi2012wireless}.

\paragraph{RFID (ISO/IEC 18000-6)}
RFID is a passive wireless technology that consists of a transponder (tag) to be read by an RFID reader. RFID based
implants are feasible in situations where battery supply is a major issue \cite{movassaghi2012wireless}. Entire transmission in passive RFID systems is powered by the reader side \cite{grosinger2013feasibility}, which in our case can be RFID reader enabled smartphone devices. The read range of RFID is 1- 100 m with the data rate of 10- 100Kbps \cite{cao2009enabling}. Battery-powered RFIDs with active radio
transmitters are available which comes with a high cost yet
low data rates. 

\begin{table*}
 \caption{Communication Mechanisms for the Internet Connectivity of Bio-Electronic Devices. }
\label{tab:communicationMechanims}
\begin{tabularx}{\textwidth}{p{3cm}p{3cm}p{3cm}p{3cm}p{3cm}}
\toprule
Technology &Data Rate &Transmission Range &RF Band &Exemplary References
  \\ 
\midrule
Bluetooth LE  &1 Mbps & 10 m    &2.4 GHz ISM &\cite{strey2013bluetooth,georgakakis2010analysis}\\ 
ZigBee &250 Kbps &10-100 m &2.4 GHz &]\cite{cao2009enabling,movassaghi2012wireless}\\
IEEE 802.15.6 &10 Kbps-10 Mbps &3 m &16-27 MHz &\cite{kurunathan2015study, ieee2010ieee,lewis2010ieee,mile2018hybrid,nie2015statistical} \\
RFID &10-100 Kbps &1-100 m &860-960 MHz &\cite{cao2009enabling,grosinger2013feasibility,movassaghi2012wireless} \\
NFC &$<424Kbps$ &$<20$ cm &13.56 MHZ &\cite{cao2009enabling,jena2019wireless,movassaghi2012wireless} \\
\bottomrule
\end{tabularx}
\end{table*}

\subsection{POWER SUPPLY}
Bio cyber interfaces need a continuous power supply for data collection, processing, and transmission. The battery is mass vise the largest unit in the bio-cyber interface and other body implants. Although considerable progress has been done in lithium rechargeable batteries, yet they cannot keep pace with the novel technology requirements and size constraints. Miniaturization of a battery source, energy harvesting methods based on human body movements, and wireless battery recharge techniques have been explored by researchers of MIT for implants in Wireless Body Area
Networks (WBANs). 

Evanescent waves have been considered in \cite{kurs2007wireless} to wirelessly power the electronic devices over a short distance. Other efforts include solar rechargeable batteries \cite{lee2013wearable}, piezoelectric nanogenerators \cite{lee2014highly}, micro-supercapacitors \cite{bae2011fiber} and endocochlear potential-based bio batteries \cite{mercier2012energy}. Implantable devices in IoBNT need a continuous power supply and prolong battery life, for this purpose mechanism has been investigated to scavenge power supply from the epidermal layer of the
wearer \cite{mercier2012energy}. These energy scavenging batteries are called
biofuel cells (BFC), which convert chemical energy into electrical energy through biocatalytic reactions \cite{meredith2012biofuel,zebda2011mediatorless,halamkova2012implanted}. Researchers from the University of California San Diego have demonstrated an epidermal BFC \cite{jia2013epidermal} to scavenge continuous energy from human preparation. Lactate is used as biofuel as it is present in human sweat in
an abundant amount \cite{jia2013epidermal}. Similarly, energy scavenging techniques from the human body can be adapted for power supply in the bio-cyber interface. A detailed energy model for nanoscale devices has been demonstrated in \cite{canovas2018nanoscale} which
discusses the power consumption of each component in the nanonetwork.

\subsection{ADHESION SYSTEM}
This section briefly discusses the patch material, as bio-cyber interface is envisioned to be a wearable/stick-able bio-electronic device. The adhesive material for the patch is under high interdisciplinary research investigation as human skin is extraordinarily stretchable ($\epsilon >100\% where \epsilon$ is the strain), highly rough (superlative height 40 $\mu$m), and generally covered with sweat and hairs \cite{baik2019bioinspired}. Hence, delivering adequate adhesion of skin patches against human skin persists to be a challenging task. Adhesion materials should be carefully chosen to prevent issues like cytotoxicity (i.e.,
condition of being toxic), skin contamination, damages, risks of infection, and loss of wet adhesion make them less effective. Adhesion materials using electronic materials are extensively reviewed in \cite{oh2019second} and bio-inspired approaches for adhesion like gecko-/beetle-inspired mushroom-shaped
architectures, endoparasite-like microneedles, octopus inspired suction cups and slug-like adhesive with energy dissipation layer, has been keenly investigated in \cite{baik2019bioinspired}.
Special measures should be taken into account while selecting the adhesive material to prevent data loss and wireless connection due to strain and physical pressure on the patch while patient movement, wrinkling of skin or
other factors.

\subsection{SUMMARY OF BIO CYBER INTERFACE}
There exist a few proposals in the literature that present interface either using nano EM or MC communication. In
the nano EM domain \cite{islam2016catch} have proposed an architecture of nanonetworks-based Coronary Heart Disease (CHD) monitoring system. The proposed model consists of nanomacro interface (NM) and nanodevice-embedded Drug Eluting Stents (DESs), termed as nanoDESs. The algorithm exploits the periodic change in mean distance between a nanoDES, inserted inside the affected coronary artery, and the NM, fitted in the intercostal space of the rib cage of a patient suffering from a CHD utilizing THz band. Another paper proposes a straight forward communication scheme
utilizing the TeraHertz Band. The architecture consists of nano nodes and nano router, deployed in the dorsum of the human hand. Nano nodes circulate in the blood vessel for the collection of medical data and transmit the aggregated
data to the nano router. Nanorouter then transmits the data to BAN (Body Area Network) device through THz band communication, which is then relayed to the Internet.\\
Some theoretical models have adopted MC paradigm for the
design of bio-cyber interface device. For example, authors in \cite{nakano2014externally} have conducted a wet-lab experiment by utilizing artificially synthesized materials as an interface between the biological and electrical world. The proposal employs optical sensitive molecules(pH rodo molecules) that activate upon expression of fluorescence through fluorescent microscopy.\\

The above-mentioned proposals that utilize nano nodes and
nanoDES in THz communication, may not be
biocompatible due to the non-biological nature of these devices. Secondly, the use of external devices like a
fluorescent microscope can also cause mobility issues for the patient, which negates the whole notion of IoBNT. A
novel hybrid approach is presented \cite{yang2019comprehensive}, which suggests utilizing MC based communication inside the human body
for its biocompatible and noninvasive properties. The MC can communicate with graphene-based nanosensor, implanted over the human body for communication with external devices. The nano micro interface will, therefore,
consist of a THz based antenna and micro/macro antenna.
The idea is to limit the non-bio devices inside the human body and to read the intra-body parameters efficiently
outside the body.

\section{CLASSIFICATION OF BIO CYBER INTERFACING TECHNOLOGIES AND THEIR SECURITY ISSUES} \label{section:CLASSIFICATION OF BIO CYBER INTERFACE}
The design of a bio-cyber interface is a major challenge in the realization of IoBNT. In the introductory paper of
IoBNT, Akyildiz et al highlighted the design challenges of bio-cyber interface and the possibility of using electronic
tattoos, and RFID based sensors as bio-cyber interface. In order to study the possibilities of electronic tattoos and RFID-based tattoos as bio-cyber interfaces, previous proposals on electronic tattoos and RFID-based tattoos are evaluated below.
Furthermore, some preliminary work on a biologically-inspired bio-cyber interface presented by the notable MC engineering research community is discussed.

\subsection{ELECTRONIC TATTOOS AND TRANSDERMAL PATCHES}
Flexible and stretchable wearable electronics are getting interdisciplinary research attention as they promise to deliver continuous patient monitoring \cite{velloso2011web,atzori2010internet} while circumventing the possible discomfort caused by regular
wearable electronics. There are a number of biomedical conditions that require frequent monitoring of patients at regular intervals, for example, glucose monitoring for diabetic patients, fitness monitoring of athletes, real-time detection of pathogens in biofluids for the plausible onset of disease \cite{bandodkar2014epidermal}. Traditional methods to measure these chemical analytes require extracting fresh blood or other bodily samples every time. This continuous and invasive blood sampling can cause discomfort to the patient’s especially elderly patients and homophobic patients.\\
To address some of these concerns, researchers have ventured into the development of wearable sensors which
provide non-invasive ways to continuously sense the patient’s vital signs and transmit the collected data to data centers for further processing. Researchers have devised non-invasive ways of drawing out samples by utilizing the epidermal layer of the human skin \cite{bandodkar2014non}.

Human skin contains several electrolytes that provide clinically useful information about patients’ overall health such as temperature, physiological, electro-physiological and
biochemical parameters \cite{lee2019skin}. Human perspiration is also an information-rich epidermal secretion that is readily available and can be used to extract information about chemical compositions like metallic ions, minerals, glucose,
lactose, lactic acid, urea, volatile organic compounds, in the human body \cite{jin2017advanced}. Variously, human epidermis experiences continuous bending, stretching, and deformation while
performing daily life physical activities.\\
The mechanical properties of wearable electronics might mismatch with unique mechanophysiology of the human skin and can cause degradation of device capability and
discomfort to the wearer. Novel fabrication techniques for developing flexible and comfortable wearable devices are trending in academic and industrial research \cite{oh2019second,baik2019bioinspired,windmiller2012electrochemical}.

To overcome the aforementioned issue, wearable electronics are now being fabricated by adapting the design
principles of temporary tattoos.
Temporary tattoos are a form of body art that firmly attaches to the skin as “secondary skin”, while their
presence is barely noticed by the wearer. Tattoos are comfortable to be worn and provide ease of performing
daily life activities to the wearer, by bearing mechanical stress, washing, and other harsh conditions. These favorable
properties of tattoos have attracted researchers of wearable electronics to develop tattoo based electronics. The
electronic tattoos typically include sensors, memory, electronic circuits, and drug reservoir units \cite{di2015stretch}. The
electronic tattoos analyze the sensed information to determine the dose and release timing of the drug contained
in the unit.

Screen printing techniques can be used to print electronic tattoos of desirable shape capable of extracting rich
chemical information from our epidermis and transmit these analytical data wirelessly to a smartphone. Moreover, dry free form cut and paste technology \cite{jeong2017nfc,jeong2019electronic} is also been introduced for electronic tattoo fabrication. A summarization of existing proposals for electronic tattoos in healthcare monitoring is presented in Table \ref{tab:electronictattoo}.

Traditional methods of drug delivery through oral intake of medicine and via injections are now quickly being replaced by transdermal drug delivery patches. These patches are good alternatives to traditional methods as they promise minimally invasive and painless drug delivery. The transdermal patch is usually made of flexible and stretchable material that contains microneedles on the sticking side of the patch. Microneedles are actually drug containers that release their payload upon external
stimulation. Fourth-generation transdermal drug delivery
systems \cite{lee2018device} are now capable to release controlled amounts of the drug upon receiving commands wirelessly through the internet. A number of wearable electronic patches based therapeutic stimulators and drug delivery patches are available now that have proven to provide accurate results when compared with regular laboratory
equipment.
{ 
\subsubsection{SECURITY CONCERNS IN ELECTRONIC TATTOOS AND TRANS DERMAL PATCHES}

Electronic tattoos and transdermal patches can be exposed to security threats as they are continuously connected with the wireless technology.The following are some of the threats that electronic tattoos and transdermal patches may encounter.\\
\paragraph{WIRELESS COMMUNICATION RISKS}
Electronic tattoos and transdermal patches need continous wireless network supply in order to receive and send information to the healthcare provider.  Mainly the electronic tattoos are connected with a gateway device (smart phone) that relays the information over the internet. In the currently published literature for electronic tattoos, it is noticed that most of the proposals have used Bluetooth or BLE as communication mechanisms between electronic tattoo and gateway device. Bluetooth technology works within close proximity and is considered safe as attacker must be present within the patient's vicinity to launch an attack. However, Bluetooth version 5.0 has a transmission range of upto 400m moreover, the range can be magnified up to a mile through omni directional antennas. Researchers from Singapore University of Technology and design have    discovered 12 different Bluetooth bugs which can affect 450 type of  IoT devices including implantable medical devices. Moreover, the security flaw named "SweynTooth" associated with BLE can crash the device, deadlock the device or bypass security to access device functionality.
Another critical threat to Bluetooth technology is Blueborne attack. In this attack, hackers infect the devices with malwares and gains access to the device to execute malicious operations \cite{hassija2020security}.

\paragraph{PHYSICAL HARM}
The electronic tattoos currently manufactured are tested rigorously to withstand mechanical deformation caused by external strain. However, in cases of long term use and extreme physical
stress conditions, the device abilities might be degraded and cause unwanted behaviour. The bio-electronic devices which are presently designed in the form of electronic tattoo and implantable
patches that do not require any surgical procedures to affix on the human body. Non-medical professionals can apply
these devices easily on human skin like stickers. This nonsurgical relaxation can be exploited by local attackers to
replace the bio-electronic device with an illegitimate device on an unconscious patient. The illegitimate device now
becomes part of the communication and can launch several passive and active attacks like eavesdropping, replay attack, and data modification, etc.\\
\textit{Countermeasures}:Some of the possible countermeasures for above mentioned attacks are
Tamper-proofing  and self destruction:Use of tamper proof package for sensing devices and enable self destruction mode upon countering an intruder attack.
Minimize information leakage: Protective measures like generating artificial noise, shielding and adding randomized delay can be taken to minimize information leakage. \cite{abdul2019comprehensive}
Run-time attestation: Generation of proof about updation of firmware by a remote entity. \cite{abdul2019comprehensive}.
}

\begin{table*}[!htpb]
\begin{center}
\setlength{\extrarowheight}{10pt}
\caption{Summarization of Electronic Tattos and Patches }\label{tab:electronictattoo}
\begin{tabu}{X[0.5]X[2]X[2]X[2]X[4]}
\hline
    Ref &
    Device Type &
    Functionality &
    Communication Unit &
    Description\\ \hline
    \cite{jia2013electrochemical} &
  Electronic Tattoo &
  Glucose Sensing, Alcohol Sensing &
  2.4 GHz BLE &
  An epidermal biosensing system for continuous sampling and analysis of two biofluids i.e., glucose and alcohol. The electronic tattoo is screen printed which facilitates sweat stimulations at anode and ISF at the cathode. \\
\cite{bandodkar2014epidermal} &
  Electronic Tattoo &
  Sodium Sensing &
  Bluetooth Radio &
  An electronic tattoo with wireless transmitting capabilities to measure and transmit sodium levels from perspiration of athletes during training. \\
\cite{kim2016noninvasive} &
  Electronic Tattoo &
  Alcohol Sensing &
  Bluetooth radio &
  An iontophoretic-biosensing system has been integrated in electronic tattoo to monitor alcohol consumptions in patients. An alcohol-oxidase enzyme and the Prussian Blue electrode transducer is used for the detection of ethanol in the induced sweat. \\
\cite{jeong2017nfc} &
  Electronic Tattoo &
  Temperature and light-sensing &
  NFC &
  The tattoo can sense the temperature and light variations in the human skin and send it wirelessly to NFC enabled through NFC technology smartphone \\
\cite{di2019remotely} &
  Drug delivery implant &
  Remote control drug delivery of enalapril and methotrexate &
  Bluetooth &
  The implant is capable to deliver two kinds of drugs namely enalapril and methotrexate for hypertension and arthritis patients respectively. The diffusion of drugs across the Nanofluidic membrane is accomplished through a low-intensity electric field. \\ 
  \bottomrule
\end{tabu}
\end{center}
\end{table*}

\subsection{RFID TAG SENSORS AS BIO-ELECTRONIC DEVICE}
RFID tags are utilized for the identification of objects in
many IoT and Cyber-Physical Systems (CPS) due to their
wireless communication capability and easy integration in
the Internet cloud system \cite{lodato2015close}. RFID tag sensors have already been extensively studied for industrial applications like food safety, logistics, healthcare, public transport, fake medicine and environmental pollution \cite{singh2017inkjet}.\\ The role of
RFID sensors in envisioned healthcare applications IoBNT is promising due to their easy integration with implanted
interfaces for communication with the Internet. RFID enabled technology is being used for several healthcare
applications like real-time tracking of patients, improving their safety, and in management and medical supplies in
hospitals \cite{chen2011silicon,kumar2016intelligent,zhang2017review,kumar2010stage,cangialosi2007leveraging, dey2016rfid, asaimi2016developing,baptista2017radio}.
RFID is a wireless technology that utilizes RF signals to identify objects with RFID tags. RFID system has two
components: a tag which is a microchip to store electronic information and radio antenna to receive signals. There are two types of RFID systems, passive that require no battery
and are powered by the reader, and an active RFID system
that has a battery to power the transmission on its own \cite{singh2017inkjet}.\\
Passive RFID tags are ideal for IoBNT applications due to their properties of cost-effectiveness, long battery life, and low power consumption. RFID technology can be used as
sensors \cite{manzari2013modeling}, for sensing applications of IoBNT and can be
used as a bio-electronic interface on its own. RFID
antennas that are sensitive to environmental changes like
gas \cite{potyrailo2013passive}, moisture \cite{virtanen2011inkjet} and temperature \cite{martinez2016design}, have already been developed. The sensing capabilities of the RFID system can be exploited to measure the chemical and
biological values of the human body. 13.56 MHz) or ultrahigh-frequency (UHF, 400 MHz) bands, has proven
adequate in powering the adaptive threshold rectifier. The
HF band is for patch sensors, whereas the UHF band supports’ implantable sensors \cite{yang2019comprehensive}. The sensor consumes 12 $\mu$ W to implement an electrocardiogram analog front end, and an analog-to-digital converter (ADC).
The envisioned Bio-electronic interface with RFID enabled technology is likely to have the following components to be
operational in the IoBNT domain. RFID sensor tags that contain chemically coated thin- film resonant sensing
antenna for sensing target chemical substances in the environment. Recent RFID tags are thin and flexible,
allowing them to be embedded easily in the human body for health monitoring purposes \cite{kassal2018wireless}. \\
RFID sensors can be integrated with sensing materials such as water-absorbing
molecules to sense humidity and carbon nanostructures as gas sensors \cite{kim2013no}. The chemical, physical, or electrical reaction of the sensing materials in the presence of the
sensed parameters modify their electrical properties (permittivity, conductivity) resulting in easy-to-observe electrical metrics, such as a shift of the resonant frequency
of the RFID tag antenna, verifying the simplicity and power
efficiency of RFID-enabled sensors \cite{vyas2011inkjet}. Moreover, RFID tag sensors contain IC microchips to store and process sensed data. Novel methods for the fabrication of
electronic devices through inkjet printing are getting
research attention in Consumer Electronics (CE).
Nanomaterials like graphene, carbon nanotubes, silver, gold
and copper nanoparticles, conductive polymer and their based 
materials are used as inkjet ink to print electronic devices. 

A comprehensive review of inkjet-printed
nanomaterial-based flexible radio frequency identification
(RFID) tag sensors for the Internet of nano things has been
presented in \cite{singh2017inkjet}. According to the review, inkjet printed nanomaterials are cheaper and flexible to be used as wearable electronics as long as they do not interfere with the biological signaling capability of the human body. In context of IoT applications, Kassal et al \cite{kassal2013wireless} demonstrates a low-power
RFID tag sensor for potentiometric sensitivity. The RFID tag has the ability to measure and  store the potential of electrode, which is then
wirelessly transferred to a smartphone by near field communication (NFC). The RFID / NFC tagged chemical sensor is suitable for detecting pH or ion selective electrodes as part of a network of chemical sensors for IoT. The practical application of the RFID / NFC tagging sensor was tested to detect deterioration of milk by monitoring the pH value of milk over a period of 6 days. The measurements
showed the fluctuation of pH value between 5.89 and 6.10
over the 5 days, averaged to a pH of 6.03. Therefore,
RFID/NFC tag sensors show potential for IoT applications.
Another smart health proposal using RFID technology has
been presented in \cite{kassal2017smart}, where smart bandage manages
chronic wounds by wirelessly transmitting wound pH status
to an external readout unit in the smartphone, using radiofrequency identification (RFID). This is a lightweight, noninvasive bandage proposal that saves the patient from
regular hospital visits for dressing change. The pH
assessment calculated in smart bandage shows high
precision and accurate results when compared with regular
pH meter in the laboratory. An RFID enabled adhesive
skin patch is demonstrated in \cite{rose2014adhesive} which monitors
important biomarkers in sweat and surface temperature. In
this demonstration, a commercial RFID chip is adapted
with minimum components to allow potentiometric sensing
of solutes in sweat, and surface temperature, as read by an
Android smartphone app with 96\% accuracy at 50 mM $Na+$
(\textit{in vitro tests}).
{
\subsubsection{SECURITY CONCERNS IN RFID based BIO CYBER INTERFACE}

RFID tags, when used with interconnected devices in environments like IoT and IoBNT can be easily compromised by attackers. Following are a few possible security threats that must be considered when using RFID based technology as bio cyber interfacing option \cite{abdul2019comprehensive}. More attack types for all types of bio cyber interfaces are presented in Section \ref{subsection:bioattacks}
\paragraph{HARDWARE TROJAN}
Hardware trojans can modify IC of the tag hardware to allow attacker an access to its software components. This attack can be launched internally at the design phase of the RFID tag and can be launched externally through some sensors or antenna.
\paragraph{SIDE CHANNEL ATTACKS}
Attacker can use specialized tool to intercept information exchange between RFID Tag and its reader and use it for malicious purpose. This attack can take place even when the messages are encrypted. 
\paragraph{TAG CLONING}
Attacker can clone the tag and steal sensitive information from the tag and even impersonate the tag for communication with the reader.
\paragraph{TAG COUNTERFEITING}
In this type of attack, attacker gains access to tag and modifies its identity using tag manipulation techniques. Unlike tag cloning this attack can be launched with very less information.
\paragraph{TAG TRACKING}
Each RFID tag has a unique identifier number. Attacker can read the tag identifier attached to a person and track his/her location at any time. 
\paragraph{TAG INVENTORYING}
RFID tags also contain meta information about the tag like product code and manufacturers code. This information contains the type and purpose of RFID tag. For example in medical applications attacker can get access to what type of disease a person is suffering from, through tag inventorying attack.
\\
\textit{Counter measures}
Some of the possible countermeasures for RFID tags are \\
Side channel analysis:Using side channel signal i.e., timing, power and spatial temperature to detect hardware Trojans and malicious firmware.\\
Isolating: Protecting RFID tags from external EM waves through active radio jammers or other isolation techniques \cite{abdul2019comprehensive}.\\ Blocking: Restricting access to the tag by public readers by using privacy bit technique. Setting the privacy bit to '1' means public scanning of tag is not possible \cite{juels2003blocker}.\\
Anonymous tag: To prevent tracking of the RFID tag from hackers,  an anonymous ID is asssigned to the RFID tag. Mapping between anonymous ID and genuine ID should be stored in a look up table \cite{chen2009low}. \\
Distance estimation:Identification of distance between tag and reader using signal to noise ratio \cite{juels2006rfid}. }
\subsection{BIOLOGICALLY INSPIRED BIO CYBER INTERFACE}
The domain of IoBNT is biologically inspired and adapts the design of devices and their communication mechanisms from nature. Therefore researchers are investigating
methods to develop skin-mounted bio-electronics that support the seamless integration of biological materials with electrical components of the IoBNT network. In this direction, some proposals have been presented by
exploiting the biochemical properties of biologically engineered materials and synthetic artificial cells to be used
as device components. The device called bio-cyber interface \cite{akyildiz2015internet,chude2016biologically} is not only capable of sensing but also contains drug reservoirs to release controlled amounts of the drug
upon commands received from external devices.
A bioresorbable device \cite{tao2014silk} was fabricated using naturally occurring silk as the first step towards the development of
remote control implantable devices. This device biodegradable device is used to eliminate Staphylococcus
aureus infection from \textit{in-vitro} environments by triggering
thermal stimulations and targeted drug delivery operations. The device has the wireless capability to be turned on
wirelessly and it disappears once it has performed the required task, thus eliminating the need to remove the
device through surgical procedures.The use of biologically inspired materials for transmission processes in bio-cyber interfaces is described in the following scholarly works. A summarization of biologically inspired bio-cyber interfaces is presented in Table \ref{tab:biocyber}.\\

\paragraph{BIOFET BASED MOLECULAR RECIEVER}
Field-effect transistors (FETs) are a type of transistors that
use an electric field to control the flow of current. FETs consist of three electrodes namely source, drain, and gate. In traditional FETs, voltage is applied to the gate electrode
which in turn modulates the conductance between source and drain electrodes. The conductance is reflected as the
voltage-current alteration in the output channel. FET based technology is now being utilized for affinity-based
electrical sensing using nanomaterials (nanowires, nanotubes, and graphene) as transducer unit \cite{curreli2008real, yang2019comprehensive}.
FET transistors can be utilized for biosensing by replacing
the gate electrode with a biofunctionalized surface called
Biorecognition Unit (BU), for the detection of target molecules in the environment \cite{kuscu2016physical}.

The BU contains receptors on the surface of the FET channel which binds
ligands with intrinsic charges which result in accumulation depletion of carriers in the semiconductor channel, and hence modulation of conductance and current. The addition of BU in conventional FETs for molecular recognition makes them bio-inspired and is therefore called BioFETs. The bioFETs work on the principle of ligand-receptor
pairing i.e., binding of a ligand (signaling molecule) to its receptor (receiving molecule) and to produce a response e.g., signal transmission. There a number of ligand-receptor
pairs that can be used in modeling BU of bioFETS e.g., antibody-antigen, aptamer-natural ligand, natural
receptor/ligand \cite{robergs2004biochemistry} depending upon the target molecule.
Semiconductor materials like NW \cite{patolsky2006nanowire}, single-walled
carbon nanotubes(SWCNT) \cite{kim2013highly}, graphene \cite{ohno2010label},
molybdenum polymers(MoS2) \cite{shan2018high} and organic
nanomaterials like conducting polymers \cite{torsi2013organic} can be used
as transducer channel of bioFETs. Among NW materials,
Silicon nanowire (SiNW) has been proven to be the best fit
for bioFETs due to their low power consumption, high-speed sampling, high integration density, and high
sensitivity \cite{chen2011silicon,duan2012quantification,tran2016toward}.\\
In this direction, Kuscu et al \cite{kuscu2016modeling} have proposed SiNW bioFET based molecular antenna which receives information molecules as biochemical signals and converts
them into equivalent electrical signals. The proposed model employs the theory of ligand-receptor binding and
considers microfluidic advection-diffusion channel for the propagation of information ligands. The receiver model consists of three functional units. The Biorecognition Unit (BU) works as the interface for sensing the concentration of
ligands. In the Transducer Unit (TU), ligand-receptor regulates the gate potential of the FET through the fieldeffect resultant from their built-in current charges. The output unit shows the current flow as a result of the modulated gate potential. Moreover, an analysis and
optimization framework has been presented by providing a closed-form expression for fundamental performance
metrics, such as SNR and SEP. The proposed SiNW bioFET is capable of providing efficient in-device, label free and continuous processing of sensed molecules.\\

\paragraph{OPTICAL TO CHEMICAL BIOLOGICAL INTERFACE}
Grebenstein et al \cite{grebenstein2018biological} proposed a microscale modulator to
transduce optical signals into chemical signals. The modulator is realized using synthetically engineered E.Coli bacteria that express protons into the environment upon stimulation from an external light source. Light-emitting
diode (LED) of the modulator uses proton pump gloeorhodopsin (GR) to express light. The E.Coli bacteria
change the pH level of their surrounding environment as a chemical reaction to an external light source. The proposed
testbed achieves higher data rates on the order of 1 bit/min as opposed to previous proposals with data rates of 1
bit/hour.\\

\paragraph{REDOX BASED CHEMICAL TO ELECTRICAL INTERFACE}
New research in biology has 
recommends the usage of redox as a global signaling modality. 
Authors in \cite{liu2017electrochemistry} 
have adopted an approach that is inspired by sonar, which access the redox information through collaborative electrochemical probing. Authors further utilize attuned
electrical inputs that are coupled with diffusible redox mediators (electron shuttles) to access redox information in a local environment and generate complex but interpretable electrical output signatures. Redox (Oxidation-Reduction)
reaction is also utilized for biochemical-electronic transduction mechanisms in a number of proposals
\cite{liu2017connecting,liu2017using,kim2019redox,kang2018redox,liu2019electrofabricated}. A wet lab interface prototype has been proposed
recently in \cite{kang2018redox} for transducing chemical signals into
electrical signals by the virtue of redox modality. The
nterface prototype device consists of a dual film with inner
film contains hydrogel-based film entrapping E.Coli
bacteria and the outer film consists of a redox capacitor to
amplify electrical signals.
These cells are engineered as
reporters, which respond to the presence of a certain
molecule (signaling molecule AI-2) by converting the redox
inactive substrate 4-Aminophenyl $\beta$-D-galactopyranoside
(PAPG) molecules into redox-active p-aminophenol (PAP).\\

\paragraph{FRET BASED UPLINK/DOWNLINK BIO CYBER INTERFACE}
FRET (Fluorescence Resonance Energy Transfer)
mechanism has been adopted by El-atty et al \cite{abd2020molcom} to model
uplink/downlink bio-cyber interface for the Internet of bio
nano things. In FRET-based optical sensing, bio-cyber
interface is designed for targeted drug delivery applications
of IoBNT. The downlink of bio-cyber interface is designed
by adopting spreading principals of SIR (Susceptible, Infected, Recovered) epidemic scheme and decode forward
(DF) basis. Three types of nanomachines are used in the downlink model to realize the targeted drug delivery
namely nanoreciever, nano transmitter (Infected) and nanorelay(Susceptible) according to SIR epidemic scheme.
The recovered nanomachines are ones that have transmitted their exciton to the nano receiver. The uplink is designed by
considering two types of nanomachines namely nanosensors and nanoactuators. Uplink signal notifies the
medical server about successful drug delivery through bioluminescence reaction. 
{ 
\subsubsection{SECURITY CONCERNS IN BIOLOGICALLY INSPIRED BIO CYBER INTERFACE}
Currently the research bio cyber interface security is immature and there is a minimum published literature in this field. \cite{bakhshi2019securing} have presented the possibility of ML based adversarial attacks for biologically inspired bio cyber interface which are presented below. Details of possible generic attacks on bio cyber interfaces can be found in Section \ref{subsection:bioattacks}. 
\paragraph{MACHINE LEARNING ADVERSARIAL ATTACKS}
In the case of bio-luminescent or thermal signaling based bio cyber interface adversaries might launch an attack by manipulating the internet-enabled parameters. By manipulating the parameters, attackers can cause inappropriate amount of drug release, initiate self-annihilation of drug molecules and modify monitoring information provide by bio chemical processes.
In redox based bio cyber interfaces, changing the input electrical signal can lead the capacitor charging and unwanted redox activity such as activating/deactivating the redox substrate to affect enzyme production. 
Bio FETs work on the principals of ligand binding through charging of electrodes. The attacker can launch sentry attack to repel required ligands or black hole attack to attract unwanted ligands to bind to the receptor to affect the current control. The changes in external current control can cause the Bio FET based bio cyber interface to exhibit unwanted behavior. \\
\textit{Countermeasures}
The attacks related to biologically inspired bio cyber interfaces can be categorized as ML-adversarial attacks. Possible countermeasures to ML adversarial attacks are \cite{bernal2019security} \\
Data sanitization: This process refers to pre-processing, validating all the input data, and rejecting the harmful samples.\\ Adversarial training: Inclusion of adversary information in the training samples to recognize attack vector.\\
Defence distillation: Creating secondary ML model with less sensitivity and more general results.\\
Differentail privacy: A cryptographic mechanism of adding noise to susceptible features of data.
Hormonic encryption. A cryptographic mechanism to perform computations over ciphered data to generate encrypted result.
}

\begin{table*}[!htpb]
\begin{center}
\setlength{\extrarowheight}{10pt}
\caption{Existing proposals for Biologically Inspired Interface Devices in Academic Literature}\label{tab:biocyber}
\begin{tabu}{X[3]X[3]X[3]}
\hline
Name of Device &
  Transduction Mechanism &
  Description \\ \hline
  Biologically Inspired Bio Cyber Interface \cite{chude2016biologically} &
  Electrical to biochemical signal through Photoresponsive biomolecules. and Thermal responsive biomolecules  -Biochemical to electrical signal through the Bioluminescence phenomenon. &
  - A theoretical model for bio cyber interface has been proposed as an interface between biological and electrical domains. The model comprises of transduction units that convert an electrical signal into biochemical signals through the bioluminescence process and biochemical signals into electrical signals through photoresponsive and thermal responsive biomolecules. \\
SiNW bioFET based Molecular receiver  \cite{kuscu2016modeling} &
  Biological to electrical signal conversion through bioFET technology. &
  - The Silicon Nanowire (SiNW) is used as the conductive nanomaterial for a molecular antenna. Microfluidic channel has been considered as a propagation medium for information molecules to flow from transmitter to receiver in unidirectionally through diffusion. \\
Biological Optical-to-Chemical Signal Conversion Interface \cite{grebenstein2018biological} &
  Optical to chemical signal through illumination effects. &
  - A biological signal conversion interface is designed based on E Coli bacteria that can change the pH of the surroundings by pumping protons in response to external light stimuli. \\
Redox based Chemical to Electrical Interface \cite{kang2018redox} &
  Chemical to electrical signal conversion through redox modality. &
  - The interface prototype device consists of a dual film with inner film contains hydrogel-based film entrapping E.Coli bacteria and the outer film consists of a redox capacitor to amplify electrical signals. \\
FRET-based biocyber interface \cite{abd2020molcom} &
  Optical to electrical and vice versa signal conversion through FRET Technology &
  - FRET technology has been utilized to model the uplink/downlink biocyber interface for targeted drug delivery applications of IoBNT. SIR epidemic model is adopted to model downlink and the bioluminescent reaction has been utilized to generate an uplink signal. \\ \hline
\end{tabu}
\end{center}
\end{table*}

\subsection{ SUMMARY OF BIO CYBER INTERFACE CLASSIFICATION}
It is obvious from the above-mentioned literature that there are a lot of possibilities for designing a bio-cyber interface. The design model of the bio-cyber interface primarily
depends on the application scenario. For example, a bio-cyber interface design and components for sensing applications will be different from drug delivery
applications. Our take on the choice of bio-cyber interface option goes in the favor of bio-inspired interfaces. The
reason being biocompatible in nature and the research in these interfaces in line with the direction of IoBNT.
Extensive in-vitro and wet-lab experiments are required to
validate the theoretical proposals for the bio-cyber interface.

\section{SECURITY IN IoBNT} \label{section:SECURITY IN IoBNT}
The advent of skin implanted bio-electronics and the IoBNT
paradigm will not only open up a plethora of novel biomedical applications but also its wireless connection capability will enable the adversaries to utilize it
malevolently. Connecting the intra-body biological environment with the cyber domain through bio-electronic
devices will provide the attackers with an apparent opportunity to devise new terrorist mechanisms to harm the
patient remotely. Maliciously accessing the human body through the internet to steal personal information or to
create new types of diseases by malevolent programming of
bio-electronic devices and intra-body nanonetworks is termed as bio-cyber terrorism \cite{akyildiz2015internet}. Bio-cyber terrorism can take advantage of wirelessly accessing the human body to launch fatal and life-threatening attacks from a remote site.
Therefore security features have to be embedded either in a
separate component of the bio-electronic device, which may enlarge the size of the device or might be infeasible in
some applications. Another possibility is to delegate security services to external devices in close proximity with
sophisticated resources as compared to bioelectronics devices. For example, in a similar field of IMD (Implantable Medical Device), some researchers propose to
assign security functionality to an external device like
Cloaker or MedMon \cite{zhang2013medmon}. Nonetheless, the bio-electronic device must execute a lightweight authentication mechanism at least once to establish a secure connection with external devices.  Bio-electronic devices will also be linked to a gateway device (such as a smartphone) to send and receive information from the healthcare practitioner.
Because of the technological differences, this section discusses the security needs for nanonetworks and bio-cyber interfaces separately. The security goals, regardless of the underlying technological variations, remain the same. The STRIDE threat approach can be used to model security threats against IoBNT. STRIDE is the acronym for Spoofing, Tampering, Repudiation, and Information Disclosure, Denial of Service and Elevation of privilege. These six categories present a broad classification of threats and can be further divided into other related threats. Each threat category is related to a security goal:
Spoofing-Authentication, Tampering-Integrity,
Repudiation, Non-Repudiation, Information Disclosur, Confidentiality, Denial of Service- Availability, and
Elevation of privileges-Authorization \cite{camara2015security}, which is
presented in Table. \ref{tab:srideThreat}.\\
\textbf{Authentication (Au):} Authentication ensures that the identity of each communicating party is established before executing any operation. Moreover, authentication also ensures that the data is coming from the authorized source 
and unauthorized users cannot access or modify the data. Authentication needs to be done on users as well as a message \cite{loscri2014security,kumar2012security}.\\ 
\textbf{Integrity (I):} Integrity ensures that the message exchanged between legitimate entities is not tampered or modified by unauthorized entities.\\
\textbf{Non- Repudiation (NR):} In the traditional networking paradigm, all the communication transactions are logged to
track the network anomalies and gain the attacker’s profile in case the attacker tries to misuse his/her privileges. Nonrepudiation can be violated if the attacker gets access to the logs and delete the records to remove traces.\\
\textbf{Confidentiality (C):} Confidentiality ensures that the
attacker should not learn the content of the message exchanged between the sender and receiver. The data must
only be accessible to authorized personnel upon authentication through some mechanism priory.\\
\textbf{Availability (A):} This goal ensures that the services and
communication of the device are always available on request. \\
\textbf{Authorization (Auth):} Authorization property ensures that
only those entities can execute a specific operation that has
privileges to order it. Authorization requires that an entity
must have been authenticated previously through the regular login (ID, password) procedure to establish the identification.

\
\begin{table*}[!htpb]
\begin{center}
\setlength{\extrarowheight}{10pt}

\caption{STRIDE Threat Model Categories, related Security Goals and Attack Possibilities in IoBNT
} \label{tab:srideThreat}
\begin{tabu}{X[3]X[3]X[5]}
\hline
Threat &Security Goal &Possible IoBNT Scenario  \\ \hline
Spoofing &Authentication &Security breach by
interception of
communication between
bio-electronic device and
the smartphone.\\
Tampering &Integrity &The dosage prescribed by
the medical server can be
altered during transmission
by the attacker and drug is
released according to
altered values.\\ 
Repudiation &Non-Repudiation &Deletion of access logs to
remove traces of malicious
user activities. \\
Information
Disclosure
&Confidentiality &Security breach by
interception of
communication between
bio-electronic device and
smartphone.\\
Denial of Service &Availability &Inability to transmit
irregular vital signs of an
elderly or unconscious
patient who is solely
dependant on the bio-electronic device for
communication with the
medical server. \\
Elevation of
Privileges
&Authorization &An internal attacker can
misuse access privileges to
steal or tamper
information. \\ 
\bottomrule
\end{tabu}
\end{center}
\end{table*}

\subsection{ATTACK TYPES IN NANONETWORKS}
This section presents the possible attacks and existing proposals related to the nanonetworks. A summarization of security proposals for nanonetworks is presented in Table \ref{tab:nanosecurity}.

\subsubsection{EAVESDROPPING}
Eavesdropping refers to passively listening to the transmission between two nodes. The listened information can be stored and later used maliciously to launch attacks. Eavesdropping in nanonetworks can take place when two legitimate nanomachines are exchanging messenger molecules and a nearby malicious nanomachine intercepts the messenger molecules silently. The passive eavesdropper can be detected in nanonetworks by a mechanism such as stochastic geometry, distance estimation techniques. The active eavesdropper might absorb the messenger molecules in the case where MC is used as a communication medium, this attack might be prevented through secrecy capacity. Other anticipated eavesdropping prevention mechanisms include beamforming, game theory (collation formation games),
and artificial noise generation.\\
Several proposals have been demonstrated to detect eavesdropper location and secure the nano communication
paradigm from eavesdropping attacks.\\
Islam et al \cite{islam2017secure} proposed a secure channel for Molecular communication. Firstly, a Diffie–Hellman algorithm-based secure key is exchanged between sender nanomachine and receiver nanomachine. Then hardware ciphering is performed using the secret key. As MC is a resource-constrained paradigm, therefore, Exclusive OR (XOR) cipher is used in this work due to its simple implementation and inexpensive computation. Moreover, hardware ciphering used in this work further reduces the associated time, instead of its software counterpart. The results are presented through simulation. In our opinion, the proposed method is simple and computationally less expensive but the overall security is compromised as MC needs more
mathematically resilient models for security.\\
Guo et al \cite{guo2016eavesdropper} have proposed a mathematical model for
eavesdropper detection and localization in a random walk channel. This is the only work in MC that considers the
detection of an absorbing malicious receiver in a random walk channel. The authors have chosen transmitter side
detection of the eavesdropper because the MC channels are 1-D and receivers cannot affect the transmission and
secondly the molecules absorbed at the receiver are quite small that do not aid in detecting eavesdropper presence.
Simulation has shown accurate results in detecting the eavesdropper’s presence.\\
Recently a proposal for physical layer authentication for 
Diffusion Based Molecular Communication (DbMC) has een proposed in \cite{zafar2019channel}. The channel impulse response of 1-D DbMC has been exploited to detect transmitting
eavesdropper in the transmission region.

\subsubsection{BLACKHOLE ATTACK}
Blackhole attack refers to the attack where malicious nodes spread attractant molecules to draw the network traffic
towards a different location from the intended target. 
Blackhole attack is similar to sinkhole attack in WSNs but the sinkhole attack disrupts the routing process while the
blackhole attack physically moves legitimate nodes away from the target. When it comes to blackhole attacks in nanonetworks, there are a variety of approaches that can be taken. For example in scenario of artificial immune system support application, the white blood cells that are responsible for detecting and tackling with the infection, can be attracted by malicious nodes to stop them from detecting an infection in the host system.\\
Blackhole attack and its countermeasure using two approaches, Bayes rule and simple threshold approach for MC have been proposed in \cite{giaretta2015security}. Blackhole attacks are type of DoS (Denial of Service) attack in which nanomachines can be drawn away for target e-g in case of targeted drug delivery the actuator nodes might not reach the target due to Blackhole attack.

\subsubsection{SENTRY ATTACK}
Sentry attacks are opposite to Blackhole attacks where legitimate nanomachines are impeded away from the target,
due to a large number of repellent molecules spread around the target location by malicious nodes. This kind of attack can be fatal in medical applications where instant lifesaving
action needs to be taken e-g nanorobots that are designated to prevent bleeding by repairing veins are attacked by a sentry to impede from reaching the target. Sentry attack and
its countermeasure using two approaches, Bayes rule, and simple threshold approach have been proposed in \cite{giaretta2015security}.
Giaretta et al \cite{giaretta2015security} have proposed a blackhole attack and
sentry attack for MC. The authors have described two scenarios where (L-BNTs) Legitimate Bio-Nano Things are
repelled from reaching the targeted site in targeted drug delivery application, thus keeping Bio-Nano Things from
performing the normal operation. In the second scenario called black hole attack, M-BNTs (Malicious Bio-Nano
Things) are attracted to the targeted site which can lead in a
delivery unwanted dosage of medication in the targeted area. Next, in the proposal, a countermeasure that enables
the Bio- NanoThings to make decisions and cooperate in order to overcome blackhole and sentry attacks during
target localization is proposed. The mechanisms are based on known cellular decision processes using Bayes’ rule as
well as artificially designed genetic circuits that evaluate chemical signal threshold (this will be known as Thresholdbased decision process), which are both lightweight enabling them to be easily implementable on resourceconstrained Bio- NanoThings. Results show that the
proposed countermeasure is effective against the attack, where L-BNTs successfully move towards the target.

\subsubsection{SPOOFED, ALTERED, REPLAY MESSAGE ATTACK}
This attack can be launched by malicious nodes by spoofing legitimate nodes identities to become trustable and enter the network. Furthermore, the malicious nodes then send fake messages in the network and alter the data. In the case of nanonetworks, consider an exemplary communication scenario where a legtimate node 'Alice' is transmitting messages to receiver 'Bob'. Attack can be launched by an intruder 'Eave" impersonates to be Alice and tries to control Bob by sending malicious commands.

A distance-dependent path loss based authentication scheme for nanonetworks using the terahertz band has been proposed in \cite{rahman2017physical} for spoofing attacks. Mahboob et al \cite{rahman2017physical} have
proposed an authentication scheme for terahertz band EMbased nanonetworks. This work exploits physical layer
attribute i.e., distance-dependent path loss for authentication at nano receiver. Moreover, an algorithmic
solution has been proposed for the authentication scheme.
Experimentation verification of authentication scheme via
tera hertz time-domain spectroscopy setup at QMUL, UK. 

\begin{table*}[!htpb]
\begin{center}
\setlength{\extrarowheight}{10pt}

\caption{SECURITY PROPOSALS FOR NANONETWORKS IN LITERATURE}
\label{tab:nanosecurity}
\begin{tabu} to \linewidth {X[3]X[3]X[3]X[3]}
Ref. &
 Communication Medium &
  Type of Attack &
  Proposal Description \\ \hline
  Secure Channel for Molecular Communications \cite{islam2017secure}. &
  Molecular Communication &
  Eavesdropping &
  Encryption using Diffie–Hellman algorithm. \\
Eavesdropper localization in random walk channels \cite{guo2016eavesdropper} &
  Molecular Communication &
  Eavesdropping &
  Eavesdropper detection and localization by reverse estimating its location. \\
Congestion Control in Molecular Cyber-Physical Systems \cite{felicetti2017congestion}. &   Molecular Communication &
  Congestion &
  Congestion detection algorithm for Diffusion Based MC. \\
Security Vulnerabilities and Countermeasures for Target Localization in Bio-NanoThings Communication Networks  \cite{giaretta2015security}. &
  Molecular Communication &
  Black hole, Sentry Attacks &
  Construction of decision process using two approaches, Bayes rule and simple threshold approach for the robustness of legitimate nanomachines. \\
Physical Layer Authentication in Nano Networks at Terahertz Frequencies for Biomedical Applications\cite{rahman2017physical} &
  Nano Electromagnetic Communications &
  Spoofed, Altered, Replay message attack. &
  A distance-dependent path loss based authentication scheme. \\
In-sequence molecule delivery over an aqueous medium \cite{nakano2010sequence} &
  Molecular Communication &
  Desynchronization &
  A formal model for in-sequence message delivery for MC which can possibly prevent Desynchronization attack. \\
Channel Impulse Response-based Physical Layer Authentication in a Diffusion-based Molecular Communication System  \cite{zafar2019channel} &
  Molecular Communication &
  Eavesdropping &
  The channel impulse response of the physical layer has been exploited to propose an authentication scheme for the detection of transmitting eavesdropper in the broadcast region. \\ \hline
\end{tabu}
\end{center}
\end{table*}

\subsection{ATTACK TYPES IN BIO CYBER INTERFACE}\label{subsection:bioattacks}
\begin{figure*}[t]
\includegraphics[width=0.6\textwidth]{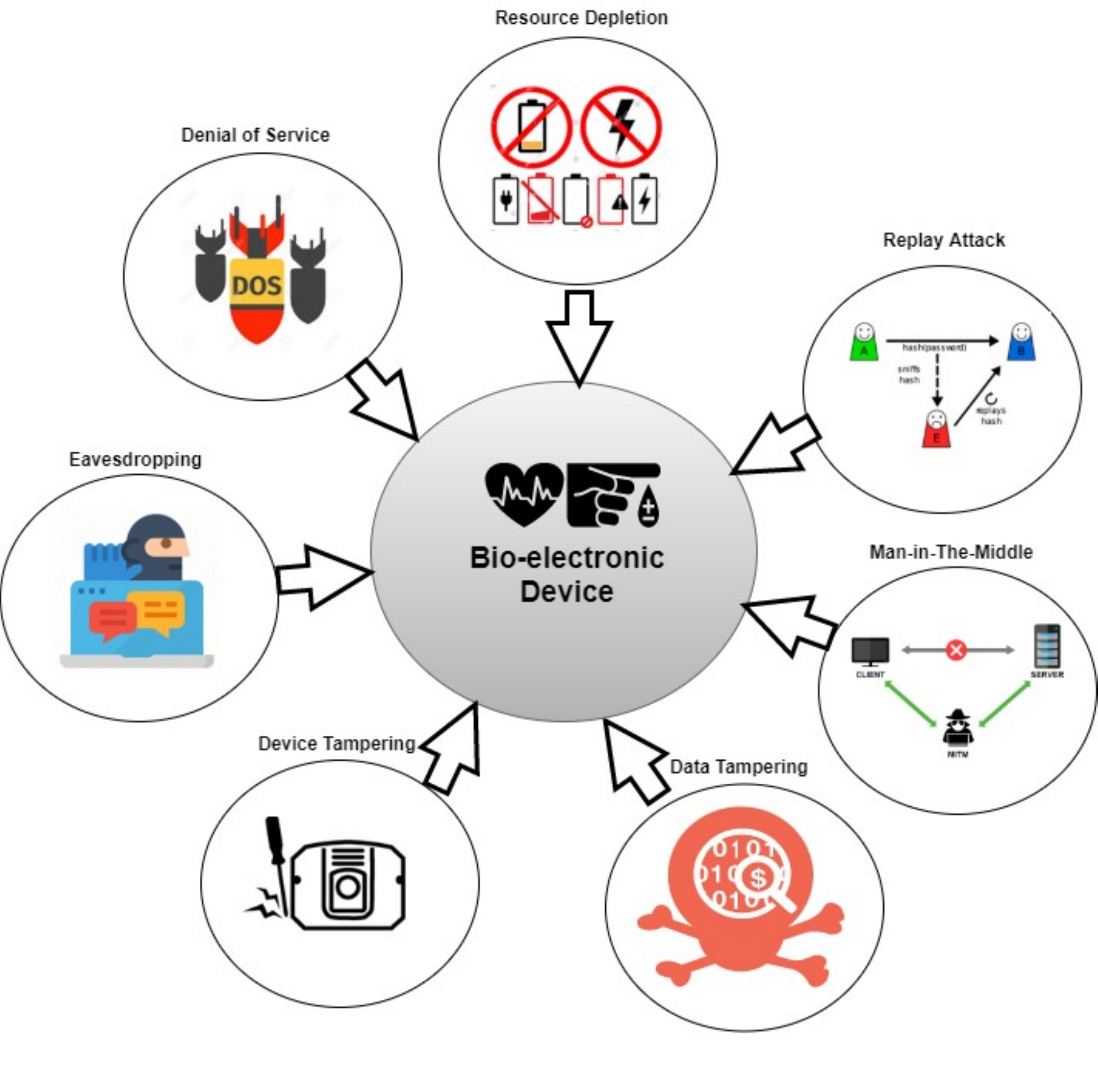}
\centering
\caption{ Types of possible attacks for Bio Cyber Interface}
\label{fig:biocyberattacks}
\end{figure*}

To ensure an end-to-end protection of IoBNT applications, the
security of these bio-electronic devices is a prerequisite. In order to pursue a preliminary investigation for types of attacks that are possible in bio-electronic devices, the attack vectors are explored in related fields like WBAN.
(Wireless Body Area Network) \cite{usman2018security}, IMD (Implantable
Medical Devices) \cite{camara2015security} and Wireless sensor networks
\cite{sharma2012security}. This investigation has helped us to individuate attacks that are likely to occur in bio-electronic devices.
The major objective of a bio-electronic device in the IoBNT healthcare application is to enable two-way communication between intra-body nanonetworks and healthcare provider.
The communication mode is divided into two categories: inbound and outbound, making it easier to recognize and categorize attacks. Figure \ref{fig:biocyberattacks} presents possible attack types in case of bio cyber interface.

\subsubsection{EAVESDROPPING}
Eavesdropping is a passive attack that enables the attacker to covertly gain access to confidential information. The bio-electronic device might possess critical information like
patient identification information, clinical history, disease
detail, treatment detail, patient location, and battery status,
etc. This confidential information can be exploited not only to breach a patient’s privacy, but also to launch other types of active attacks. Eavesdropping can be realized during the
communication of bio-electronic devices with biological nanonetworks and during communication with the gateway
devices.\\
\textit{Countermeasures:}Traditional networking paradigms employ encryption schemes like RSA and DES, to prevent against eavesdropping attacks. These encryption schemes are
effective for the prevention of eavesdropping attacks, but these are computationally expensive and must be adopted after analyzing the resources of bio-electronic devices
\cite{zhang2014trustworthiness}. Lightweight encryption schemes like Elliptic Curve cryptography has been proven to be effective for resource-constrained devices \cite{ye2014efficient}, also a review on other lightweight encryption schemes has been presented in \cite{singh2017advanced}.

\subsubsection{REPLAY ATTACK}
This type of attack can be launched after a successful
eavesdropping attempt. The eavesdropped authentication sequence of the legitimate devices can be replayed to get illegitimate access into the communication channel.
Moreover, an attacker can copy previously sent commands
from legitimate users to replay the message again which in
the case of bio-electronic can have a number of
consequences like drug release multiple times which harm
patient's health, unnecessary and repeated queries for
patients physiological values to engage and deplete the
resources, etc. \\
\textit{Countermeasures:} Authentication schemes, Intrusion detection, delegate authentication to external devices.
(See Section \ref{section:POTENTIAL MITIGATION STRATEGIES})
\subsubsection{MAN-IN-THE-MIDDLE (MITM) ATTACK}
A MITM attack is projected through illegitimate devices when they become part of communication between legitimate transmitting devices, and legitimate devices are
spoofed to believe that they are communicating with the
authorized device. In the case of bio-electronic devices, the
attacker must be in the vicinity of the patient to launch the
MITM attack. From the communication
perspective, the MITM attack can be achieved by replaying the legitimate authentication sequence to get access to the communication channel. MITM is the type of an active eavesdropping attack where the attacker not only listens to the communication but also alters the data and communication sequence.\\
\textit{Countermeasures}:Traditionally MITM is prevented using encryption schemes. A lightweight scheme to prevent implantable medical devices from MITM attacks has been
proposed that utilize chaotic generators for randomness and
a signature algorithm to prevent third-party interference
\cite{belkhouja2018light}.
\subsubsection{RESOURCE DEPLETION (RD)}
Bio-electronic devices are essentially capable of performing
\textit{in-situ} \cite{kuscu2019transmitter} operations and for this purpose they contain
built-in processing unit, memory unit, and power unit.
However, these units are only able to perform trivial tasks
due to resource limitation in terms of space, power consumption and computation complexity, which comes
with their minute size. The attacker can cause resource depletion by sending multiple authentication messages with
the wrong credentials to occupy the processer. Each authentication request is processed which exploits the
memory to create access logs and drains the battery. Moreover, bogus communication packets sent by MITM
can utilize and drain the resources. 
\textit{Countermeasures:} Access control mechanisms,Anomaly detection system,user notification,and use of rechargeable batteries. Moreover, mitigation from resource depletion attacks can be achieved by using passive
wireless communication media like RF signals and the use of novel energy preserving techniques like ZPD (ZeroPower Defence) \cite{siddiqi2019towards}.
{
\subsubsection{INJECTION ATTACKS}
Injection attack can be performed by the illegitimate users in three ways, insertion, alteration and replication \cite{abdul2019comprehensive}. In insertion attack, hacker generates new seemingly legitimate data packet into the communication channel. In alteration attack, hacker captures the data packet from communication link, tamper the values of dosage commands from the inbound channel and alter the
values of bodily parameters in the outbound channel. In replication attack, attacker re-sends previously executed commands into the system.
Insecure communication between bio-electronic and
gateway devices can result in an injection attack. This attack can have fatal consequences on the patient. For example in a scenario where a diabetic patient is fully dependent on wirelessly controlled insulin and glucose monitoring pump \cite{abbasi2017information}, the alteration in dosage values received from health care provider can result in hazardous consequences like
underdose can cause hyperglycemia and overdose can result in hypoglycemia both of which can be fatal \cite{anhalt2010insulin}. .
\textit{Countermeasures:} Careful
mitigation strategies are needed to be designed which include authentication protocols \cite{al2017survey}, access control mechanisms \cite{camara2015security}Intrusion Detection,
Input validation,and 
Authorization techniques.
}
\subsubsection{DEVICE TAMPERING}
This attack is majorly launched in the bio-electronic devices by local attackers, due to low transmission range of bio-electronic devices. The device can be reprogrammed or physically replaced to perform the malicious tasks that are not intended by the device in the first place. This attack can also be launched by accessing the device remotely by sending fake firmware update and device upgrade message
that alter the software configuration of the device to perform maliciously \cite{kintzlinger2019keep}. This attacker can alter the original treatment prescribed by the healthcare provider and replace it with false treatment which can cause life threatening consequences.
\textit{Countermeasures:}Tamper proofing  and self destruction, device hardening,
and Physically unclonable Function (PUF).

\subsubsection{DENIAL OF SERVICE (DOS) ATTACK}
Denial of service attack causes disruption and blockage of
information flow between legitimate communicating parties. An attacker tries to suspend services of the bio-electronic device to make it unavailable for communication and processing. DoS can be launched in the form of any of the above-mentioned attacks. The DoS attacks such as battery drainage, sleep deprivation and outage attacks can be performed by hackers after getting access to the devices. As the power source of these devices is limited and attackers can send an unlimited number of messages, causing the battery to drain. In outage attacks the device can be made unavailable by stopping it from carrying its essential functions, behaving abnormally or premature shutdown. These types of attacks can pose serious consequences for the patients \cite{abdul2019comprehensive} . 
Moreover, the
hacker can intentionally drop the forwarded packets, making the recipient deprived of requested data.
\textit{Countermeasures:} Intrusion Detection
{ 
\subsubsection{MALWARE ATTACKS}
Malware attack is another type of network attack that is common in IoT based applications. Malware attack is used to remote control a distant device maliciously, steal sensitive information from a device and use it to launch further malware attacks. A system infected with malware attack may start running certain programs automatically like sending messages, re-configuring internal software and turning of anti virus. In case of IoBNT application an infected bio cyber interface might start dispensing inappropriate amount of drugs, thus causing harm to the patient.
Some of the popular malware attacks that are still active in IoT environments are Mirai, Echobot, Reaper, Emotet, Gamut and Nucer \cite{wazid2019iomt}. There different types of malware attacks, but malware attacks specific to IoBNT applications can be botnets, rootkit, ransomware,and keylogger attack.  \\
\textbf{Botnets:} In this type of attack, attacker gains access to a number interconnected devices by launching malware and further control them to steal information, launch DDoS (Destributed Denial of Service) attack to launch unplanned system downtime and even sell network access to other cyber criminals.\\
\textbf{Ransomeware:} The attacker gets hold of the device by encrypting user's data and locks the device,thus, restricting access of the owner to the device. The attacker than demands for some ransome amount by displaying messages on the device screen.\\   
\textbf{Keylogger:} It is a malicious piece of code that records the keystrokes of user to gain access to ID and passwords. This attack is dangerous than brute force attacks and strong passwords does not provide prtection against this malware. \\
\textbf{Rootkit:} A malicious piece of code is installed on IoT device that hides its identity to steal data, reconfigure the device, or control the system by executing malicious commands. This kind of attack is most dangerous as it bypasses all the security mechanisms and successfully hides its presence.  
\textit{Countermeasures: } Access control, device hardening and system monitoring, Antivirus,Intrusion Detection. (See Section \ref{section:POTENTIAL MITIGATION STRATEGIES})
\subsubsection{FIRMWARE ATTACKS}
Firmware updates are necessary in order to ensure proper device functioning. The firmware updates can be done remotely or directly(through usb port). An advertisement is usually broadcast on the network whenever a new version of the firmware update is available. Hackers can send false firmware update requests to the users in order to access the firmware and program it with malicious code. Fake firmware updates can cause fatal consequences to the patients.
\textit{Countermeasures:} Firmware Encryption, periodic firmware updates,malicious firmware detection
}

\subsection{SUMMARY ON THE SECURITY OF IOBNT}
The protection of IoBNT based systems is crucial as the consequences of security compromise can be detrimental. A summarization of existing security proposals for nanonetworks is presented in Table \ref{tab:nanosecurity}. Apart from the software attacks mentioned in the above
section, there exist some other attributes can effect the
security of IoBNT. The origin from where the attack has been instantiated can help in providing some insights into
the profile and goals of the attacker. Local attacks require the attacker to be close to the physical location of the
patient. Local attackers can replace the bio-electronics device with a maliciously programmed device, can
eavesdrop the communication between bio-electronic devices and gateway devices to gather information to
launch an active attack remotely. Remote attacks are launched outside the premises of the bio-electronic devices.
These attacks can be launched by accessing the smartphone of the patient that is delegated a gateway device, to send malware and reconfiguration requests to reprogram the device.
In wireless communication, attackers can evade the system in two possible ways: active and passive.
Passive attack is launched by an attacker silently without revealing their presence in the system. A passive attack can violate the privacy of the patient and can reveal confidential
information about the device. By just intercepting the communication, a passive attacker can gather information
like the patient’s location, diagnosed disease, type of treatment, etc. This information can be used to further
mount active attacks. Active attacker not only enables the attacker to read the transmission but also uses this information to disrupt the
communication. The active attacker can modify the messages in transit, drop the messages so they do not arrive at the destination, block the communication, reprogram the device and even induce a shock to the patient by manipulating the device. As the major application of area of IoBNT is bio medicine, therefore, breach in security means life threatening consequences.

\section{POTENTIAL MITIGATION STRATEGIES} \label{section:POTENTIAL MITIGATION STRATEGIES}

This section discusses the network security methods employed in typical wireless networks, as well as the potential for each measure to be applied to the domain of bio-electronic devices.

\subsection{CRYPTOGRAPHIC MECHANISMS}
 Cryptographic primitives are efficient security mechanisms
that protect the wireless channel from attacks that cause device tampering and information disclosure. Moreover,
cryptographic mechanisms enforce access control to ensure that information is accessible to authorized entities only.
Cryptographic solutions depend upon key management and distribution. There are three possible ways in which key
distribution can take place i.e., unkeyed, symmetric key and
public-key cryptography. Unkeyed cryptography does not use any key and is mainly implemented as hash
functions or one-way permutations. This scheme can be implemented using Message Authentication Codes, pseudorandom sequences, block ciphers, and identification
primitives \cite{camara2015security}. In asymmetric key distribution schemes, a
secret key is shared between communicating parties which is used to generate an authentication token. The
authentication token is used to access the device and to encrypt the communication. In symmetric cryptography key distribution can take in two ways; pre key distribution and
on-demand key distribution. In pre key distribution, key is pre-loaded inside the communicating entities. Pre key
distribution is efficient in the environments where number
of communication parties is less and fixed. On demand key distribution is suitable for distributed and scalable
environments. In asymmetric or public cryptography two sets of keys i.e.,
public and private, need to be shared among communicating entities. This scheme depends upon high
message exchange for authentication procedure which
makes it expensive in terms of computation and power
consumption. Additionally, ciphers in this scheme result in
complex circuits that demand high computation resources
\cite{gollakota2011they, lee2008elliptic}. Due to these limitations, public-key
cryptography is not feasible for resource-constrained
devices. The feasibility and application of cryptographic schemes
for bio-electronic device security are questionable. The envisioned bio-electronic devices have limited memory and maximum of which will be occupied by biomedical
functionality. Inclusion of expensive security primitives is
not possible in the current situation; however additional
memory chips can be integrated into the device for security
functionality. This might also not be recommendable as it
will increase the size of the device \cite{camara2015security}.
{
\subsection{DEVICE HARDENING}
Device hardening and system monitoring has been pointed out in \cite{choi2018system} to mitigate malware attacks. The device hardening refers to restrict unauthorized access to the device by blocking unused communication ports and denying access to unknown IP addresses, blocking reboot from alternate media. In security monitoring the event logs of network traffic to and from bio cyber interface must be continuously monitored to detect an anomaly at early stages.   
}

\subsection{DELEGATION TO EXTERNAL DEVICES}
As discussed above that limited memory of bio-electronic devices restraint them from executing sophisticated security
solutions. Some authors in a similar domain of resource-constrained devices have suggested delegating a part or all
of the security functionality to external devices. These devices will not be part of the human body but can be
thought of as wearable bracelets, watches or even a smartphone that patient carries all the time. Delegation of security to an external device will allow using expensive
encryption, cryptographic and accessing control mechanisms. These external devices will work as an authenticating device that checks the incoming requests
before transmitting them to the bio-electronic device. Some
proposals in this direction can be found for security in IMD
like MedMon \cite{zhang2013medmon}, IMD Cloaker \cite{son2010u} and RFID
Guardian \cite{rieback2005rfid}. The idea of using an external device has
some drawbacks as well. The external device can itself be
compromised and impersonated by malicious entities to
communicate with the bio-electronic device. Moreover, if
the device is lost the bio-electronic device will become
unavailable to queries and commands from remote
healthcare providers.

\subsection{INTRUSION DETECTION (ID)}
There are two types of intrusion detection schemes; signature-based intrusion detection and anomaly-based
intrusion detection. In signature-based ID schemes new data is confronted with the pre-recorded intrusions dataset
and in case of match countermeasure mechanisms are activated. Intrusion detection techniques are used as a
classical security solution to thwart DoS attacks. DoS attack
can be launched in bio-electronic devices by depleting the
limited resources and ultimately making the device unavailable. Anomaly detection in wireless communication
is generally performed by evaluating observable wireless
patterns (time, frequency, location, etc.). Hei et. al. \cite{hei2013resource} have
proposed a Support Vector Machine (SVM) based algorithm to defend against RD and DoS attacks. The application runs on the patient’s smartphone and takes five types of input data to carry out detection: reader action, a time interval of reader action, time of the day, location, day of the week. The classifier is trained with regular input
patterns. When the new request comes in, SVM validates the input data and reply accordingly. Access is granted if
the data is valid, access is denied if the credential is wrong
and if a request is repeated frequently with wrong credentials, the application sends sleep command to IMD
and turns off to external requests. Similar to proposals of
IMD and other resource-constrained devices, intrusion
detection techniques can be implemented for bio-electronic that employ active jamming techniques to deny access to an
adversary. Research efforts in security are now utilizing Artificial Intelligence (AI) for intrusion detection systems. There are three main catagories of AI that can be utilized to
detect intrusions in a system. Supervised learning category
can include techniques based on SVM and naïve bayes classifier, the second category is the semi supervised
learning that includes techniques such as K-means and K-nearest neighbours and the third category id deep learning.
A review on AI based ID techniques for software defined
wireless sensor netwroks can be found in \cite{umba2019review}. An
intrusion detection system for wireless nanosensors has been proposed \cite{rahim2013intrusion} that aims at detecting malicious nodes
and data. This proposal is based on a literature survey and
does not provide satisfactory experimental results.

\subsection{ACCESS CONTROL} \label{subsectionac}
Access control mechanisms enforce rules and policies to prevent unauthorized access to the devices. Most of the
access control schemes revolve around the concepts of a subject, object, and rights. A subject is an entity that
requests to access the device, an object is an entity that needs to be accessed e.g., file, data, image, etc. Rights are
the permissible operations that can be performed by a subject e.g., read, write, modify, etc. Access control
policies are applied to the subjects after authentication, which provides the identification of requester and access
grant decision is made on the basis of requestor identity.
Two popular access control methods are ACLs (Access Control Lists) and PKI (Public Key Infrastructure) \cite{camara2015security}.
ACL is based on discretionary access control that provides a matrix containing subjects, objects they can access and rights that they possess over the objects. While PKI is a certificate-based solution, the requester receives an authentication certificate in each session of connection establishment with the object. 
{ The most popular authentication schemes in access control are RBAC(Role based Access Control) and ABAC(Role based Access Control). In RBAC, access is granted over the object according to the rights of roles. Where as in ABAC access to an objected is granted after evaluating attributes of the requester. Another light weight authentication sceme for IoT environments is ACE (Authentication and Authorization
for Constrained Environments)\cite{aragon2018ace}.}  
Access control mechanisms can be used in parallel with other encryption or cryptographic measures.

\subsection{COMMUNICATION TECHNOLOGY SPECIFIC SECURITY}
Bio cyber interfaces can communicate with gateway devices using a number of communication technologies such as Bluetooth, 4G and other next generation technologies,BLE, and Wi-Fi, to name a few. Each communication technology poses security threat to user's. Therefore, security in communication technologies must be preserved.  
\subsection{HARDWARE BASED SOLUTIONS}
To counter the device tampering attack, some hardware based solution can be implemented \cite{abdul2019comprehensive}. 
Tamper proofing and self
destruction: Use of tamper proof packages for bio electronic devices and execution of self destruction mode upon encountering a physical attack.
PUF(Physically Unclonable Function): PUF is a function in which a noisy fuction is added to the IC(Intergrated Circuit) of the device. PUF is unclonable and tamper-proof. Moreover, PUFs has unique object identification and authentication which can detect unintentended changes in the IC.  
\subsection{CONSUMER TRAINING}
End users i.e., patients,clinicians or care takers must be given training and demo sessions about the possible risks of this technology \cite{bernal2019security}. Moreover, users must be notified if an attack is detected to take part in defense by turning of the device.  
\begin{table*}[!htpb]
\begin{center}
\setlength{\extrarowheight}{10pt}

\caption{Attack types, Counter measures and compromised security goals in Bio cyber interface}
\label{tab:attackscountermeasures}
\begin{tabu}{X[3]X[4]X[4]}
\hline
Attack Type &   Possible Countermeasure &   Compromised goal \\ \hline
  Eavesdropping &   Access Control Strong cryptographic schemes &   Non-Repudition Confidentiality Authorization   \\
Replay &
  Intrusion detection Delegate authentication to external devices &
  Authentication Confidentiality
   \\
MITM &
  Cryptographic primitives &
  Availability Authentication
   \\
Resource Depletion &
  Access control mechanisms  Anomaly detection system User notification Use rechargeable batteries &
  Availability
   \\
Device Tempering &
  Tamper proofing  and self destruction Device Hardening  Physically unclonable Function (PUF) &
  All
   \\
DoS &
  Intrusion Detection Analysis of the physical medium Use of directional antennas &
  Intergrity Availability Non-Repudition Authorization   \\
Malware &
  Access control Device hardening and system monitoring Antivirus Intrusion Detection &
  Confidentiality Integrity Availability Authentication    \\
Firmware attack &
  Firmware Encryption Periodic Firmware updates Malicious Firmware detection &
  All    \\
Injection Attacks &   Intrusion Detection Input Validation Authorization Techniques &  Integrity Non-Repudiation Authorization \\ \hline
\end{tabu}
\end{center}
\end{table*}

\subsection{SUMMARY OF ATTACK MITIGATION TECHNIQUES}
Along with the implementation of countermeasures, below are listed some of the general security guidelines for a secure holistic IoBNT eco-system \cite{abdul2019comprehensive}.
\begin{itemize}
    \item IoBNT is envisioned to use the IoT communnication backbone for communication \textit{via} internet. The already established field of IoT provides security solutions for each layer i.e., link layer, transport layer and network layer \cite{abdul2019comprehensive}. These security solutions are equally application for IoBNT applications.
    \item Secure boot process for the device is a critical security requirement as hackers might add a malicious patch which leads to replacing device's firmware with mailicious one.
    \item Multi-factor authentication schemes like biometric information of the patient and strong access control mechanisms will make the access challenging to the hackers. 
    \item Creating logs of device activity and access events will aid in detecting the anomalies both malicious and unintentional. 
    \item Initiatives for user awareness and risk factors of this novel technology will definitely aid in a secure IoBNT system.  

\end{itemize}

\section{ CHALLENGES AND OPPORTUNITIES}
This section provides a detailed discussion on the challenges and opportunities for IoBNT and its security.

\subsection{CHALLENGES FOR IOBNT}
The knowledge of computer scientists is far from being complete in the field of biology, electrochemistry,
mechanical engineering, and physics. The seamless interface between biological and electrical world is only
possible when researchers form interdisciplinary fieldwork
in close integration. These devices need a collaboration of researchers for the above-mentioned disciplines to make
them a reality. In this direction, research groups from chemical engineering, mechanical engineering, biological sciences, and computer sciences are already working individually.

\subsubsection{NANO NETWORKS CHALLENGES}
The functionality of nanomachines can be heavily affected by changes in surrounding environmental factors such as pH levels and temperature etc. An attacker might alter the
environmental factors with malicious intent to make the communication ineffective. Moreover, if an attacker gains physical access to the human body there are some possible attacks that an attacker can launch to disturb the
functionality of nanomachines. For example, attackers might inject reprogrammed nanomachines to destroy the
previous therapeutic nanonetwork. The physical medium in which the nanonetworks are deployed poses unique fluidic medium physical attacks that are specific to the surrounding
environment where nanonetwork is deployed. For example
attacks such as:
\begin{itemize}
    \item Viruses in the human body might compromise the
communication between deployed nanomachines.
\item Inherent Human Immune System of the patient’s
body may try to destroy the nanomachines by
interpreting them as intruders.
\item Data might become unreliable due to signal
propagation and changes in the propagation
medium.
\end{itemize}

Other specific nanonetworking challenges are listed below:
\paragraph{TRANSMITTER AND RECIEVER DESIGN} 
The design of nano
transceivers is the preliminary step towards the realization
of fully operational nanonetworks. There are two approaches found in theory that can be adopted for the
development of envisioned nano transceivers; biological
structures enabled by synthetic biology and nanomaterials based structures. The MC theory still relies on assumptions
as there is no implementation of artificial nano/microsystem
to date, yet the theory is loaded with a plethora of design options for nano transmitters and receivers, the feasibility of which could not be validated. There is a clear discrepancy between theory and practice, further research, in theory, should be coupled with experimental validation. Particularly, \textit{in-vitro} wet labs should be envisioned for healthcare applications that will be subjected to further clinical trials.
\paragraph{ADDRESSING IN NANONETWORK} Another major challenge related to nanonetworks is the naming and addressing of nanomachines. There are a huge number of nanomachines in a nano network, addressing each nanomachine with a unique identity is not feasible. However, novel addressing
schemes such as FCNN (Function Centric Nano Networking) \cite{stelzner2017function} is an option for dealing with the addressing issue. Instead of addressing each and every
node, FCNN based approach assigns addresses to functions
and location of the body.\
\paragraph{POWERING NANOMACHINES} The issue of power source is
very crucial in nanomachines, as nanomachines are not connected to an external power source. Moreover, the
battery unit is mass wise the largest unit which may increase the size of nanomachines which is undesirable.
Therefore, energy scavenging schemes for drawing energy from inside the body needs to be investigated.\\
\paragraph{SAFETY CHALLENGES} Physical protection of patients is very
crucial so that the attacker might not be able to access the patient’s body with malicious intent. Physical attacks can be launched via administering the medicine orally that contains maliciously programmed nanomachines or
injecting malicious nanomachines into the patient's body. Moreover, the use of THz wave on the human skin can
cause heat and effect the tissue of skin, therefore a communication and sensing standard should be maintained.

\subsubsection{BIO CYBER INTERFACE CHALLENGES}
Despite having made tremendous research, wearable bioelectronic devices face challenges in the way of wide
acceptance and adoption. There are just a few examples of fully functional, successfully deployed and FDA (Food and
Drug Administration) approved bio-electronic devices found in the literature. Some proposals are demonstrated through theoretical models, models of some devices are
being simulated and some are under clinical trials. The research in these devices needs collaborations between university research teams, clinical organizations, community health service providers, small companies emerging with specific technology offerings, and large
worldwide corporations interested to influence their existing technologies into new market offerings. This
section provides challenges and their possible solutions towards the overall challenges in IoBNT domain and its
security.
Main goal of bio cyber interface is to accurately read and convert chemical and biological processes from inside body into electrical signals. For this goal, possible solutions might be found in chemical, physical and biological sensors which promise unprecedented sensing abilities. Therefore, for biologically inspired bio cyber interfaces challanges like ligand-receptor selection,realistic ICT-based modeling of artificial structures and, biological circuits complexity must be addressed.
Moreover,these devices are intended to operate directly in contact with the epidermal biological environment. The issues of biocompatibility, biodegradability, stretchability, and biofouling must be taken into account while designing these devices. When these devices are implanted the body may
perceive it as a foreign agent and biological plaque starts building up around the sensor device, which causes degradation of sensor performance. Therefore, bio-inspired approaches and material must be considered while designing these devices. 
Enhancing comfort while minimizing the size of the device will increase the likelihood of device adoption. Flexible
materials based on CNT \cite{liu2019carbon,wang2019advanced} and graphene \cite{huang2019graphene},
\cite{kabiri2017graphene} are constantly being investigated for the fabrication
of these devices. Electronic tattoos are body-compliant wearable devices that combine an attractive performance of
electrochemical devices with a favorable substrate-skin elasticity of temporary tattoos and resistance to mechanical
stress. The realization of these electronic tattoos and their
integration with a domain like IoBNT will open up a plethora of applications in the healthcare domain. The
currently developed electronic tattoos are able to withstand strain and deformations without deteriorating the device performance. The tattoos are used temporarily as compared to other counterpart bio-electronic devices; hence leaching reagents (i.e., detaching from its carrier) is not a critical
issue. However to use these electronic tattoos for long term use issues like leaching reagents, need for re-calibration, effects of washing and increased stretchability for wearers
comfort must be revisited and addressed. The devices currently manufactured are tested rigorously to withstand
mechanical deformation caused by external strain. However, in cases of long term use and extreme physical
stress conditions, the device abilities might be degraded. To counter this deformation effect, self-healing \cite{cordier2008self, tee2012electrically,yang2015flexible}
devices are being manufactured \cite{bandodkar2016wearable} that utilize healing agents encapsulated in microcapsules \cite{bandodkar2015self}. The healing agents quickly dissolve the crack in the device by filling it with agents like hexyl acetate.
Summing up,  the interface design encompasses expertise from the different disciplines such as wireless communication, physics, biology and optogenetics. Moreover, there is a need to take a step ahead from theoratical research and computer simulations towards experimental research. Wet labs are currently not available or ae extremely costly to perform experimental validation of healthcare applications of IoBNT.  

\subsubsection{BIG DATA ANALYTICS}
Realization of Internet of Bio-Nano Things leads to the interconnection of millions of devices from nano to macro
scale. The interconnection of these devices means the generation of big data in large volume, velocity, and variety \cite{rizwan2018review}. Efficient monitoring and diagnostic in IoBNT
critically depend upon the quality of collected sensory data and management of this big data is important to obtain
clean data for further analysis. Currently, bio-electronics are being developed using divergent technology,
equipment, and services. This diversity of technology yields in data generation of having diverse formats and transfer
protocols. A unified solution is needed which provides some standard for light data formats and protocols for better
efficiency of real-time health monitoring applications \cite{shirazi2012protocol}.
Another issue with an increased amount of data is increased noisy data and erroneous and missing values. This bad
quality of data affects performance analytics and may lead to incorrect findings. New data validation and big data analytics algorithms are the need of time to improve the quality of raw data.\\
 Data aggregation of intra-body nanosensors and data collected from subsequent smart hops is also an issue. New rules of data aggregation are also needed to be developed to
ensure smooth processing, sharing, and analysis of data. Early detection and prevention of disease can only be
achieved when intelligent knowledge extraction techniques are applied to collected sensory big data. For real-time monitoring, bio-electronic devices must provide high data
rates, which require powerful data storage and processing ability. Keeping in mind the small size of these devices, the above-mentioned capabilities cannot be implemented in these devices. Therefore researchers are exploring cloud-enabled integration \cite{doukas2011managing, liu2015external} with these devices to store and process raw sensory data in cloud platforms.

\subsection{CHALLENGES IN SECURITY OF IOBNT}
The leverage to connect and control the human body through the internet not only opens up exciting IoBNT applications but also open up gateways for hackers to use
the same with malicious intent. This event will pose a serious security threat to the patients termed as “Bio cyber terrorism” in \cite{akyildiz2015internet}. Security is, therefore, the most important and critical challenge for the implementation and adoption of bio-electronic devices. Data generated by these devices is highly confidential as it reveals the most sensitive
information of the patient. Hackers and adversaries can launch a number of attacks by misusing this data which can
lead to fatal consequences. Data tampering and alteration by the middle man (hacker) can lead to incorrect
physiological values, which in turn results in an erroneous analysis of the patient’s condition and flawed prescriptions from healthcare provider. The overall communication route
from device to healthcare provider needs protection from security breaches down the way. Solutions to security issues
of IoBNT cannot be found in existing mechanisms of traditional wireless networking. As the networking domain
of IoBNT is itself novel and has diverse system requirements. The security mechanisms and access control \cite{zhang2015secure} methods for IoBNT must be lightweight and
compatible with resource-constrained \cite{lin2016differential} IoBNT devices
and nanonetwork.
\paragraph{CRYPTOGRAPHIC PRIMITIVES}
As IoBNT deals with sensitive physiological parameters of patients, strong cryptographic mechanisms must be implemented to ensure user privacy. Considering the resource-constrained environment of IoBNT, symmetric and asymmetric cryptographic systems are computationally very expensive.
The selection of a feasible cryptographic system for IoBNT depends upon the following characteristics of
cryptosystems; Energy, memory and execution time. A novel field of bio-chemical cryptography has been proposed by Dressler and Kargl \cite{dressler2012towards} which seems promising for nanonetworks. Biochemical cryptography techniques proposed so far use DNA molecules for encryption. There a number of proposals using DNA cryptography as an encryption scheme \cite{Dhawan2012SecureDT,cui2008encryption,rusia2014review,zhang2012research, jacob2013dna,jacob2013encryption,gao2008new,gehani2003dna}.
\paragraph{AUTHENTICATION} Authentication and access control
schemes might not be possible for intra body nanonetworks. However, the source of data and messges communicated to
the gateway devices must be authenticated before forwarding to intra body nanonetworks.\\
 Nano communication mediums i.e., MC, nanoelectromagnetic communication, and nano-acoustic communication should be considered separately when
designing authentication schemes. An authentication scheme for nano-electromagnetic communication has been proposed \cite{rahman2017physical} recently which exploits distance-dependent path loss as a device fingerprint to discriminate the data
sent by legitimate transmitter and intruder. In the case of MC, this authentication scheme might not be feasible due to underlying communication differences. For MC the emerging field of biochemical cryptography \cite{dressler2012towards} i.e., the
use of biological molecules such as DNA and Ribonucleic Acid (RNA) as a source of encryption might be explored to
solve authentication problems of molecular communications at nano-scale. However, it comes with entirely new challenges for the researchers and
collaboration of biologists, computer scientists and biochemists might be required for investigation innovative
ways \cite{usman2018security}.\\
\paragraph{ENCRYPTION} 
Key management is still a major challenge in
nanonetworks whether it be pre-distribution before network
deployment or pro-active distribution before data transmission. Both the schemes have their drawbacks at the
nano-scale, pre-distribution is not feasible as storing a large
number of pairwise keys is not possible considering the limited storage capacity of nano-devices. Pro-active or ondemand key distribution needs online access to the devices, which are not feasible \textit{in-vivo} nanonetwork applications where online access to nanonetwork is limited. 

\hak{
Some researchers have proposed ECC (Elliptic Curve Cryptography) \cite{malasri2009design,seo2009tinyecck16,malasri2006snap} for resource-constrained devices, but in real-time ECC based primitives are still expensive in terms of time complexity. Moreover, quantum based cryptographic algorithms in resource constrained devices have proven to efficient \cite{fernandez2019pre}. 
}
In nanocommunication networks, the use of public-key cryptography is not very realistic due to the very high
resource limitations, however biochemical cryptography \cite{dressler2012towards} or DNA cryptography \cite{jacob2013dna}
might be potentially suitable cryptography mechanisms at
the nano-scale.\\
 Bio-inspired algorithms are now being adopted for computing and networking problems, which work on the
principle of nature-inspired solutions \cite{dressler2010bio,dressler2010survey}. Bio-inspired computing is a research field that finds principles
of problem-solving in nature and maps them into algorithms for computing domain. A subclass of this domain is bio-inspired algorithms for security, which provides algorithms involving techniques adopted by nature to defend against intruders \cite{rauf2018taxonomy,dressler2005bio}. The most popular natural defense system is the human immune system, which
fights against intruders and external viruses to prevent them from entering the body. In this direction, some proposals have been presented to improve cyber security in the field of
wireless sensor networks. Bio-inspired algorithms such as Artificial Immune System (AIS) \cite{noeparast2012cognitive}, Genetic mutation \cite{fulp2015evolutionary} and Ant Colony Optimization (ACO) \cite{bitam2016bio} are
utilized to mitigate against certain cyber-attacks.

The bio-inspired algorithms are a promising novel security approach that is scalable and robust and provides plenty of algorithms that can be adopted to mitigate against attacks in IoBNT. One of the goals for future work is to use bio-inspired algorithms to provide security solutions for IoBNT.

\section{Conclusions}
\hak{IoBNT is an emerging research domain that holds significant promises in various healthcare applications. These applications range from early symptom detection to remote diagnosis to treatments of patients (e.g., via targeted drug delivery), and so on. The design and integration of device components are dependent on individual IoBNT deployment requirements and settings (e.g., in sensing applications, relevant sensors must be integrated to sense individual biochemical substances like pH, sodium, and calcium). However, there are a number of challenges such as those identified and discussed in this paper and the challenges moving forward are interdisciplinary (e.g., physics, chemical engineering, biology, mechanical engineering, and computer science). In addition to these challenges, this research highlights prospective research avenues for future researchers as well as IoBNT hardware and software designers and engineers.}
\begin{table*}[!htpb]
\begin{center}
\caption{Summarization of challenges in Internet of Bio Nano Things\label{tab:challanges}}
\setlength{\extrarowheight}{10pt}
\begin{tabu}{X[3]X[4]X[4]}
\hline
Area of Concentration 	&Brief Description 	&Research Initiatives \\  \hline
Nano Network Challanges   & -Transmitter and Receiver Design
\newline -Addressing in nanonetwork
\newline -Powering Nanomachines
\newline -Safety Challenges
& -Stochastic simulations
and experiments to validate the performance.
\newline -Realistic ICT-based modeling of artificial structures.
\newline -Function Centric Nano Networking(FCNN) based approach assigns addresses to functions
and location of the body.\\ 
Bio Cyber Interface Challanges      &-Need to accurately read and convert chemical and biological processes into chemical signals.
\newline -Interdisciplinary research
\newline -Experimental validation.
\newline -Bio compatible design
&-Exploration of potential materials and mechanisms for interfacing between nano,micro and macro communications 
\newline Needs collaborations between university research teams(Wireless communication, physics, biotechnology disciplines), clinical organizations, community health service providers.
\newline -Wet lab experiments
\newline Use of biological materials to prevent biofouling and formation of plaque around the device surface.
\\ 
Big Data Analytics  &-Unavailability of realistic data for feature selection in ML algorithms. 
\newline -Data produced is in diverse formats and transfer protocols
&-Use of deep learning techniques to automate feature selection
\newline -A unified solution is needed which provides some standard for light data formats and protocols for better
efficiency of real-time health monitoring applications 
\\
Security in IoBNT     &-Compromise of devices and data 
\newline -Compromise of patient privacy and safety
&-Design and validation procedures that address Confidentiality, Integrity, and Availability.
\newline -Access control methods to limit unauthorized access.
-newline -Development of novel security frameworks for resource constraint devices. \\
\bottomrule
\end{tabu}
\end{center}
\end{table*}

\hak{The field of security in to IoBNT technologies is not yet mature, generating opportunities for attackers, even non-sophisticated attacks can have a significant impact on both IoBNT technologies and users’ safety. There is a current opportunity for standardization initiatives to unify IoBNT in terms of information security. The well established fields like IoT, WBANs and IMDs have been helpful in identifying attack types in IoBNT. Apart from software and hardware security mechanisms, there is a need for user's training and awareness for wide adoption of these novel technologies. One of the future plans is to build and implement systems that can detect and mitigate security attacks affecting the theragnostics process in real time.}

\begin{table}
 \caption{Summary of Notations}
\label{tab:notations}
\begin{tabularx}{\linewidth}{p{2cm}p{6cm}}
\toprule
Abbr. &Full Form  \\ 
\midrule
AIS &Artificial Immune System\\
ATP &Adenosine Triphosphate\\ 
BAN & Body Area Network\\
BFC & Biofuel Cell\\
BRET & Bioluminescence Resonance Energy Transfer\\
CHD & Coronary Heart Disease\\
CNT &Carbon Nano Tubes\\
CPS &Cyber Physical System\\
DbMC &Diffusion  Based  Molecular  Communication\\
DES & Drug Eluting Stents\\
DNA & Deoxyribonucleic Acid\\
DoS & Denial of Service\\
ECC & Elliptic Curve Cryptography\\
EM & Electromagnetic Communication\\
FES & Functional Electric Stimulations\\
FET & Field Effect Transistor\\
FRET &Fluorescence Resonance Energy Transfer\\
HIS &Human Immune System\\
IC & Integrated Circuit\\
ICT &Information and Communication Technology\\
IMD &Implantable Medical Device\\
IoBNT & Internet of Bio Nano Things\\
IoMt &Internet of Medical Things\\
IoNT & Internet of Nano Things\\
IoT & Internet of Things\\
L & Luciferin \\
LU &Luciferase\\
MICS &Medical  Implant  Communications  Service\\
MITM &Man-In-The-Middle\\
MC &Molecular Communication\\
NFC &Near-Field-Communication\\
OSI &Open Systems Interconnection \\
PDA &Personal Digital Assistant\\
RFID &Radio-frequency IDentification\\
Rx &Reciever\\
SIR &Susceptible,  Infected,Recovered\\
SPR &  Surface Plasmon Resonance\\
SVM &Support Vector Machine\\
TCP/IP  &Transmission Control Protocol/Internet Protocol\\
TDD & Targeted Drug Delivery\\
THz & Tera Hertz\\
Tx &Transmitter\\
WBAN & Wireless Body Area Network\\
WCS &Wireless Chemical Sensor\\
\bottomrule
\end{tabularx}
\end{table}

\section{Acknowledgment} 
This research work is intended to be included in the Ph.D. thesis of Sidra Zafar, supervised by Dr.Mohsin Nazir and Dr.Aneeqa Sabah. We thank all the other co-authors for their contributions in this research work.

\bibliographystyle{IEEEtran}
\bibliography{Ref}

\begin{thebibliography}{100}
\providecommand{\url}[1]{#1}
\csname url@samestyle\endcsname
\providecommand{\newblock}{\relax}
\providecommand{\bibinfo}[2]{#2}
\providecommand{\BIBentrySTDinterwordspacing}{\spaceskip=0pt\relax}
\providecommand{\BIBentryALTinterwordstretchfactor}{4}
\providecommand{\BIBentryALTinterwordspacing}{\spaceskip=\fontdimen2\font plus
\BIBentryALTinterwordstretchfactor\fontdimen3\font minus
  \fontdimen4\font\relax}
\providecommand{\BIBforeignlanguage}[2]{{%
\expandafter\ifx\csname l@#1\endcsname\relax
\typeout{** WARNING: IEEEtran.bst: No hyphenation pattern has been}%
\typeout{** loaded for the language `#1'. Using the pattern for}%
\typeout{** the default language instead.}%
\else
\language=\csname l@#1\endcsname
\fi
#2}}
\providecommand{\BIBdecl}{\relax}
\BIBdecl

\bibitem{steager2013automated}
E.~B. Steager, M.~Selman~Sakar, C.~Magee, M.~Kennedy, A.~Cowley, and V.~Kumar,
  ``Automated biomanipulation of single cells using magnetic microrobots,''
  \emph{The International Journal of Robotics Research}, vol.~32, no.~3, pp.
  346--359, 2013.

\bibitem{kim2009microengineered}
D.-H. Kim, P.~K. Wong, J.~Park, A.~Levchenko, and Y.~Sun, ``Microengineered
  platforms for cell mechanobiology,'' \emph{Annual review of biomedical
  engineering}, vol.~11, pp. 203--233, 2009.

\bibitem{ergeneman2008magnetically}
O.~Ergeneman, G.~Dogangil, M.~P. Kummer, J.~J. Abbott, M.~K. Nazeeruddin, and
  B.~J. Nelson, ``A magnetically controlled wireless optical oxygen sensor for
  intraocular measurements,'' \emph{IEEE Sensors Journal}, vol.~8, no.~1, pp.
  29--37, 2008.

\bibitem{nakamura2008capsule}
T.~Nakamura and A.~Terano, ``Capsule endoscopy: past, present, and future,''
  \emph{Journal of gastroenterology}, vol.~43, no.~2, pp. 93--99, 2008.

\bibitem{dubach2007fluorescent}
J.~M. Dubach, D.~I. Harjes, and H.~A. Clark, ``Fluorescent ion-selective
  nanosensors for intracellular analysis with improved lifetime and size,''
  \emph{Nano Letters}, vol.~7, no.~6, pp. 1827--1831, 2007.

\bibitem{chen2005development}
C.-J. Chen, Y.~Haik, and J.~Chatterjee, ``Development of nanotechnology for
  biomedical applications,'' in \emph{Conference, Emerging Information
  Technology 2005.}\hskip 1em plus 0.5em minus 0.4em\relax Taipei, Taiwan:
  IEEE, 2005, pp. 4--pp.

\bibitem{carlsen2014bio}
R.~W. Carlsen and M.~Sitti, ``Bio-hybrid cell-based actuators for
  microsystems,'' \emph{Small}, vol.~10, no.~19, pp. 3831--3851, 2014.

\bibitem{aylott2003optical}
J.~W. Aylott, ``Optical nanosensors—an enabling technology for intracellular
  measurements,'' \emph{Analyst}, vol. 128, no.~4, pp. 309--312, 2003.

\bibitem{timko2010remotely}
B.~P. Timko, T.~Dvir, and D.~S. Kohane, ``Remotely triggerable drug delivery
  systems,'' \emph{Advanced materials}, vol.~22, no.~44, pp. 4925--4943, 2010.

\bibitem{pan2011swallowable}
G.~Pan and L.~Wang, ``Swallowable wireless capsule endoscopy: Progress and
  technical challenges,'' \emph{Gastroenterology research and practice}, vol.
  2012, pp. 841\,691--841\,691, 2011.

\bibitem{fernandez2007magnetic}
R.~Fern{\'a}ndez-Pacheco, C.~Marquina, J.~G. Valdivia, M.~Guti{\'e}rrez, M.~S.
  Romero, R.~Cornudella, A.~Laborda, A.~Viloria, T.~Higuera, A.~Garc{\'\i}a
  \emph{et~al.}, ``Magnetic nanoparticles for local drug delivery using
  magnetic implants,'' \emph{Journal of Magnetism and Magnetic Materials}, vol.
  311, no.~1, pp. 318--322, 2007.

\bibitem{liao2010indications}
Z.~Liao, R.~Gao, C.~Xu, and Z.-S. Li, ``Indications and detection, completion,
  and retention rates of small-bowel capsule endoscopy: a systematic review,''
  \emph{Gastrointestinal endoscopy}, vol.~71, no.~2, pp. 280--286, 2010.

\bibitem{li2003cholesterol}
J.~Li, T.~Peng, and Y.~Peng, ``A cholesterol biosensor based on entrapment of
  cholesterol oxidase in a silicic sol-gel matrix at a prussian blue modified
  electrode,'' \emph{Electroanalysis: An International Journal Devoted to
  Fundamental and Practical Aspects of Electroanalysis}, vol.~15, no.~12, pp.
  1031--1037, 2003.

\bibitem{kawahara2013chip}
T.~Kawahara, M.~Sugita, M.~Hagiwara, F.~Arai, H.~Kawano, I.~Shihira-Ishikawa,
  and A.~Miyawaki, ``On-chip microrobot for investigating the response of
  aquatic microorganisms to mechanical stimulation,'' \emph{Lab on a Chip},
  vol.~13, no.~6, pp. 1070--1078, 2013.

\bibitem{fluckiger2007ultrasound}
M.~Fluckiger and B.~J. Nelson, ``Ultrasound emitter localization in
  heterogeneous media,'' in \emph{2007 29th Annual International Conference of
  the IEEE Engineering in Medicine and Biology Society}.\hskip 1em plus 0.5em
  minus 0.4em\relax ,: IEEE, 2007, pp. 2867--2870.

\bibitem{freitas2006pharmacytes}
R.~A. Freitas, ``Pharmacytes: An ideal vehicle for targeted drug delivery,''
  \emph{Journal of Nanoscience and Nanotechnology}, vol.~6, no. 9-10, pp.
  2769--2775, 2006.

\bibitem{akyildiz2008nanonetworks}
I.~F. Akyildiz, F.~Brunetti, and C.~Bl{\'a}zquez, ``Nanonetworks: A new
  communication paradigm,'' \emph{Computer Networks}, vol.~52, no.~12, pp.
  2260--2279, 2008.

\bibitem{marzo2019nanonetworks}
J.~L. Marzo, J.~M. Jornet, and M.~Pierobon, ``Nanonetworks in biomedical
  applications,'' \emph{Current drug targets}, vol.~20, no.~8, pp. 800--807,
  2019.

\bibitem{felicetti2014tcp}
L.~Felicetti, M.~Femminella, G.~Reali, T.~Nakano, and A.~V. Vasilakos,
  ``Tcp-like molecular communications,'' \emph{IEEE Journal on Selected Areas
  in Communications}, vol.~32, no.~12, pp. 2354--2367, 2014.

\bibitem{nakano2012molecular}
T.~Nakano, M.~J. Moore, F.~Wei, A.~V. Vasilakos, and J.~Shuai, ``Molecular
  communication and networking: Opportunities and challenges,'' \emph{IEEE
  transactions on nanobioscience}, vol.~11, no.~2, pp. 135--148, 2012.

\bibitem{akyildiz2010electromagnetic}
I.~F. Akyildiz and J.~M. Jornet, ``Electromagnetic wireless nanosensor
  networks,'' \emph{Nano Communication Networks}, vol.~1, no.~1, pp. 3--19,
  2010.

\bibitem{akyildiz2010internet}
{I. F. Akyildiz and J. M. Jornet,}, ``The internet of nano-things,'' \emph{IEEE
  Wireless Communications}, vol.~17, no.~6, pp. 58--63, 2010.

\bibitem{velloso2011web}
E.~Velloso, D.~Cardador, K.~Vega, W.~Ugulino, A.~Bulling, H.~Gellersen, and
  H.~Fuks, ``The web of things as an infrastructure for improving users’
  health and wellbeing,'' in \emph{II Workshop of the Brazilian Institute for
  Web Science Research, Rio de Janeiro, Brazil}, 2011.

\bibitem{michel2017optogenerapy}
F.~Michel and M.~Folcher, ``Optogenerapy: When bio-electronic implant enters
  the modern syringe era,'' \emph{Porto Biomedical Journal}, vol.~2, no.~5, pp.
  145--149, 2017.

\bibitem{akyildiz2015internet}
I.~F. Akyildiz, M.~Pierobon, S.~Balasubramaniam, and Y.~Koucheryavy, ``The
  internet of bio-nano things,'' \emph{IEEE Communications Magazine}, vol.~53,
  no.~3, pp. 32--40, 2015.

\bibitem{kirichek2016live}
R.~Kirichek, R.~Pirmagomedov, R.~Glushakov, and A.~Koucheryavy, ``Live
  substance in cyberspace—biodriver system,'' in \emph{2016 18th
  International Conference on Advanced Communication Technology (ICACT)}.\hskip
  1em plus 0.5em minus 0.4em\relax Pyeongchang, South Korea: IEEE, 2016, pp.
  274--278.

\bibitem{kulakowski2020nano}
P.~Kulakowski, K.~Turbic, and L.~M. Correia, ``From nano-communications to body
  area networks: A perspective on truly personal communications,'' \emph{IEEE
  Access}, vol.~8, pp. 159\,839--159\,853, 2020.

\bibitem{usman2018security}
M.~Usman, M.~R. Asghar, I.~S. Ansari, and M.~Qaraqe, ``Security in wireless
  body area networks: From in-body to off-body communications,'' \emph{IEEE
  Access}, vol.~6, pp. 58\,064--58\,074, 2018.

\bibitem{loscri2014security}
V.~Loscri, C.~Marchal, N.~Mitton, G.~Fortino, and A.~V. Vasilakos, ``Security
  and privacy in molecular communication and networking: Opportunities and
  challenges,'' \emph{IEEE transactions on nanobioscience}, vol.~13, no.~3, pp.
  198--207, 2014.

\bibitem{yang2019comprehensive}
K.~Yang, D.~Bi, Y.~Deng, R.~Zhang, M.~Rahman, N.~A. Ali, M.~A. Imran, J.~M.
  Jornet, Q.~H. Abbasi, and A.~Alomainy, ``A comprehensive survey on hybrid
  communication for internet of nano-things in context of body-centric
  communications,'' \emph{arXiv preprint arXiv:1912.09424}, 2019.

\bibitem{al2020intelligence}
F.~Al-Turjman, ``Intelligence and security in big 5g-oriented iont: An
  overview,'' \emph{Future Generation Computer Systems}, vol. 102, pp.
  357--368, 2020.

\bibitem{akyildiz20206g}
I.~F. Akyildiz, A.~Kak, and S.~Nie, ``6g and beyond: The future of wireless
  communications systems,'' \emph{IEEE Access}, vol.~8, pp. 133\,995--134\,030,
  2020.

\bibitem{silva2020domiciliary}
A.~F. Silva and M.~Tavakoli, ``Domiciliary hospitalization through wearable
  biomonitoring patches: Recent advances, technical challenges, and the
  relation to covid-19,'' \emph{Sensors}, vol.~20, no.~23, p. 6835, 2020.

\bibitem{williams2020electronic}
N.~X. Williams and A.~D. Franklin, ``Electronic tattoos: A promising approach
  to real-time theragnostics,'' \emph{Journal of Dermatology and Skin Science},
  vol.~2, no.~1, 2020.

\bibitem{oh2019second}
J.~Y. Oh and Z.~Bao, ``Second skin enabled by advanced electronics,''
  \emph{Advanced Science}, vol.~6, no.~11, p. 1900186, 2019.

\bibitem{kassal2013wireless}
P.~Kassal, I.~M. Steinberg, and M.~D. Steinberg, ``Wireless smart tag with
  potentiometric input for ultra low-power chemical sensing,'' \emph{Sensors
  and actuators B: chemical}, vol. 184, pp. 254--259, 2013.

\bibitem{kassal2018wireless}
P.~Kassal, M.~D. Steinberg, and I.~M. Steinberg, ``Wireless chemical sensors
  and biosensors: A review,'' \emph{Sensors and Actuators B: Chemical}, vol.
  266, pp. 228--245, 2018.

\bibitem{singh2017inkjet}
R.~Singh, E.~Singh, and H.~S. Nalwa, ``Inkjet printed nanomaterial based
  flexible radio frequency identification (rfid) tag sensors for the internet
  of nano things,'' \emph{RSC advances}, vol.~7, no.~77, pp. 48\,597--48\,630,
  2017.

\bibitem{bandodkar2015tattoo}
A.~J. Bandodkar, W.~Jia, and J.~Wang, ``Tattoo-based wearable electrochemical
  devices: A review,'' \emph{Electroanalysis}, vol.~27, no.~3, pp. 562--572,
  2015.

\bibitem{qadri2020future}
Y.~A. Qadri, A.~Nauman, Y.~B. Zikria, A.~V. Vasilakos, and S.~W. Kim, ``The
  future of healthcare internet of things: A survey of emerging technologies,''
  \emph{IEEE Communications Surveys \& Tutorials}, vol.~22, no.~2, pp.
  1121--1167, 2020.

\bibitem{pramanik2020advancing}
P.~K.~D. Pramanik, A.~Solanki, A.~Debnath, A.~Nayyar, S.~El-Sappagh, and K.-S.
  Kwak, ``Advancing modern healthcare with nanotechnology, nanobiosensors, and
  internet of nano things: Taxonomies, applications, architecture, and
  challenges,'' \emph{IEEE Access}, vol.~8, pp. 65\,230--65\,266, 2020.

\bibitem{bernal2019security}
S.~L. Bernal, A.~H. Celdr{\'a}n, G.~M. P{\'e}rez, M.~T. Barros, and
  S.~Balasubramaniam, ``Security in brain-computer interfaces:
  State-of-the-art, opportunities, and future challenges,'' \emph{arXiv
  preprint arXiv:1908.03536}, 2019.

\bibitem{dissanayak2021exact}
M.~B. Dissanayak and N.~Ekanayake, ``On the exact performance analysis of
  molecular communication via diffusion for internet of bio-nano things.''
  \emph{IEEE Transactions on Nanobioscience}, 2021.

\bibitem{al2017cognitive}
F.~Al-Turjman, ``A cognitive routing protocol for bio-inspired networking in
  the internet of nano-things (iont),'' \emph{Mobile Networks and
  Applications}, vol.~:, pp. 1--15, 2017.

\bibitem{roman2004single}
C.~Roman, F.~Ciontu, and B.~Courtois, ``Single molecule detection and
  macromolecular weighting using an all-carbon-nanotube nanoelectromechanical
  sensor,'' in \emph{4th IEEE Conference on Nanotechnology, 2004.}\hskip 1em
  plus 0.5em minus 0.4em\relax ,: IEEE, 2004, pp. 263--266.

\bibitem{schedin2007detection}
F.~Schedin, A.~K. Geim, S.~V. Morozov, E.~Hill, P.~Blake, M.~Katsnelson, and
  K.~S. Novoselov, ``Detection of individual gas molecules adsorbed on
  graphene,'' \emph{Nature materials}, vol.~6, no.~9, pp. 652--655, 2007.

\bibitem{tallury2010nanobioimaging}
P.~Tallury, A.~Malhotra, L.~M. Byrne, and S.~Santra, ``Nanobioimaging and
  sensing of infectious diseases,'' \emph{Advanced drug delivery reviews},
  vol.~62, no. 4-5, pp. 424--437, 2010.

\bibitem{yeh2009real}
H.-Y. Yeh, M.~V. Yates, W.~Chen, and A.~Mulchandani, ``Real-time molecular
  methods to detect infectious viruses,'' in \emph{Seminars in cell \&
  developmental biology}, vol.~20.\hskip 1em plus 0.5em minus 0.4em\relax ,:
  Elsevier, 2009, pp. 49--54.

\bibitem{buther2017formal}
F.~B{\"u}ther, F.-L. Lau, M.~Stelzner, and S.~Ebers, ``A formal definition for
  nanorobots and nanonetworks,'' in \emph{Internet of Things, Smart Spaces, and
  Next Generation Networks and Systems}.\hskip 1em plus 0.5em minus 0.4em\relax
  St. Petersburg,Russia: Springer, 2017, pp. 214--226.

\bibitem{pereira2011electronic}
L.~F. Pereira and M.~S. Ferreira, ``Electronic transport on carbon nanotube
  networks: A multiscale computational approach,'' \emph{Nano Communication
  Networks}, vol.~2, no.~1, pp. 25--38, 2011.

\bibitem{yoo2011bio}
J.-W. Yoo, D.~J. Irvine, D.~E. Discher, and S.~Mitragotri, ``Bio-inspired,
  bioengineered and biomimetic drug delivery carriers,'' \emph{Nature reviews
  Drug discovery}, vol.~10, no.~7, pp. 521--535, 2011.

\bibitem{nishimura2014genetic}
Y.~Nishimura, J.~Ishii, C.~Ogino, and A.~Kondo, ``Genetic engineering of
  bio-nanoparticles for drug delivery: A review,'' \emph{Journal of Biomedical
  Nanotechnology}, vol.~10, no.~9, pp. 2063--2085, 2014.

\bibitem{tan2015cell}
S.~Tan, T.~Wu, D.~Zhang, and Z.~Zhang, ``Cell or cell membrane-based drug
  delivery systems,'' \emph{Theranostics}, vol.~5, no.~8, p. 863, 2015.

\bibitem{lockney2011viruses}
D.~Lockney, S.~Franzen, and S.~Lommel, ``Viruses as nanomaterials for drug
  delivery,'' in \emph{Biomedical Nanotechnology}.\hskip 1em plus 0.5em minus
  0.4em\relax ,: Springer, 2011, pp. 207--221.

\bibitem{esfandiari2016new}
N.~Esfandiari, M.~K. Arzanani, M.~Soleimani, M.~Kohi-Habibi, and W.~E.
  Svendsen, ``A new application of plant virus nanoparticles as drug delivery
  in breast cancer,'' \emph{Tumor Biology}, vol.~37, no.~1, pp. 1229--1236,
  2016.

\bibitem{steidler2004live}
L.~Steidler, ``Live genetically modified bacteria as drug delivery tools: at
  the doorstep of a new pharmacology?'' 2004.

\bibitem{yacoby2007targeted}
I.~Yacoby, H.~Bar, and I.~Benhar, ``Targeted drug-carrying bacteriophages as
  antibacterial nanomedicines,'' \emph{Antimicrobial agents and chemotherapy},
  vol.~51, no.~6, pp. 2156--2163, 2007.

\bibitem{batrakova2011cell}
E.~V. Batrakova, H.~E. Gendelman, and A.~V. Kabanov, ``Cell-mediated drug
  delivery,'' \emph{Expert opinion on drug delivery}, vol.~8, no.~4, pp.
  415--433, 2011.

\bibitem{su2015design}
Y.~Su, Z.~Xie, G.~B. Kim, C.~Dong, and J.~Yang, ``Design strategies and
  applications of circulating cell-mediated drug delivery systems,'' \emph{ACS
  biomaterials science \& engineering}, vol.~1, no.~4, pp. 201--217, 2015.

\bibitem{strobel2021richard}
G.~Strobel and J.~Mittnacht, ``Richard, are we there yet?-an internet of nano
  things information system architecture,'' in \emph{Proceedings of the 54th
  Hawaii International Conference on System Sciences}, 2021, p. 4578.

\bibitem{koucheryavy2021review}
Y.~Koucheryavy, A.~Yastrebova, D.~P. Martins, and S.~Balasubramaniam, ``A
  review on bio-cyber interfaces for intrabody molecular communications
  systems,'' \emph{arXiv preprint arXiv:2104.14944}, 2021.

\bibitem{kassal2017smart}
P.~Kassal, M.~Zubak, G.~Scheipl, G.~J. Mohr, M.~D. Steinberg, and I.~M.
  Steinberg, ``Smart bandage with wireless connectivity for optical monitoring
  of ph,'' \emph{Sensors and Actuators B: Chemical}, vol. 246, pp. 455--460,
  2017.

\bibitem{chude2016biologically}
U.~A. Chude-Okonkwo, R.~Malekian, and B.~T. Maharaj, ``Biologically inspired
  bio-cyber interface architecture and model for internet of bio-nanothings
  applications,'' \emph{IEEE Transactions on Communications}, vol.~64, no.~8,
  pp. 3444--3455, 2016.

\bibitem{nakano2014externally}
T.~Nakano, S.~Kobayashi, T.~Suda, Y.~Okaie, Y.~Hiraoka, and T.~Haraguchi,
  ``Externally controllable molecular communication,'' \emph{IEEE Journal on
  Selected Areas in Communications}, vol.~32, no.~12, pp. 2417--2431, 2014.

\bibitem{hu2014enzyme}
Q.~Hu, P.~S. Katti, and Z.~Gu, ``Enzyme-responsive nanomaterials for controlled
  drug delivery,'' \emph{Nanoscale}, vol.~6, no.~21, pp. 12\,273--12\,286,
  2014.

\bibitem{wu2009genetically}
Y.~I. Wu, D.~Frey, O.~I. Lungu, A.~Jaehrig, I.~Schlichting, B.~Kuhlman, and
  K.~M. Hahn, ``A genetically encoded photoactivatable rac controls the
  motility of living cells,'' \emph{Nature}, vol. 461, no. 7260, pp. 104--108,
  2009.

\bibitem{ellis2007caged}
G.~C. Ellis-Davies, ``Caged compounds: photorelease technology for control of
  cellular chemistry and physiology,'' \emph{Nature methods}, vol.~4, no.~8,
  pp. 619--628, 2007.

\bibitem{dykman2012gold}
L.~Dykman and N.~Khlebtsov, ``Gold nanoparticles in biomedical applications:
  recent advances and perspectives,'' \emph{Chemical Society Reviews}, vol.~41,
  no.~6, pp. 2256--2282, 2012.

\bibitem{needham2000new}
D.~Needham, G.~Anyarambhatla, G.~Kong, and M.~W. Dewhirst, ``A new
  temperature-sensitive liposome for use with mild hyperthermia:
  characterization and testing in a human tumor xenograft model,'' \emph{Cancer
  research}, vol.~60, no.~5, pp. 1197--1201, 2000.

\bibitem{dicheva2013cationic}
B.~M. Dicheva, T.~L.~t. Hagen, L.~Li, D.~Schipper, A.~L. Seynhaeve, G.~C.~v.
  Rhoon, A.~M. Eggermont, L.~H. Lindner, and G.~A. Koning, ``Cationic
  thermosensitive liposomes: a novel dual targeted heat-triggered drug delivery
  approach for endothelial and tumor cells,'' \emph{Nano letters}, vol.~13,
  no.~6, pp. 2324--2331, 2013.

\bibitem{kono2007temperature}
K.~Kono, T.~Miyoshi, Y.~Haba, E.~Murakami, C.~Kojima, and A.~Harada,
  ``Temperature sensitivity control of alkylamide-terminated poly (amidoamine)
  dendrimers induced by guest molecule binding,'' \emph{Journal of the American
  Chemical Society}, vol. 129, no.~23, pp. 7222--7223, 2007.

\bibitem{hamad2002remote}
K.~Hamad-Schifferli, J.~J. Schwartz, A.~T. Santos, S.~Zhang, and J.~M.
  Jacobson, ``Remote electronic control of dna hybridization through inductive
  coupling to an attached metal nanocrystal antenna,'' \emph{Nature}, vol. 415,
  no. 6868, pp. 152--155, 2002.

\bibitem{abbasi2017information}
N.~A. Abbasi and O.~B. Akan, ``An information theoretical analysis of human
  insulin-glucose system toward the internet of bio-nano things,'' \emph{IEEE
  transactions on nanobioscience}, vol.~16, no.~8, pp. 783--791, 2017.

\bibitem{nakano2014molecular}
T.~Nakano, T.~Suda, Y.~Okaie, M.~J. Moore, and A.~V. Vasilakos, ``Molecular
  communication among biological nanomachines: A layered architecture and
  research issues,'' \emph{IEEE transactions on nanobioscience}, vol.~13,
  no.~3, pp. 169--197, 2014.

\bibitem{rudsari2021tdma}
H.~K. Rudsari, M.~R. Javan, M.~Orooji, N.~Mokari, and E.~A. Jorswieck,
  ``Tdma-mtmr-based molecular communication with ligand-binding reception,''
  \emph{IEEE Transactions on Molecular, Biological and Multi-Scale
  Communications}, 2021.

\bibitem{farsad2016comprehensive}
N.~Farsad, H.~B. Yilmaz, A.~Eckford, C.-B. Chae, and W.~Guo, ``A comprehensive
  survey of recent advancements in molecular communication,'' \emph{IEEE
  Communications Surveys \& Tutorials}, vol.~18, no.~3, pp. 1887--1919, 2016.

\bibitem{kuscu2019transmitter}
M.~Kuscu, E.~Dinc, B.~A. Bilgin, H.~Ramezani, and O.~B. Akan, ``Transmitter and
  receiver architectures for molecular communications: A survey on physical
  design with modulation, coding, and detection techniques,'' \emph{Proceedings
  of the IEEE}, vol. 107, no.~7, pp. 1302--1341, 2019.

\bibitem{kuscu2016modeling}
M.~Kuscu and O.~B. Akan, ``Modeling and analysis of sinw fet-based molecular
  communication receiver,'' \emph{IEEE Transactions on Communications},
  vol.~64, no.~9, pp. 3708--3721, 2016.

\bibitem{kuscu2016physical}
{Kuscu, Murat and Akan, Ozgur B}, ``On the physical design of molecular
  communication receiver based on nanoscale biosensors,'' \emph{IEEE Sensors
  Journal}, vol.~16, no.~8, pp. 2228--2243, 2016.

\bibitem{kim2013novel}
N.-R. Kim and C.-B. Chae, ``Novel modulation techniques using isomers as
  messenger molecules for nano communication networks via diffusion,''
  \emph{IEEE Journal on Selected Areas in Communications}, vol.~31, no.~12, pp.
  847--856, 2013.

\bibitem{kuran2010energy}
M.~{\c{S}}. Kuran, H.~B. Yilmaz, T.~Tugcu, and B.~{\"O}zerman, ``Energy model
  for communication via diffusion in nanonetworks,'' \emph{Nano Communication
  Networks}, vol.~1, no.~2, pp. 86--95, 2010.

\bibitem{garralda2011diffusion}
N.~Garralda, I.~Llatser, A.~Cabellos-Aparicio, E.~Alarc{\'o}n, and M.~Pierobon,
  ``Diffusion-based physical channel identification in molecular
  nanonetworks,'' \emph{Nano Communication Networks}, vol.~2, no.~4, pp.
  196--204, 2011.

\bibitem{kuran2011modulation}
M.~S. Kuran, H.~B. Yilmaz, T.~Tugcu, and I.~F. Akyildiz, ``Modulation
  techniques for communication via diffusion in nanonetworks,'' in \emph{2011
  IEEE international conference on communications (ICC)}.\hskip 1em plus 0.5em
  minus 0.4em\relax Kyoto, Japan: IEEE, 2011, pp. 1--5.

\bibitem{farsad2014channel}
N.~Farsad, N.-R. Kim, A.~W. Eckford, and C.-B. Chae, ``Channel and noise models
  for nonlinear molecular communication systems,'' \emph{IEEE Journal on
  Selected Areas in Communications}, vol.~32, no.~12, pp. 2392--2401, 2014.

\bibitem{llatser2011exploring}
I.~Llatser, I.~Pascual, N.~Garralda, A.~Cabellos-Aparicio, M.~Pierobon,
  E.~Alarc{\'o}n, and J.~Sol{\'e}-Pareta, ``Exploring the physical channel of
  diffusion-based molecular communication by simulation,'' in \emph{2011 IEEE
  Global Telecommunications Conference-GLOBECOM 2011}.\hskip 1em plus 0.5em
  minus 0.4em\relax Houston, TX, USA, USA: IEEE, 2011, pp. 1--5.

\bibitem{chou2013extended}
C.~T. Chou, ``Extended master equation models for molecular communication
  networks,'' \emph{IEEE transactions on nanobioscience}, vol.~12, no.~2, pp.
  79--92, 2013.

\bibitem{noel2014improving}
A.~Noel, K.~C. Cheung, and R.~Schober, ``Improving receiver performance of
  diffusive molecular communication with enzymes,'' \emph{IEEE Transactions on
  NanoBioscience}, vol.~13, no.~1, pp. 31--43, 2014.

\bibitem{pierobon2010physical}
M.~Pierobon and I.~F. Akyildiz, ``A physical end-to-end model for molecular
  communication in nanonetworks,'' \emph{IEEE Journal on Selected Areas in
  Communications}, vol.~28, no.~4, pp. 602--611, 2010.

\bibitem{felicetti2016applications}
L.~Felicetti, M.~Femminella, G.~Reali, and P.~Li{\`o}, ``Applications of
  molecular communications to medicine: A survey,'' \emph{Nano Communication
  Networks}, vol.~7, pp. 27--45, 2016.

\bibitem{chude2017molecular}
U.~A. Chude-Okonkwo, R.~Malekian, B.~T. Maharaj, and A.~V. Vasilakos,
  ``Molecular communication and nanonetwork for targeted drug delivery: A
  survey,'' \emph{IEEE Communications Surveys \& Tutorials}, vol.~19, no.~4,
  pp. 3046--3096, 2017.

\bibitem{chude2016molecular}
U.~A. Chude-Okonkwo, R.~Malekian, and B.~S. Maharaj, ``Molecular communication
  model for targeted drug delivery in multiple disease sites with diversely
  expressed enzymes,'' \emph{IEEE transactions on nanobioscience}, vol.~15,
  no.~3, pp. 230--245, 2016.

\bibitem{chude2019nanosystems}
U.~Chude-Okonkwo, R.~Malekian, and B.~Maharaj, ``Nanosystems and devices for
  advanced targeted nanomedical applications,'' in \emph{Advanced Targeted
  Nanomedicine}.\hskip 1em plus 0.5em minus 0.4em\relax ,: Springer, 2019, pp.
  39--58.

\bibitem{chahibi2014antibody}
Y.~Chahibi, I.~F. Akyildiz, and S.~O. Song, ``Antibody-based molecular
  communication for targeted drug delivery systems,'' in \emph{2014 36th Annual
  International Conference of the IEEE Engineering in Medicine and Biology
  Society}.\hskip 1em plus 0.5em minus 0.4em\relax Chicago, IL, USA: IEEE,
  2014, pp. 5707--5710.

\bibitem{chahibi2014molecular}
Y.~Chahibi and I.~F. Akyildiz, ``Molecular communication noise and capacity
  analysis for particulate drug delivery systems,'' \emph{IEEE Transactions on
  Communications}, vol.~62, no.~11, pp. 3891--3903, 2014.

\bibitem{chahibi2015molecular}
Y.~Chahibi, I.~F. Akyildiz, S.~Balasubramaniam, and Y.~Koucheryavy, ``Molecular
  communication modeling of antibody-mediated drug delivery systems,''
  \emph{IEEE Transactions on Biomedical Engineering}, vol.~62, no.~7, pp.
  1683--1695, 2015.

\bibitem{chahibi2015pharmacokinetic}
Y.~Chahibi, M.~Pierobon, and I.~F. Akyildiz, ``Pharmacokinetic modeling and
  biodistribution estimation through the molecular communication paradigm,''
  \emph{IEEE Transactions on Biomedical Engineering}, vol.~62, no.~10, pp.
  2410--2420, 2015.

\bibitem{piro2015terahertz}
G.~Piro, K.~Yang, G.~Boggia, N.~Chopra, L.~A. Grieco, and A.~Alomainy,
  ``Terahertz communications in human tissues at the nanoscale for healthcare
  applications,'' \emph{IEEE Transactions on Nanotechnology}, vol.~14, no.~3,
  pp. 404--406, 2015.

\bibitem{jornet2011channel}
J.~M. Jornet and I.~F. Akyildiz, ``Channel modeling and capacity analysis for
  electromagnetic wireless nanonetworks in the terahertz band,'' \emph{IEEE
  Transactions on Wireless Communications}, vol.~10, no.~10, pp. 3211--3221,
  2011.

\bibitem{abbasi2016nano}
Q.~H. Abbasi, K.~Yang, N.~Chopra, J.~M. Jornet, N.~A. Abuali, K.~A. Qaraqe, and
  A.~Alomainy, ``Nano-communication for biomedical applications: A review on
  the state-of-the-art from physical layers to novel networking concepts,''
  \emph{IEEE Access}, vol.~4, pp. 3920--3935, 2016.

\bibitem{zhang2016biosensors}
D.~Zhang and Q.~Liu, ``Biosensors and bioelectronics on smartphone for portable
  biochemical detection,'' \emph{Biosensors and Bioelectronics}, vol.~75, pp.
  273--284, 2016.

\bibitem{movassaghi2012wireless}
S.~Movassaghi, P.~Arab, and M.~Abolhasan, ``Wireless technologies for body area
  networks: Characteristics and challenges,'' in \emph{2012 International
  Symposium on Communications and Information Technologies (ISCIT)}.\hskip 1em
  plus 0.5em minus 0.4em\relax Gold Coast, QLD, Australia: IEEE, 2012, pp.
  42--47.

\bibitem{strey2013bluetooth}
H.~H. Strey, P.~Richman, R.~Rozensky, S.~Smith, and L.~Endee, ``Bluetooth low
  energy technologies for applications in health care: proximity and
  physiological signals monitors,'' in \emph{2013 10th International Conference
  and Expo on Emerging Technologies for a Smarter World (CEWIT)}.\hskip 1em
  plus 0.5em minus 0.4em\relax Melville, NY, USA: IEEE, 2013, pp. 1--4.

\bibitem{georgakakis2010analysis}
E.~Georgakakis, S.~A. Nikolidakis, D.~D. Vergados, and C.~Douligeris, ``An
  analysis of bluetooth, zigbee and bluetooth low energy and their use in
  wbans,'' in \emph{International Conference on Wireless Mobile Communication
  and Healthcare}.\hskip 1em plus 0.5em minus 0.4em\relax ,: Springer, 2010,
  pp. 168--175.

\bibitem{cao2009enabling}
H.~Cao, V.~Leung, C.~Chow, and H.~Chan, ``Enabling technologies for wireless
  body area networks: A survey and outlook,'' \emph{IEEE Communications
  Magazine}, vol.~47, no.~12, pp. 84--93, 2009.

\bibitem{kurunathan2015study}
J.~H. Kurunathan, ``Study and overview on wban under ieee 802.15. 6,'' \emph{U.
  Porto Journal of Engineering}, vol.~1, no.~1, pp. 11--21, 2015.

\bibitem{ieee2010ieee}
I.~.~W. Group \emph{et~al.}, ``Ieee standard for information
  technology--telecommunications and information exchange between
  systems--local and metropolitan area networks--specific requirements--part
  11: Wireless lan medium access control (mac) and physical layer (phy)
  specifications amendment 6: Wireless access in vehicular environments,''
  IEEE, Tech. Rep.~11, 2010.

\bibitem{lewis2010ieee}
D.~Lewis, ``Ieee p802. 15.6/d0 draft standard for body area network (vol. 6),''
  IEEE, Tech. Rep., 2010.

\bibitem{mile2018hybrid}
A.~Mile, G.~Okeyo, and A.~Kibe, ``Hybrid ieee 802.15. 6 wireless body area
  networks interference mitigation model for high mobility interference
  scenarios,'' \emph{Wireless Engineering and Technology}, vol.~9, no.~2, pp.
  34--48, 2018.

\bibitem{nie2015statistical}
Z.~Nie, Z.~Li, R.~Huang, Y.~Liu, J.~Li, and L.~Wang, ``A statistical frame
  based tdma protocol for human body communication,'' \emph{Biomedical
  engineering online}, vol.~14, no.~1, p.~65, 2015.

\bibitem{jena2019wireless}
S.~Jena, A.~Gupta, R.~K. Pippara, P.~Pal \emph{et~al.}, ``Wireless sensing
  systems: A review,'' in \emph{Sensors for Automotive and Aerospace
  Applications}.\hskip 1em plus 0.5em minus 0.4em\relax ,: Springer, 2019, pp.
  143--192.

\bibitem{grosinger2013feasibility}
J.~Grosinger, ``Feasibility of backscatter rfid systems on the human body,''
  \emph{EURASIP journal on embedded systems}, vol. 2013, no.~1, pp. 1--10,
  2013.

\bibitem{kurs2007wireless}
A.~Kurs, A.~Karalis, R.~Moffatt, J.~D. Joannopoulos, P.~Fisher, and
  M.~Solja{\v{c}}i{\'c}, ``Wireless power transfer via strongly coupled
  magnetic resonances,'' \emph{science}, vol. 317, no. 5834, pp. 83--86, 2007.

\bibitem{lee2013wearable}
Y.-H. Lee, J.-S. Kim, J.~Noh, I.~Lee, H.~J. Kim, S.~Choi, J.~Seo, S.~Jeon,
  T.-S. Kim, J.-Y. Lee \emph{et~al.}, ``Wearable textile battery rechargeable
  by solar energy,'' \emph{Nano letters}, vol.~13, no.~11, pp. 5753--5761,
  2013.

\bibitem{lee2014highly}
J.-H. Lee, K.~Y. Lee, M.~K. Gupta, T.~Y. Kim, D.-Y. Lee, J.~Oh, C.~Ryu, W.~J.
  Yoo, C.-Y. Kang, S.-J. Yoon \emph{et~al.}, ``Highly stretchable
  piezoelectric-pyroelectric hybrid nanogenerator,'' \emph{Advanced Materials},
  vol.~26, no.~5, pp. 765--769, 2014.

\bibitem{bae2011fiber}
J.~Bae, M.~K. Song, Y.~J. Park, J.~M. Kim, M.~Liu, and Z.~L. Wang, ``Fiber
  supercapacitors made of nanowire-fiber hybrid structures for
  wearable/flexible energy storage,'' \emph{Angewandte Chemie International
  Edition}, vol.~50, no.~7, pp. 1683--1687, 2011.

\bibitem{mercier2012energy}
P.~P. Mercier, A.~C. Lysaght, S.~Bandyopadhyay, A.~P. Chandrakasan, and K.~M.
  Stankovic, ``Energy extraction from the biologic battery in the inner ear,''
  \emph{Nature biotechnology}, vol.~30, no.~12, pp. 1240--1243, 2012.

\bibitem{meredith2012biofuel}
M.~T. Meredith and S.~D. Minteer, ``Biofuel cells: enhanced enzymatic
  bioelectrocatalysis,'' \emph{Annual review of analytical chemistry}, vol.~5,
  pp. 157--179, 2012.

\bibitem{zebda2011mediatorless}
A.~Zebda, C.~Gondran, A.~Le~Goff, M.~Holzinger, P.~Cinquin, and S.~Cosnier,
  ``Mediatorless high-power glucose biofuel cells based on compressed carbon
  nanotube-enzyme electrodes,'' \emph{Nature communications}, vol.~2, no.~1,
  pp. 1--6, 2011.

\bibitem{halamkova2012implanted}
L.~Hal{\'a}mkov{\'a}, J.~Hal{\'a}mek, V.~Bocharova, A.~Szczupak, L.~Alfonta,
  and E.~Katz, ``Implanted biofuel cell operating in a living snail,''
  \emph{Journal of the American Chemical Society}, vol. 134, no.~11, pp.
  5040--5043, 2012.

\bibitem{jia2013epidermal}
W.~Jia, G.~Vald{\'e}s-Ram{\'\i}rez, A.~J. Bandodkar, J.~R. Windmiller, and
  J.~Wang, ``Epidermal biofuel cells: energy harvesting from human
  perspiration,'' \emph{Angewandte Chemie International Edition}, vol.~52,
  no.~28, pp. 7233--7236, 2013.

\bibitem{canovas2018nanoscale}
S.~Canovas-Carrasco, A.-J. Garcia-Sanchez, and J.~Garcia-Haro, ``A nanoscale
  communication network scheme and energy model for a human hand scenario,''
  \emph{Nano communication networks}, vol.~15, pp. 17--27, 2018.

\bibitem{baik2019bioinspired}
S.~Baik, H.~J. Lee, D.~W. Kim, J.~W. Kim, Y.~Lee, and C.~Pang, ``Bioinspired
  adhesive architectures: from skin patch to integrated bioelectronics,''
  \emph{Advanced Materials}, vol.~31, no.~34, p. 1803309, 2019.

\bibitem{islam2016catch}
N.~Islam and S.~Misra, ``“catch the pendulum”: The problem of asymmetric
  data delivery in electromagnetic nanonetworks,'' \emph{IEEE transactions on
  nanobioscience}, vol.~15, no.~6, pp. 576--584, 2016.

\bibitem{atzori2010internet}
L.~Atzori, A.~Iera, and G.~Morabito, ``The internet of things: A survey,''
  \emph{Computer networks}, vol.~54, no.~15, pp. 2787--2805, 2010.

\bibitem{bandodkar2014epidermal}
A.~J. Bandodkar, D.~Molinnus, O.~Mirza, T.~Guinovart, J.~R. Windmiller,
  G.~Vald{\'e}s-Ram{\'\i}rez, F.~J. Andrade, M.~J. Sch{\"o}ning, and J.~Wang,
  ``Epidermal tattoo potentiometric sodium sensors with wireless signal
  transduction for continuous non-invasive sweat monitoring,'' \emph{Biosensors
  and bioelectronics}, vol.~54, pp. 603--609, 2014.

\bibitem{bandodkar2014non}
A.~J. Bandodkar and J.~Wang, ``Non-invasive wearable electrochemical sensors: a
  review,'' \emph{Trends in biotechnology}, vol.~32, no.~7, pp. 363--371, 2014.

\bibitem{lee2019skin}
E.~K. Lee, M.~K. Kim, and C.~H. Lee, ``Skin-mountable biosensors and
  therapeutics: a review,'' \emph{Annual Review of Biomedical Engineering},
  vol.~21, pp. 299--323, 2019.

\bibitem{jin2017advanced}
H.~Jin, Y.~S. Abu-Raya, and H.~Haick, ``Advanced materials for health
  monitoring with skin-based wearable devices,'' \emph{Advanced healthcare
  materials}, vol.~6, no.~11, p. 1700024, 2017.

\bibitem{windmiller2012electrochemical}
J.~R. Windmiller, A.~J. Bandodkar, G.~Vald{\'e}s-Ram{\'\i}rez, S.~Parkhomovsky,
  A.~G. Martinez, and J.~Wang, ``Electrochemical sensing based on printable
  temporary transfer tattoos,'' \emph{Chemical Communications}, vol.~48,
  no.~54, pp. 6794--6796, 2012.

\bibitem{di2015stretch}
J.~Di, S.~Yao, Y.~Ye, Z.~Cui, J.~Yu, T.~K. Ghosh, Y.~Zhu, and Z.~Gu,
  ``Stretch-triggered drug delivery from wearable elastomer films containing
  therapeutic depots,'' \emph{ACS nano}, vol.~9, no.~9, pp. 9407--9415, 2015.

\bibitem{jeong2017nfc}
H.~Jeong, T.~Ha, I.~Kuang, L.~Shen, Z.~Dai, N.~Sun, and N.~Lu, ``Nfc-enabled,
  tattoo-like stretchable biosensor manufactured by “cut-and-paste”
  method,'' in \emph{2017 39th Annual International Conference of the IEEE
  Engineering in Medicine and Biology Society (EMBC)}.\hskip 1em plus 0.5em
  minus 0.4em\relax ,: IEEE, 2017, pp. 4094--4097.

\bibitem{jeong2019electronic}
H.~Jeong and N.~Lu, ``Electronic tattoos: the most multifunctional but
  imperceptible wearables,'' in \emph{Smart Biomedical and Physiological Sensor
  Technology XV}, vol. 11020.\hskip 1em plus 0.5em minus 0.4em\relax ,:
  International Society for Optics and Photonics, 2019, p. 110200P.

\bibitem{lee2018device}
H.~Lee, C.~Song, S.~Baik, D.~Kim, T.~Hyeon, and D.-H. Kim, ``Device-assisted
  transdermal drug delivery,'' \emph{Advanced drug delivery reviews}, vol. 127,
  pp. 35--45, 2018.

\bibitem{hassija2020security}
V.~Hassija, V.~Chamola, B.~C. Bajpai, S.~Zeadally \emph{et~al.}, ``Security
  issues in implantable medical devices: Fact or fiction?'' \emph{Sustainable
  Cities and Society}, p. 102552, 2020.

\bibitem{abdul2019comprehensive}
H.~A. Abdul-Ghani and D.~Konstantas, ``A comprehensive study of security and
  privacy guidelines, threats, and countermeasures: An iot perspective,''
  \emph{Journal of Sensor and Actuator Networks}, vol.~8, no.~2, p.~22, 2019.

\bibitem{jia2013electrochemical}
W.~Jia, A.~J. Bandodkar, G.~Valdes-Ramirez, J.~R. Windmiller, Z.~Yang,
  J.~Ram{\'\i}rez, G.~Chan, and J.~Wang, ``Electrochemical tattoo biosensors
  for real-time noninvasive lactate monitoring in human perspiration,''
  \emph{Analytical chemistry}, vol.~85, no.~14, pp. 6553--6560, 2013.

\bibitem{kim2016noninvasive}
J.~Kim, I.~Jeerapan, S.~Imani, T.~N. Cho, A.~Bandodkar, S.~Cinti, P.~P.
  Mercier, and J.~Wang, ``Noninvasive alcohol monitoring using a wearable
  tattoo-based iontophoretic-biosensing system,'' \emph{Acs Sensors}, vol.~1,
  no.~8, pp. 1011--1019, 2016.

\bibitem{di2019remotely}
N.~Di~Trani, A.~Silvestri, G.~Bruno, T.~Geninatti, C.~Y.~X. Chua, A.~Gilbert,
  G.~Rizzo, C.~S. Filgueira, D.~Demarchi, and A.~Grattoni, ``Remotely
  controlled nanofluidic implantable platform for tunable drug delivery,''
  \emph{Lab on a Chip}, vol.~19, no.~13, pp. 2192--2204, 2019.

\bibitem{lodato2015close}
R.~Lodato and G.~Marrocco, ``Close integration of a uhf-rfid transponder into a
  limb prosthesis for tracking and sensing,'' \emph{IEEE Sensors Journal},
  vol.~16, no.~6, pp. 1806--1813, 2015.

\bibitem{chen2011silicon}
K.-I. Chen, B.-R. Li, and Y.-T. Chen, ``Silicon nanowire field-effect
  transistor-based biosensors for biomedical diagnosis and cellular recording
  investigation,'' \emph{Nano today}, vol.~6, no.~2, pp. 131--154, 2011.

\bibitem{kumar2016intelligent}
N.~Kumar, K.~Kaur, S.~C. Misra, and R.~Iqbal, ``An intelligent rfid-enabled
  authentication scheme for healthcare applications in vehicular mobile
  cloud,'' \emph{Peer-to-Peer Networking and Applications}, vol.~9, no.~5, pp.
  824--840, 2016.

\bibitem{zhang2017review}
J.~Zhang, G.~Y. Tian, A.~M. Marindra, A.~I. Sunny, and A.~B. Zhao, ``A review
  of passive rfid tag antenna-based sensors and systems for structural health
  monitoring applications,'' \emph{Sensors}, vol.~17, no.~2, p. 265, 2017.

\bibitem{kumar2010stage}
S.~Kumar, G.~Livermont, and G.~Mckewan, ``Stage implementation of rfid in
  hospitals,'' \emph{Technology and Health Care}, vol.~18, no.~1, pp. 31--46,
  2010.

\bibitem{cangialosi2007leveraging}
A.~Cangialosi, J.~E. Monaly, and S.~C. Yang, ``Leveraging rfid in hospitals:
  Patient life cycle and mobility perspectives,'' \emph{IEEE Communications
  Magazine}, vol.~45, no.~9, pp. 18--23, 2007.

\bibitem{dey2016rfid}
A.~Dey, B.~Vijayaraman, and J.~H. Choi, ``{RFID in US hospitals: an exploratory
  investigation of technology adoption},'' \emph{Management Research Review},
  vol.~39, no.~4, pp. 399--424, 2016.

\bibitem{asaimi2016developing}
A.~A.~A. Asaimi, M.~I. Razzak, and R.~Alshammari, ``Developing ontology for
  rfid based infant protection system,'' \emph{Advanced Science Letters},
  vol.~22, no.~10, pp. 2759--2763, 2016.

\bibitem{baptista2017radio}
R.~Baptista, G.~Morris, R.~Jones, N.~Ridley, G.~Kushinga, A.~Arshad,
  J.~Douglas, A.~Ricketts, and D.~Moss, ``Radio frequency identification (rfid)
  as a solution for track and trace of cryogenically-stored cell and gene
  therapies,'' \emph{Cytotherapy}, vol.~19, no.~5, p. S122, 2017.

\bibitem{manzari2013modeling}
S.~Manzari and G.~Marrocco, ``Modeling and applications of a chemical-loaded
  uhf rfid sensing antenna with tuning capability,'' \emph{IEEE transactions on
  antennas and propagation}, vol.~62, no.~1, pp. 94--101, 2013.

\bibitem{potyrailo2013passive}
R.~A. Potyrailo and C.~Surman, ``A passive radio-frequency identification
  (rfid) gas sensor with self-correction against fluctuations of ambient
  temperature,'' \emph{Sensors and Actuators B: Chemical}, vol. 185, pp.
  587--593, 2013.

\bibitem{virtanen2011inkjet}
J.~Virtanen, L.~Ukkonen, T.~Bjorninen, A.~Z. Elsherbeni, and L.~Syd{\"a}nheimo,
  ``Inkjet-printed humidity sensor for passive uhf rfid systems,'' \emph{IEEE
  Transactions on Instrumentation and Measurement}, vol.~60, no.~8, pp.
  2768--2777, 2011.

\bibitem{martinez2016design}
J.~J. Mart{\'\i}nez-Mart{\'\i}nez, F.~J. Herraiz-Mart{\'\i}nez, and
  G.~Galindo-Romera, ``Design and characterization of a passive temperature
  sensor based on a printed miw delay line,'' \emph{IEEE Sensors Journal},
  vol.~16, no.~22, pp. 7884--7891, 2016.

\bibitem{kim2013no}
S.~Kim, C.~Mariotti, F.~Alimenti, P.~Mezzanotte, A.~Georgiadis, A.~Collado,
  L.~Roselli, and M.~M. Tentzeris, ``No battery required: Perpetual
  rfid-enabled wireless sensors for cognitive intelligence applications,''
  \emph{IEEE Microwave magazine}, vol.~14, no.~5, pp. 66--77, 2013.

\bibitem{vyas2011inkjet}
R.~Vyas, V.~Lakafosis, H.~Lee, G.~Shaker, L.~Yang, G.~Orecchini, A.~Traille,
  M.~M. Tentzeris, and L.~Roselli, ``Inkjet printed, self powered, wireless
  sensors for environmental, gas, and authentication-based sensing,''
  \emph{IEEE Sensors Journal}, vol.~11, no.~12, pp. 3139--3152, 2011.

\bibitem{rose2014adhesive}
D.~P. Rose, M.~E. Ratterman, D.~K. Griffin, L.~Hou, N.~Kelley-Loughnane, R.~R.
  Naik, J.~A. Hagen, I.~Papautsky, and J.~C. Heikenfeld, ``Adhesive rfid sensor
  patch for monitoring of sweat electrolytes,'' \emph{IEEE Transactions on
  Biomedical Engineering}, vol.~62, no.~6, pp. 1457--1465, 2014.

\bibitem{juels2003blocker}
A.~Juels, R.~L. Rivest, and M.~Szydlo, ``The blocker tag: Selective blocking of
  rfid tags for consumer privacy,'' in \emph{Proceedings of the 10th ACM
  conference on Computer and communications security}, 2003, pp. 103--111.

\bibitem{chen2009low}
Y.-Y. Chen, J.-C. Lu, S.-I. Chen, and J.-K. Jan, ``A low-cost rfid
  authentication protocol with location privacy protection,'' in \emph{2009
  Fifth International Conference on Information Assurance and Security},
  vol.~2.\hskip 1em plus 0.5em minus 0.4em\relax IEEE, 2009, pp. 109--113.

\bibitem{juels2006rfid}
A.~Juels, ``Rfid security and privacy: A research survey,'' \emph{IEEE journal
  on selected areas in communications}, vol.~24, no.~2, pp. 381--394, 2006.

\bibitem{tao2014silk}
H.~Tao, S.-W. Hwang, B.~Marelli, B.~An, J.~E. Moreau, M.~Yang, M.~A. Brenckle,
  S.~Kim, D.~L. Kaplan, J.~A. Rogers \emph{et~al.}, ``Silk-based resorbable
  electronic devices for remotely controlled therapy and in vivo infection
  abatement,'' \emph{Proceedings of the National Academy of Sciences}, vol.
  111, no.~49, pp. 17\,385--17\,389, 2014.

\bibitem{curreli2008real}
M.~Curreli, R.~Zhang, F.~N. Ishikawa, H.-K. Chang, R.~J. Cote, C.~Zhou, and
  M.~E. Thompson, ``Real-time, label-free detection of biological entities
  using nanowire-based fets,'' \emph{IEEE Transactions on Nanotechnology},
  vol.~7, no.~6, pp. 651--667, 2008.

\bibitem{robergs2004biochemistry}
R.~A. Robergs, F.~Ghiasvand, and D.~Parker, ``Biochemistry of exercise-induced
  metabolic acidosis,'' \emph{American Journal of Physiology-Regulatory,
  Integrative and Comparative Physiology}, vol. 287, no.~3, pp. R502--R516,
  2004.

\bibitem{patolsky2006nanowire}
F.~Patolsky, G.~Zheng, and C.~M. Lieber, ``Nanowire-based biosensors,''
  \emph{Analytical chemistry (Washington, DC)}, vol.~78, no.~13, pp.
  4260--4269, 2006.

\bibitem{kim2013highly}
B.~Kim, H.~S. Song, H.~J. Jin, E.~J. Park, S.~H. Lee, B.~Y. Lee, T.~H. Park,
  and S.~Hong, ``Highly selective and sensitive detection of neurotransmitters
  using receptor-modified single-walled carbon nanotube sensors,''
  \emph{Nanotechnology}, vol.~24, no.~28, p. 285501, 2013.

\bibitem{ohno2010label}
Y.~Ohno, K.~Maehashi, and K.~Matsumoto, ``Label-free biosensors based on
  aptamer-modified graphene field-effect transistors,'' \emph{Journal of the
  American Chemical Society}, vol. 132, no.~51, pp. 18\,012--18\,013, 2010.

\bibitem{shan2018high}
J.~Shan, J.~Li, X.~Chu, M.~Xu, F.~Jin, X.~Wang, L.~Ma, X.~Fang, Z.~Wei, and
  X.~Wang, ``High sensitivity glucose detection at extremely low concentrations
  using a mos 2-based field-effect transistor,'' \emph{RSC advances}, vol.~8,
  no.~15, pp. 7942--7948, 2018.

\bibitem{torsi2013organic}
L.~Torsi, M.~Magliulo, K.~Manoli, and G.~Palazzo, ``Organic field-effect
  transistor sensors: a tutorial review,'' \emph{Chemical Society Reviews},
  vol.~42, no.~22, pp. 8612--8628, 2013.

\bibitem{duan2012quantification}
X.~Duan, Y.~Li, N.~K. Rajan, D.~A. Routenberg, Y.~Modis, and M.~A. Reed,
  ``Quantification of the affinities and kinetics of protein interactions using
  silicon nanowire biosensors,'' \emph{Nature nanotechnology}, vol.~7, no.~6,
  p. 401, 2012.

\bibitem{tran2016toward}
D.~P. Tran, M.~A. Winter, B.~Wolfrum, R.~Stockmann, C.-T. Yang,
  M.~Pourhassan-Moghaddam, A.~Offenhausser, and B.~Thierry, ``Toward
  intraoperative detection of disseminated tumor cells in lymph nodes with
  silicon nanowire field effect transistors,'' \emph{ACS nano}, vol.~10, no.~2,
  pp. 2357--2364, 2016.

\bibitem{grebenstein2018biological}
L.~Grebenstein, J.~Kirchner, R.~S. Peixoto, W.~Zimmermann, F.~Irnstorfer,
  W.~Wicke, A.~Ahmadzadeh, V.~Jamali, G.~Fischer, R.~Weigel \emph{et~al.},
  ``Biological optical-to-chemical signal conversion interface: A small-scale
  modulator for molecular communications,'' \emph{IEEE transactions on
  nanobioscience}, vol.~18, no.~1, pp. 31--42, 2018.

\bibitem{liu2017electrochemistry}
Y.~Liu, E.~Kim, J.~Li, M.~Kang, W.~E. Bentley, and G.~F. Payne,
  ``Electrochemistry for bio-device molecular communication: The potential to
  characterize, analyze and actuate biological systems,'' \emph{Nano
  Communication Networks}, vol.~11, pp. 76--89, 2017.

\bibitem{liu2017connecting}
Y.~Liu, J.~Li, T.~Tschirhart, J.~L. Terrell, E.~Kim, C.-Y. Tsao, D.~L. Kelly,
  W.~E. Bentley, and G.~F. Payne, ``Connecting biology to electronics:
  Molecular communication via redox modality,'' \emph{Advanced Healthcare
  Materials}, vol.~6, no.~24, p. 1700789, 2017.

\bibitem{liu2017using}
Y.~Liu, C.-Y. Tsao, E.~Kim, T.~Tschirhart, J.~L. Terrell, W.~E. Bentley, and
  G.~F. Payne, ``Using a redox modality to connect synthetic biology to
  electronics: Hydrogel-based chemo-electro signal transduction for molecular
  communication,'' \emph{Advanced healthcare materials}, vol.~6, no.~1, p.
  1600908, 2017.

\bibitem{kim2019redox}
E.~Kim, J.~Li, M.~Kang, D.~L. Kelly, S.~Chen, A.~Napolitano, L.~Panzella,
  X.~Shi, K.~Yan, S.~Wu \emph{et~al.}, ``Redox is a global biodevice
  information processing modality,'' \emph{Proceedings of the IEEE}, vol. 107,
  no.~7, pp. 1402--1424, 2019.

\bibitem{kang2018redox}
M.~Kang, E.~Kim, J.~Li, W.~E. Bentley, and G.~F. Payne, ``Redox: Electron-based
  approach to bio-device molecular communication,'' in \emph{2018 IEEE 19th
  International Workshop on Signal Processing Advances in Wireless
  Communications (SPAWC)}.\hskip 1em plus 0.5em minus 0.4em\relax ,: IEEE,
  2018, pp. 1--5.

\bibitem{liu2019electrofabricated}
Y.~Liu, J.~S. McGrath, J.~H. Moore, G.~L. Kolling, J.~A. Papin, and N.~S.
  Swami, ``Electrofabricated biomaterial-based capacitor on nanoporous gold for
  enhanced redox amplification,'' \emph{Electrochimica Acta}, vol. 318, pp.
  828--836, 2019.

\bibitem{abd2020molcom}
S.~M. Abd El-atty, R.~Bidar, and E.-S.~M. El-Rabaie, ``Molcom system with
  downlink/uplink biocyber interface for internet of bio-nanothings,''
  \emph{International Journal of Communication Systems}, vol.~33, no.~1, p.
  e4171, 2020.

\bibitem{bakhshi2019securing}
T.~Bakhshi and S.~Shahid, ``Securing internet of bio-nano things: Ml-enabled
  parameter profiling of bio-cyber interfaces,'' in \emph{2019 22nd
  International Multitopic Conference (INMIC)}.\hskip 1em plus 0.5em minus
  0.4em\relax IEEE, 2019, pp. 1--8.

\bibitem{zhang2013medmon}
M.~Zhang, A.~Raghunathan, and N.~K. Jha, ``Medmon: Securing medical devices
  through wireless monitoring and anomaly detection,'' \emph{IEEE Transactions
  on Biomedical circuits and Systems}, vol.~7, no.~6, pp. 871--881, 2013.

\bibitem{camara2015security}
C.~Camara, P.~Peris-Lopez, and J.~E. Tapiador, ``Security and privacy issues in
  implantable medical devices: A comprehensive survey,'' \emph{Journal of
  biomedical informatics}, vol.~55, pp. 272--289, 2015.

\bibitem{kumar2012security}
P.~Kumar and H.-J. Lee, ``Security issues in healthcare applications using
  wireless medical sensor networks: A survey,'' \emph{sensors}, vol.~12, no.~1,
  pp. 55--91, 2012.

\bibitem{islam2017secure}
S.~R. Islam, F.~Ali, H.~Moon, and K.-S. Kwak, ``Secure channel for molecular
  communications,'' in \emph{2017 International Conference on Information and
  Communication Technology Convergence (ICTC)}.\hskip 1em plus 0.5em minus
  0.4em\relax ,: IEEE, 2017, pp. 1--4.

\bibitem{guo2016eavesdropper}
W.~Guo, Y.~Deng, B.~Li, C.~Zhao, and A.~Nallanathan, ``Eavesdropper
  localization in random walk channels,'' \emph{IEEE Communications Letters},
  vol.~20, no.~9, pp. 1776--1779, 2016.

\bibitem{zafar2019channel}
S.~Zafar, W.~Aman, M.~M.~U. Rahman, A.~Alomainy, and Q.~H. Abbasi, ``Channel
  impulse response-based physical layer authentication in a diffusion-based
  molecular communication system,'' in \emph{2019 UK/China Emerging
  Technologies (UCET)}.\hskip 1em plus 0.5em minus 0.4em\relax Glasgow, United
  Kingdom, United Kingdom: IEEE, 2019, pp. 1--2.

\bibitem{giaretta2015security}
A.~Giaretta, S.~Balasubramaniam, and M.~Conti, ``Security vulnerabilities and
  countermeasures for target localization in bio-nanothings communication
  networks,'' \emph{IEEE Transactions on Information Forensics and Security},
  vol.~11, no.~4, pp. 665--676, 2015.

\bibitem{rahman2017physical}
M.~M.~U. Rahman, Q.~H. Abbasi, N.~Chopra, K.~Qaraqe, and A.~Alomainy,
  ``Physical layer authentication in nano networks at terahertz frequencies for
  biomedical applications,'' \emph{IEEE Access}, vol.~5, pp. 7808--7815, 2017.

\bibitem{felicetti2017congestion}
L.~Felicetti, M.~Femminella, and G.~Reali, ``Congestion control in molecular
  cyber-physical systems,'' \emph{IEEE Access}, vol.~5, pp. 10\,000--10\,011,
  2017.

\bibitem{nakano2010sequence}
T.~Nakano and M.~Moore, ``In-sequence molecule delivery over an aqueous
  medium,'' \emph{Nano Communication Networks}, vol.~1, no.~3, pp. 181--188,
  2010.

\bibitem{sharma2012security}
G.~Sharma, S.~Bala, and A.~K. Verma, ``Security frameworks for wireless sensor
  networks-review,'' \emph{Procedia Technology}, vol.~6, pp. 978--987, 2012.

\bibitem{zhang2014trustworthiness}
M.~Zhang, A.~Raghunathan, and N.~K. Jha, ``Trustworthiness of medical devices
  and body area networks,'' \emph{Proceedings of the IEEE}, vol. 102, no.~8,
  pp. 1174--1188, 2014.

\bibitem{ye2014efficient}
N.~Ye, Y.~Zhu, R.-c. Wang, R.~Malekian, and L.~Qiao-min, ``An efficient
  authentication and access control scheme for perception layer of internet of
  things,'' \emph{Appl. Math}, vol.~8, no.~4, pp. 1--59, 2014.

\bibitem{singh2017advanced}
S.~Singh, P.~K. Sharma, S.~Y. Moon, and J.~H. Park, ``Advanced lightweight
  encryption algorithms for iot devices: survey, challenges and solutions,''
  \emph{Journal of Ambient Intelligence and Humanized Computing}, vol.~4, pp.
  1--18, 2017.

\bibitem{belkhouja2018light}
T.~Belkhouja, A.~Mohamedz, A.~K. Al-Aliz, X.~Dux, and M.~Guizani,
  ``{Light-Weight Solution to Defend Implantable Medical Devices against
  Man-In-The-Middle Attack},'' in \emph{IEEE Global Communications Conference
  (GLOBECOM)}.\hskip 1em plus 0.5em minus 0.4em\relax Abu Dhabi, United Arab
  Emirates, United Arab Emirates: IEEE, 2018, pp. 1--5.

\bibitem{siddiqi2019towards}
M.~A. Siddiqi and C.~Strydis, ``Towards realistic battery-dos protection of
  implantable medical devices,'' in \emph{Proceedings of the 16th ACM
  international conference on computing frontiers}.\hskip 1em plus 0.5em minus
  0.4em\relax Alghero, Italy: Association for Computing Machinery, 2019, pp.
  42--49.

\bibitem{anhalt2010insulin}
H.~Anhalt and N.~J. Bohannon, ``Insulin patch pumps: their development and
  future in closed-loop systems,'' \emph{Diabetes technology \& therapeutics},
  vol.~12, no.~S1, pp. S--51, 2010.

\bibitem{al2017survey}
S.~Al-Janabi, I.~Al-Shourbaji, M.~Shojafar, and S.~Shamshirband, ``Survey of
  main challenges (security and privacy) in wireless body area networks for
  healthcare applications,'' \emph{Egyptian Informatics Journal}, vol.~18,
  no.~2, pp. 113--122, 2017.

\bibitem{kintzlinger2019keep}
M.~Kintzlinger and N.~Nissim, ``Keep an eye on your personal belongings! the
  security of personal medical devices and their ecosystems,'' \emph{Journal of
  biomedical informatics}, vol.~95, p. 103233, 2019.

\bibitem{wazid2019iomt}
M.~Wazid, A.~K. Das, J.~J. Rodrigues, S.~Shetty, and Y.~Park, ``Iomt malware
  detection approaches: Analysis and research challenges,'' \emph{IEEE Access},
  vol.~7, pp. 182\,459--182\,476, 2019.

\bibitem{gollakota2011they}
S.~Gollakota, H.~Hassanieh, B.~Ransford, D.~Katabi, and K.~Fu, ``{They can hear
  your heartbeats: non-invasive security for implantable medical devices},'' in
  \emph{SIGCOMM '11 Proceedings of the ACM SIGCOMM}.\hskip 1em plus 0.5em minus
  0.4em\relax Toronto: ACM, 2011, pp. 2--13.

\bibitem{lee2008elliptic}
Y.~K. Lee, K.~Sakiyama, L.~Batina, and I.~Verbauwhede, ``Elliptic-curve-based
  security processor for rfid,'' \emph{IEEE Transactions on Computers},
  vol.~57, no.~11, pp. 1514--1527, 2008.

\bibitem{choi2018system}
S.-K. Choi, C.-H. Yang, and J.~Kwak, ``System hardening and security monitoring
  for iot devices to mitigate iot security vulnerabilities and threats.''
  \emph{KSII Transactions on Internet \& Information Systems}, vol.~12, no.~2,
  2018.

\bibitem{son2010u}
S.~Son, K.~Lee, D.~Won, and S.~Kim, ``U-healthcare system protecting privacy
  based on cloaker,'' in \emph{2010 IEEE International Conference on
  Bioinformatics and Biomedicine Workshops (BIBMW)}.\hskip 1em plus 0.5em minus
  0.4em\relax Hong, Kong, China: IEEE, 2010, pp. 417--423.

\bibitem{rieback2005rfid}
M.~R. Rieback, B.~Crispo, and A.~S. Tanenbaum, ``Rfid guardian: A
  battery-powered mobile device for rfid privacy management,'' in
  \emph{Australasian Conference on Information Security and Privacy}.\hskip 1em
  plus 0.5em minus 0.4em\relax Brisbane, Australia: Springer, 2005, pp.
  184--194.

\bibitem{hei2013resource}
X.~Hei and X.~Du, ``The resource depletion attack and defense scheme,'' in
  \emph{Security for Wireless Implantable Medical Devices}.\hskip 1em plus
  0.5em minus 0.4em\relax ,: Springer, 2013, pp. 9--18.

\bibitem{umba2019review}
S.~M.~W. Umba, A.~M. Abu-Mahfouz, T.~Ramotsoela, and G.~P. Hancke, ``A review
  of artificial intelligence based intrusion detection for software-defined
  wireless sensor networks,'' in \emph{2019 IEEE 28th International Symposium
  on Industrial Electronics (ISIE)}.\hskip 1em plus 0.5em minus 0.4em\relax
  ancouver, BC, Canada: IEEE, 2019, pp. 1277--1282.

\bibitem{rahim2013intrusion}
A.~Rahim and P.~Malone, ``Intrusion detection system for wireless nano sensor
  networks,'' in \emph{8th International Conference for Internet Technology and
  Secured Transactions (ICITST-2013)}.\hskip 1em plus 0.5em minus 0.4em\relax
  London, UK: IEEE, 2013, pp. 327--330.

\bibitem{aragon2018ace}
S.~Aragon, M.~Tiloca, M.~Maass, M.~Hollick, and S.~Raza, ``Ace of spades in the
  iot security game: A flexible ipsec security profile for access control,'' in
  \emph{2018 IEEE Conference on Communications and Network Security
  (CNS)}.\hskip 1em plus 0.5em minus 0.4em\relax IEEE, 2018, pp. 1--9.

\bibitem{stelzner2017function}
M.~Stelzner, F.~Dressler, and S.~Fischer, ``Function centric nano-networking:
  Addressing nano machines in a medical application scenario,'' \emph{Nano
  communication networks}, vol.~14, pp. 29--39, 2017.

\bibitem{liu2019carbon}
L.~Liu, L.~Ding, D.~Zhong, J.~Han, S.~Wang, Q.~Meng, C.~Qiu, X.~Zhang, L.-M.
  Peng, and Z.~Zhang, ``Carbon nanotube complementary gigahertz integrated
  circuits and their applications on wireless sensor interface systems,''
  \emph{ACS nano}, vol.~13, no.~2, pp. 2526--2535, 2019.

\bibitem{wang2019advanced}
C.~Wang, K.~Xia, H.~Wang, X.~Liang, Z.~Yin, and Y.~Zhang, ``Advanced carbon for
  flexible and wearable electronics,'' \emph{Advanced materials}, vol.~31,
  no.~9, p. 1801072, 2019.

\bibitem{huang2019graphene}
H.~Huang, S.~Su, N.~Wu, H.~Wan, S.~Wan, H.~Bi, and L.~Sun, ``Graphene-based
  sensors for human health monitoring,'' \emph{Frontiers in chemistry}, vol.~7,
  p. 399, 2019.

\bibitem{kabiri2017graphene}
S.~Kabiri~Ameri, R.~Ho, H.~Jang, L.~Tao, Y.~Wang, L.~Wang, D.~M. Schnyer,
  D.~Akinwande, and N.~Lu, ``Graphene electronic tattoo sensors,'' \emph{ACS
  nano}, vol.~11, no.~8, pp. 7634--7641, 2017.

\bibitem{cordier2008self}
P.~Cordier, F.~Tournilhac, C.~Souli{\'e}-Ziakovic, and L.~Leibler,
  ``Self-healing and thermoreversible rubber from supramolecular assembly,''
  \emph{Nature}, vol. 451, no. 7181, pp. 977--980, 2008.

\bibitem{tee2012electrically}
B.~C. Tee, C.~Wang, R.~Allen, and Z.~Bao, ``An electrically and mechanically
  self-healing composite with pressure-and flexion-sensitive properties for
  electronic skin applications,'' \emph{Nature nanotechnology}, vol.~7, no.~12,
  pp. 825--832, 2012.

\bibitem{yang2015flexible}
Y.~Yang, B.~Zhu, D.~Yin, J.~Wei, Z.~Wang, R.~Xiong, J.~Shi, Z.~Liu, and Q.~Lei,
  ``Flexible self-healing nanocomposites for recoverable motion sensor,''
  \emph{Nano Energy}, vol.~17, pp. 1--9, 2015.

\bibitem{bandodkar2016wearable}
A.~J. Bandodkar, I.~Jeerapan, and J.~Wang, ``Wearable chemical sensors: Present
  challenges and future prospects,'' \emph{Acs Sensors}, vol.~1, no.~5, pp.
  464--482, 2016.

\bibitem{bandodkar2015self}
A.~J. Bandodkar, V.~Mohan, C.~S. L{\'o}pez, J.~Ram{\'\i}rez, and J.~Wang,
  ``Self-healing inks for autonomous repair of printable electrochemical
  devices,'' \emph{Advanced Electronic Materials}, vol.~1, no.~12, p. 1500289,
  2015.

\bibitem{rizwan2018review}
A.~Rizwan, A.~Zoha, R.~Zhang, W.~Ahmad, K.~Arshad, N.~A. Ali, A.~Alomainy,
  M.~A. Imran, and Q.~H. Abbasi, ``A review on the role of nano-communication
  in future healthcare systems: A big data analytics perspective,'' \emph{IEEE
  Access}, vol.~6, pp. 41\,903--41\,920, 2018.

\bibitem{shirazi2012protocol}
A.~Z. Shirazi, S.~M. Mazinani, and S.~K. Eghbal, ``Protocol stack for nano
  networks,'' in \emph{2012 International Symposium on Computer, Consumer and
  Control}.\hskip 1em plus 0.5em minus 0.4em\relax Taichung, Taiwan: IEEE,
  2012, pp. 849--853.

\bibitem{doukas2011managing}
C.~Doukas and I.~Maglogiannis, ``Managing wearable sensor data through cloud
  computing,'' in \emph{2011 IEEE Third International Conference on Cloud
  Computing Technology and Science}.\hskip 1em plus 0.5em minus 0.4em\relax
  Athen, Greece: IEEE, 2011, pp. 440--445.

\bibitem{liu2015external}
C.~Liu, C.~Yang, X.~Zhang, and J.~Chen, ``External integrity verification for
  outsourced big data in cloud and iot: A big picture,'' \emph{Future
  generation computer systems}, vol.~49, pp. 58--67, 2015.

\bibitem{zhang2015secure}
C.~Zhang, C.~Li, and J.~Zhang, ``A secure privacy-preserving data aggregation
  model in wearable wireless sensor networks,'' \emph{Journal of Electrical and
  Computer Engineering}, vol. 2015, pp. 1--9, 2015.

\bibitem{lin2016differential}
C.~Lin, P.~Wang, H.~Song, Y.~Zhou, Q.~Liu, and G.~Wu, ``A differential privacy
  protection scheme for sensitive big data in body sensor networks,''
  \emph{Annals of Telecommunications}, vol.~71, no. 9-10, pp. 465--475, 2016.

\bibitem{dressler2012towards}
F.~Dressler and F.~Kargl, ``Towards security in nano-communication: Challenges
  and opportunities,'' \emph{Nano communication networks}, vol.~3, no.~3, pp.
  151--160, 2012.

\bibitem{Dhawan2012SecureDT}
S.~Dhawan and A.~Saini, ``Secure data transmission techniques based on dna
  cryptography,'' \emph{International Journal of Emerging Technologies in
  Computational and Applied Sciences}, vol.~2, pp. 95--100, 2012.

\bibitem{cui2008encryption}
G.~Cui, L.~Qin, Y.~Wang, and X.~Zhang, ``An encryption scheme using dna
  technology,'' in \emph{2008 3rd International Conference on Bio-Inspired
  Computing: Theories and Applications}.\hskip 1em plus 0.5em minus 0.4em\relax
  ,: IEEE, 2008, pp. 37--42.

\bibitem{rusia2014review}
M.~Rusia and R.~H. Makwana, ``Review on dna based encryption algorithm for text
  and image data,'' \emph{International Journal of Engineering Research \&
  Technology (IJERT)}, vol.~3, no.~1, pp. 3182--3186, 2014.

\bibitem{zhang2012research}
Y.~Zhang and L.~B. Fu, ``Research on dna cryptography,'' in \emph{Applied
  cryptography and network security}, vol. 357.\hskip 1em plus 0.5em minus
  0.4em\relax ,: Rijeka, 2012, pp. 10--5772.

\bibitem{jacob2013dna}
G.~Jacob, ``Dna based cryptography: An overview and analysis,''
  \emph{International Journal of Emerging Sciences}, vol.~3, no.~1, p.~36,
  2013.

\bibitem{jacob2013encryption}
G.~Jacob and A.~Murugan, ``An encryption scheme with dna technology and jpeg
  zigzag coding for secure transmission of images,'' \emph{arXiv preprint
  arXiv:1305.1270}, 2013.

\bibitem{gao2008new}
T.~Gao and Z.~Chen, ``A new image encryption algorithm based on hyper-chaos,''
  \emph{Physics Letters A}, vol. 372, no.~4, pp. 394--400, 2008.

\bibitem{gehani2003dna}
A.~Gehani, T.~LaBean, and J.~Reif, ``Dna-based cryptography,'' in \emph{Aspects
  of Molecular Computing}.\hskip 1em plus 0.5em minus 0.4em\relax ,: Springer,
  2003, pp. 167--188.

\bibitem{malasri2009design}
K.~Malasri and L.~Wang, ``Design and implementation of a securewireless
  mote-based medical sensor network,'' \emph{Sensors}, vol.~9, no.~8, pp.
  6273--6297, 2009.

\bibitem{seo2009tinyecck16}
S.~C. Seo, D.-G. Han, and S.~Hong, ``Tinyecck16: An efficient field
  multiplication algorithm on 16-bit environment and its application to tmote
  sky sensor motes,'' \emph{IEICE transactions on information and systems},
  vol.~92, no.~5, pp. 918--928, 2009.

\bibitem{malasri2006snap}
K.~Malasri and L.~Wang, ``Snap: an architecture for secure medical sensor
  networks,'' in \emph{2006 2nd IEEE Workshop on Wireless Mesh Networks}.\hskip
  1em plus 0.5em minus 0.4em\relax Reston, VA, USA: IEEE, 2006, pp. 160--162.

\bibitem{fernandez2019pre}
T.~M. Fern{\'a}ndez-Caram{\'e}s, ``From pre-quantum to post-quantum iot
  security: A survey on quantum-resistant cryptosystems for the internet of
  things,'' \emph{IEEE Internet of Things Journal}, vol.~7, no.~7, pp.
  6457--6480, 2019.

\bibitem{dressler2010bio}
F.~Dressler and O.~B. Akan, ``Bio-inspired networking: from theory to
  practice,'' \emph{IEEE Communications Magazine}, vol.~48, no.~11, pp.
  176--183, 2010.

\bibitem{dressler2010survey}
{Dressler, Falko and Akan, Ozgur B}, ``A survey on bio-inspired networking,''
  \emph{Computer networks}, vol.~54, no.~6, pp. 881--900, 2010.

\bibitem{rauf2018taxonomy}
U.~Rauf, ``A taxonomy of bio-inspired cyber security approaches: existing
  techniques and future directions,'' \emph{Arabian Journal for Science and
  Engineering}, vol.~43, no.~12, pp. 6693--6708, 2018.

\bibitem{dressler2005bio}
F.~Dressler, ``Bio-inspired mechanisms for efficient and adaptive network
  security mechanisms,'' in \emph{Service Management and Self-Organization in
  IP-based Networks}, ser. Dagstuhl Seminar Proceedings, M.~Bossardt, G.~Carle,
  D.~Hutchison, H.~de~Meer, and B.~Plattner, Eds., no. 04411.\hskip 1em plus
  0.5em minus 0.4em\relax Dagstuhl, Germany: Internationales Begegnungs- und
  Forschungszentrum f{\"u}r Informatik (IBFI), Schloss Dagstuhl, Germany, 2005,
  pp. 1--6.

\bibitem{noeparast2012cognitive}
E.~B. Noeparast and T.~Banirostam, ``A cognitive model of immune system for
  increasing security in distributed systems,'' in \emph{2012 UKSim 14th
  International Conference on Computer Modelling and Simulation}.\hskip 1em
  plus 0.5em minus 0.4em\relax Cambridge, UK: IEEE, 2012, pp. 181--186.

\bibitem{fulp2015evolutionary}
E.~W. Fulp, H.~D. Gage, D.~J. John, M.~R. McNiece, W.~H. Turkett, and X.~Zhou,
  ``An evolutionary strategy for resilient cyber defense,'' in \emph{2015 IEEE
  Global Communications Conference (GLOBECOM)}.\hskip 1em plus 0.5em minus
  0.4em\relax ,: IEEE, 2015, pp. 1--6.

\bibitem{bitam2016bio}
S.~Bitam, S.~Zeadally, and A.~Mellouk, ``Bio-inspired cybersecurity for
  wireless sensor networks,'' \emph{IEEE Communications Magazine}, vol.~54,
  no.~6, pp. 68--74, 2016.

\end{thebibliography}

\begin{IEEEbiography}[{\includegraphics[width=1in,height=1.5in,clip,keepaspectratio]{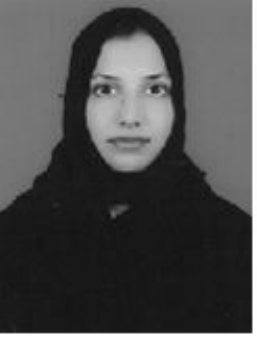}}]{SIDRA ZAFAR} is a Ph.D. scholar at Lahore College for Women University (LCWU), Lahore, Pakistan. She received her MS (Computer Science) from LCWU in 2012. She did her BS (Software Engineering) from International Islamic University Islamabad, Pakistan. She has presented her work at the International Conference. Her research interests include the Internet of Nano Things, Nano Communication Network, Security in Nanonetworks, Access Control.
\end{IEEEbiography}

\begin{IEEEbiography}[{\includegraphics[width=1in,height=1.5in,clip,keepaspectratio]{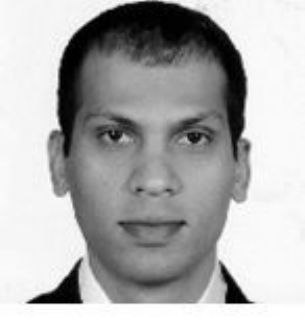}}]{Mohsin Nazir} did his Ph.D. from Asian Institute of Technology (AIT), Thailand in 2011. He is working in Lahore College for Women University as Associate Professor since 2012. He is a member of various renowned research societies. His research interests include Cooperative Cognitive Wireless Communications, Game Theory, Evolutionary Game Theory, Graph Theory, Smart Energy Grids, Nano Communications, Cloud Computing, QoS, and 4G network.
\end{IEEEbiography}

\begin{IEEEbiography}[{\includegraphics[width=1in,height=1.5in,clip,keepaspectratio]{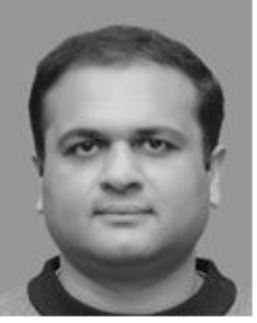}}]{Taimur Bakhshi} holds a Ph.D. in computer science, specializing in software defined networking from the Department of Computing \& Mathematics, University of Plymouth, UK. His academic credentials also include an M.Sc. in network systems engineering completed with distinction from University of Plymouth and a B.Sc. in electronic engineering from Pakistan. Prior to starting his academic career, Dr. Bakhshi worked for more than five years in IT incident management and network engineering roles for BlackBerry Ltd. and Orange SA (France Telecom). He holds Associate Fellowship of the Higher Education Academy (HEA), U.K and is a professional member of the British Computer Society (BCS) and the Institute of Electrical and Electronic Engineers (IEEE). His present research focuses on software defined networking, network measurements, Internet traffic classification, and systems security.
\end{IEEEbiography}

\begin{IEEEbiography}[{\includegraphics[width=1in,height=1.5in,clip,keepaspectratio]{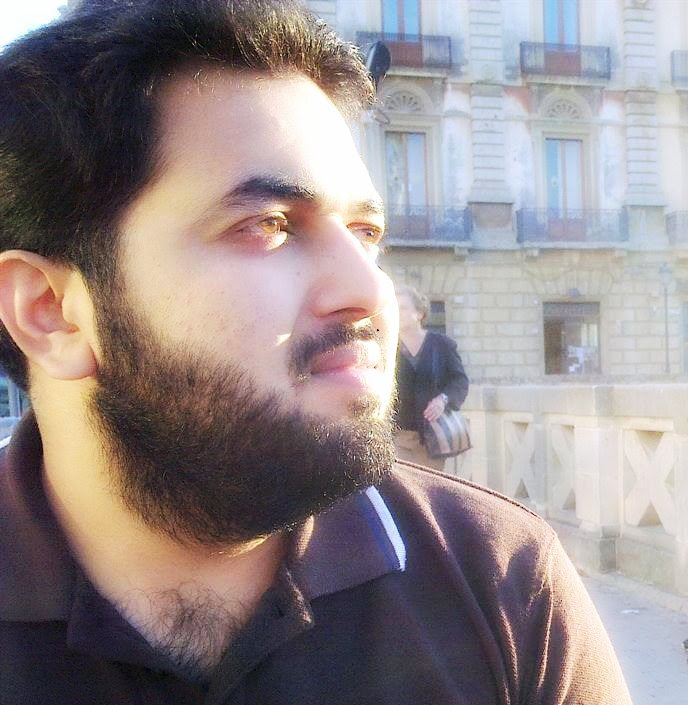}}]{Hasan Ali Khattak} received his Ph.D. in Electrical and Computer Engineering degree from Politecnico di Bari, Bari, Italy on April 2015, Master's degree in Information Engineering from Politecnico di Torino, Torino, Italy, in 2011, and B.CS. degree in Computer Science from the University of Peshawar, Peshawar, Pakistan in 2006. He is currently serving as Associate Professor - School of Electrical Engineering and Computer Science, National University of Sciences and Technology - Pakistan since October 2020. His current research interests focus on Future Internet Architectures such as the Web of Things and leveraging Data Sciences and Social Engineering for Future Smart Cities. 
\end{IEEEbiography}

\begin{IEEEbiography}[{\includegraphics[width=1in,height=1.5in,clip,keepaspectratio]{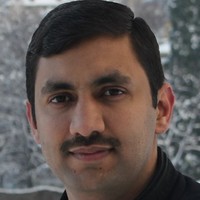}}]{Sarmadullah Khan} received the M.Sc. and Ph.D. degrees from Politecnico di Torino, Italy, in 2009 and 2013, respectively, all in electronics and telecommunication engineering. He was an Assistant Professor with the CECOS University of IT and Emerging Science, Peshawar, Pakistan, from 2013 to 2017. He is currently a Lecturer with De Montfort University, U.K. He is also a Program Leader of M.Sc. cyber technology and its pathways. Dr. Khan is a member of academic board of studies, curriculum revision committee, program management board, and school management group. His current research interests include the Internet of things security, content centric networks security, cryptographic key establishment and management, and wireless sensor networks security. He is a guest editor of two special issues in the journal of wireless communication and mobile computing.
\end{IEEEbiography}

\begin{IEEEbiography}[{\includegraphics[width=1in,height=1.5in,clip,keepaspectratio]{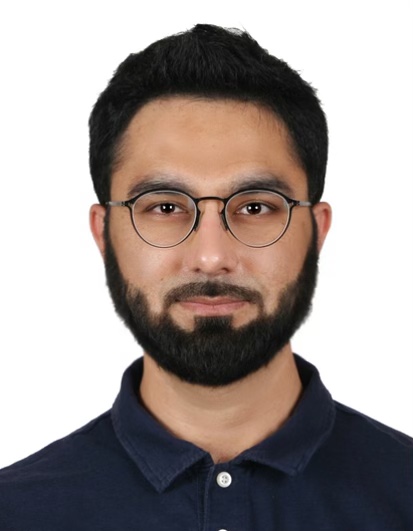}}]{Muhammad Bilal} received the B.Sc. degree in computer systems engineering from the University of Engineering and Technology, Peshawar, Pakistan, in 2008, the M.S. degree in computer engineering from the Chosun University, Gwangju, South Korea, in 2012, and the Ph.D. degree in information and communication network engineering from the School of Electronics and Telecommunications Research Institute (ETRI), Korea University of Science and Technology, in 2017.  He is an Assistant Professor with the Division of Computer and Electronic Systems Engineering, Hankuk University of Foreign Studies, Yongin, South Korea. Prior to joining Hankuk University of Foreign Studies, he was a Postdoctoral Research Fellow at Smart Quantum Communication Center, Korea University, Seoul, South Korea, in 2017. His research interests include design and analysis of network protocols, network architecture, network security, IoT, named data networking, Blockchain, cryptology, and Neural Networks. He is an editor of IEEE Future Directions Ethics and Policy in Technology Newsletter and IEEE Internet Policy Newsletter.
\end{IEEEbiography}

\begin{IEEEbiography}[{\includegraphics[width=1in,height=1.5in,clip,keepaspectratio]{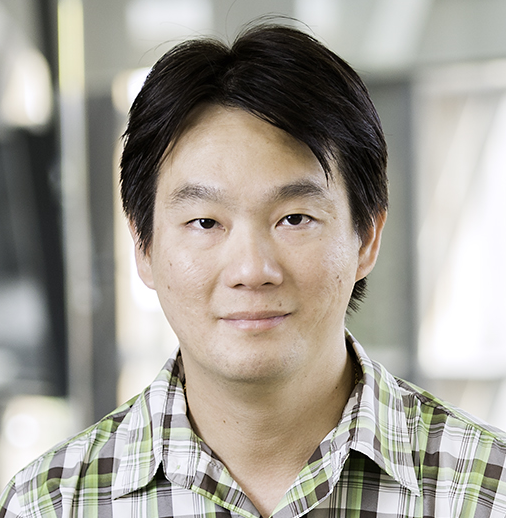}}]{Kim-Kwang Raymond Choo}  (Senior Member, IEEE) received the Ph.D. degree in information security from the Queensland University of Technology, Australia, in 2006. He currently holds the Cloud Technology Endowed Professorship at The University of Texas at San Antonio (UTSA). He is the founding co-editor-in-chief of ACM's \emph{Distributed Ledger Technologies: Research \& Practice}, and founding chair of IEEE Technology and Engineering Management Society's Technical Committee on Blockchain and Distributed Ledger Technologies. He is also the recipient of the 2019 IEEE Technical Committee on Scalable Computing Award for Excellence in Scalable Computing (Middle Career Researcher), the 2018 UTSA College of Business Col. Jean Piccione and Lt. Col. Philip Piccione Endowed Research Award for Tenured Faculty, the Outstanding Associate Editor of 2018 for \emph{IEEE Access}, the British Computer Society's 2019 Wilkes Award Runner-up, the 2014 Highly Commended Award by the Australia New Zealand Policing Advisory Agency, the Fulbright Scholarship in 2009, the 2008 Australia Day Achievement Medallion, and the British Computer Society's Wilkes Award in 2008. He has  received best paper awards from the \emph{IEEE Systems Journal} in 2021, \emph{IEEE Consumer Electronics Magazine} for 2020, \emph{EURASIP Journal on Wireless Communications and Networking} in 2019, IEEE TrustCom 2018, and ESORICS 2015; the Korea Information Processing Society's \emph{Journal of Information Processing Systems} Outstanding Research Award (Most-cited Paper) for 2020 and Survey Paper Award (Gold) in 2019; the IEEE Blockchain 2019 Outstanding Paper Award; and Best Student Paper Awards from Inscrypt 2019 and ACISP 2005.
\end{IEEEbiography}

\begin{IEEEbiography}[{\includegraphics[width=1in,height=1.5in,clip,keepaspectratio]{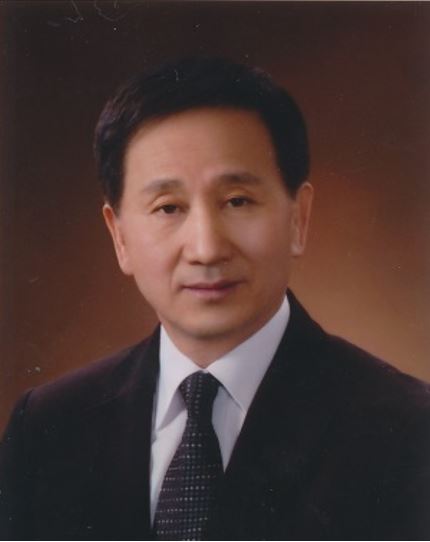}}]{KYUNG-SUP KWAK} (Life Senior Member, IEEE) received the Ph.D. degree from the University of California. He was with Hughes Network Systems and IBM Network Analysis Center, USA. He was with Inha University, South Korea, as a Professor. He was also the Dean of the Graduate School of Information Technology and Telecommunications and the Director of the UWB Wireless Communications Research Center. In 2008, he was an Inha Fellow Professor (IFP). He is currently an Inha Hanlim Fellow Professor. His research interests include UWB radio systems, wireless body area network and nano and molecular communications. In 2006, he was the President of the Korean Institute of Communication Sciences (KICS) and the Korea Institute of Intelligent Transport Systems (KITS) in 2009. He received the official commendations for achievements of UWB radio technology R/D from the Korean President in 2009.
\end{IEEEbiography}

\begin{IEEEbiography}[{\includegraphics[width=1in,height=1.5in,clip,keepaspectratio]{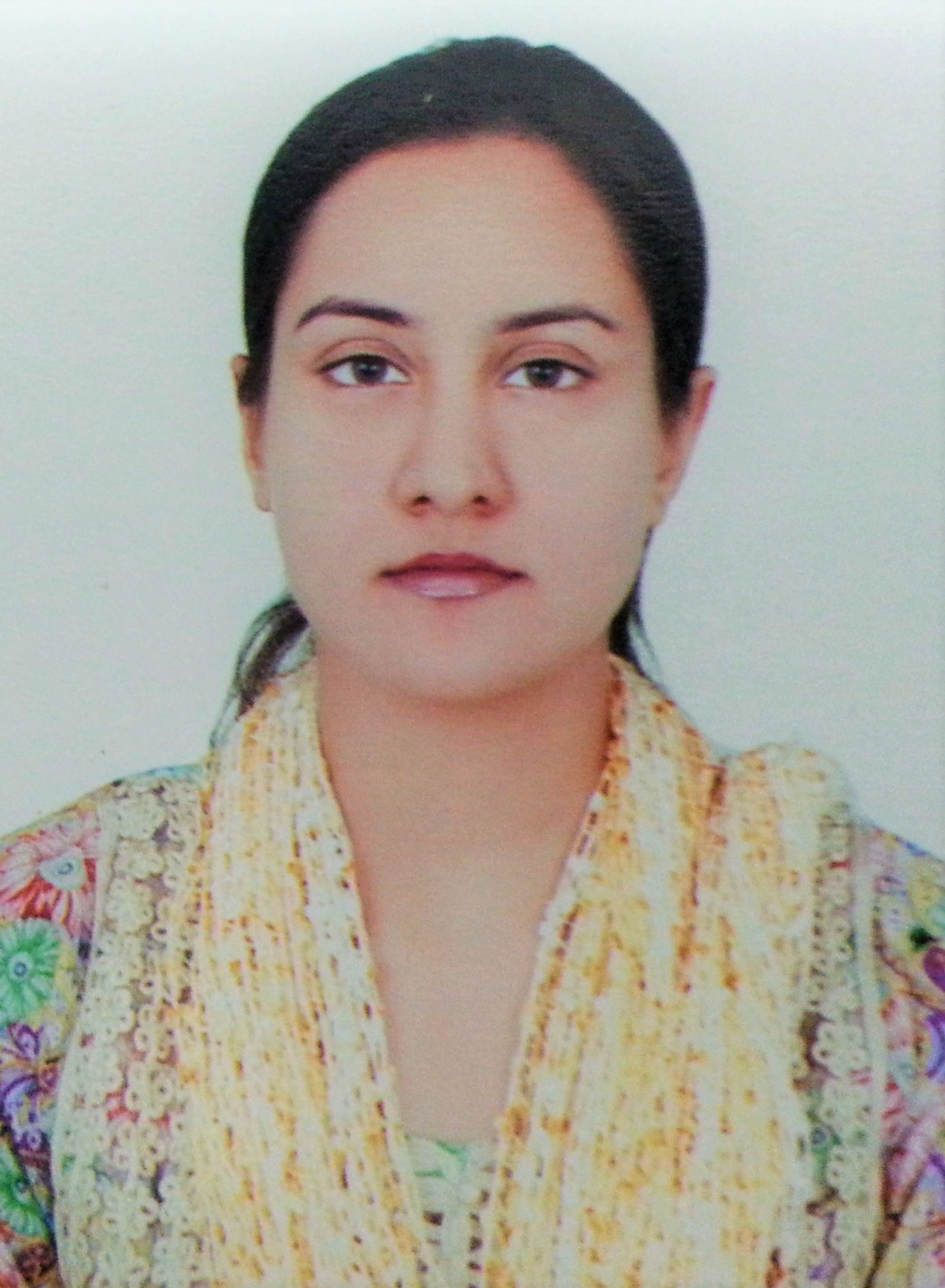}}]{ANEEQA SABAH} is currently working as an Assistant professor in physics department LCWU. She has been Awarded for Overseas Scholarship for PhD degree in the field of Nanotechnology by HEC (Higher education commission), Pakistan. It was a successful project. She has been supervising the several projects on BS, MS, and PhD level, based on nano fabrication, nano films and membranes, quantum dots, hybrid materials, spectroscopy, composites, and metal doped sensing. She has published many research papers in high impact factor journals, journal of physical chemistry and ACS, and involved in participating and organizing seminar and conferences in respective departments. Her expertise is nano synthesis, self-assembly and organization, colloids, one dimensional nano materials, green chemistry, and quantum dots. 
\end{IEEEbiography}

\EOD
\end{document}